\documentclass[a4paper,11pt]{article}
\pdfoutput=1 

\usepackage{jheppub} 

\usepackage[T1]{fontenc} 
\usepackage[utf8]{inputenc}

\newcommand{\AdS}{$\text{AdS}_{3}$ }

\newcommand{\arctanh}{\text{arctanh}}

\newcommand{\dd}[1]{\mathrm{d} #1}
\newcommand{\sh}[1]{\sinh\left[#1\right]}
\newcommand{\ch}[1]{\cosh\left[#1\right]}

\newcommand{\Poincare}{Poincar\'e }

\newcommand{\be}{\begin{eqnarray}}
\newcommand{\ee}{\end{eqnarray}}

\def\>{\rangle}
\def\<{\langle}

\newcommand{\executeiffilenewer}[3]{%
	\ifnum\pdfstrcmp{\pdffilemoddate{#1}}%
	{\pdffilemoddate{#2}}>0%
	{\immediate\write18{#3}}\fi%
}

\graphicspath{{pics/}}

\title{\boldmath Time evolution of entanglement for holographic steady state formation }

\author{Johanna Erdmenger$^{\dagger +}$, Daniel Fern\'andez$^{\dagger\lozenge}$, Mario Flory$^{\dagger\blacklozenge}$, Eugenio Meg\'ias$^{\dagger\Box}$, Ann-Kathrin Straub$^{\dagger}$ and Piotr Witkowski$^{\dagger \star}$}

\affiliation{$^{\dagger}$Max-Planck-Institut f\"ur Physik (Werner-Heisenberg-Institut), \\
F\"{o}hringer Ring 6, 80805 M\"unchen, Germany}
\affiliation{$^+$Institut f\"ur Theoretische Physik und Astrophysik, Julius-Maximilians-Universit\"at W\"urzburg, \\ 
Am Hubland, 97074 W\"urzburg, Germany} 
\affiliation{$^{\lozenge}$University of Iceland, Science Institute, Dunhaga 3, 107 Reykjav\'ik, Iceland}
\affiliation{$^{\blacklozenge}$Institute of Physics, Jagiellonian University, \\
\L{}ojasiewicza 11, 30-348 Krak\'ow, Poland}
\affiliation{$^{\Box}$Departamento de F\'{\i}sica Te\'orica, Universidad del Pa\'{\i}s Vasco UPV/EHU, \\
Apartado 644,  48080 Bilbao, Spain}
\affiliation{$^\star$Max-Planck-Institut f\"ur Physik komplexer Systeme, \\
N\"othnitzer Stra{\ss}e 38, D-01187 Dresden, Germany} 

\emailAdd{jke@mpp.mpg.de}
\emailAdd{danielf@mpp.mpg.de}
\emailAdd{mflory@mpp.mpg.de}
\emailAdd{emegias@mpp.mpg.de}
\emailAdd{astraub@mpp.mpg.de}
\emailAdd{piotr@pks.mpg.de}

\abstract{Within gauge/gravity duality, we consider the local quench-like time evolution
  obtained by joining two 1+1-dimensional heat
  baths at different temperatures at time $t=0$. A steady state
  forms and expands in space. For the 2+1-dimensional 
gravity dual, we find that the ``shockwaves'' expanding the
steady-state region  are of spacelike nature in the bulk despite being
null at the boundary.  However, they do not transport information.
  Moreover, by adapting the time-dependent Hubeny-Rangamani-Takayanagi
  prescription, we holographically calculate the
  entanglement entropy and also the  mutual information for different entangling regions. For general temperatures, we find that the entanglement entropy increase rate satisfies the same bound as in the `entanglement tsunami' setups. For small temperatures of the two baths, we derive an
  analytical formula for the time dependence of the entanglement
  entropy. This replaces the entanglement tsunami-like behaviour seen for high temperatures. 
  Finally, we check that strong subadditivity holds in
  this time-dependent system, as well as further
  more general entanglement inequalities for five or more regions
  recently derived for the static case.
}


\preprint{}

\begin{document}

\hfill MPP-2017-87

\maketitle


\section{Introduction}
\label{sec:Introduction}

In recent years, the application of holography to the study of far-from-equilibrium physics has been successfully implemented (see \cite{Danielsson:1999fa,Grumiller:2008va,Bhattacharyya:2008xc,Chesler:2008hg} for early work and  \cite{Berti:2016rij,Zhang:2016coy,Cardoso:2012qm} for reviews). The usefulness of this approach lies in the fact that real-time dynamics of strongly correlated systems are directly computable, and its collective responses can easily be found. This provides a new approach to studying quantum quenches in strongly coupled systems. Such quenches can be roughly divided into global quenches and local  quenches. In quantum field theory, `global' refers to changes of the Lagrangian and `local' to changes of the ground state. In holography however, `global' quenches refer to the evolution of the entire gravity dual from an initial configuration, while `local' holographic quenches involve a sudden change of the geometry at a region localized in space.

Following \cite{AbajoArrastia:2010yt,Balasubramanian:2010ce}, important results on holographic global quenches have been obtained using the AdS-Vaidya metric, see for example \cite{Balasubramanian:2011ur,Balasubramanian:2011at,Liu:2013iza,Li:2013sia,Liu:2013qca,Alishahiha:2014cwa,Kundu:2016cgh}. Quenches of a local type can be holographically studied in a variety of manners. These include investigating sudden local excitations of bulk fields  \cite{Caputa:2014vaa,Caputa:2014eta,Caputa:2015waa,Rangamani:2015agy,David:2016pzn,Rozali:2017bco}, specific bulk spacetimes describing a massive point particle dropping from the boundary into the bulk \cite{Nozaki:2013wia,Jahn:2017xsg}, or the sudden joining of two previously separated boundary CFTs (BCFTs) \cite{Astaneh:2014fga}. Local quenches can also be naturally studied in holographic models of defect or interface CFTs \cite{Erdmenger:2016msd}. Formulae for the evolution of holographic entanglement entropy were recently  used in \cite{Ageev:2017wet} to obtain an explicit description for different regimes of a holographic heating process. Analytic progress in this direction was made in \cite{Pedraza:2014moa}, where the late-time behavior of two-point functions, Wilson loops and entanglement entropy was studied perturbatively in a boost-invariant system. For calculating these correlations, a useful approach is to consider two-point functions given by lengths of spatial geodesics anchored at the boundary. In particular, \cite{Balasubramanian:2011ur} gives an early comprehensive study of correlations in the geodesic approximation, and \cite{Ecker:2016thn} contains a study in a background of colliding shockwaves.

An important conclusion  \cite{Heller:2012km,Heller:2011ju,Chesler:2010bi,Fernandez:2014fua} is the fact that a transition to a hydrodynamic regime can take place very early in the time evolution, before reaching thermodynamic equilibrium. This is also the case in non-conformal systems \cite{Attems:2017zam}. In some cases, equilibrium is never reached, and instead a steady state is obtained at late times. Such a state involves a constant flow of energy or charge between two reservoirs \cite{Bernard:2012je,Bernard:2013bqa,Chang:2013gba,Bernard2015,Bernard:2016nci,Hoogeveen:2014bqa}. The study of steady states is particularly interesting in the presence of emergent collective behavior, since it provides insight into the interplay between quantum effects and out-of-equilibrium physics. As an example for the formation of  a thermal steady state,  we consider in this paper, following the work given above,  the time evolution of  a 1+1-dimensional field theory system which is initially separated into two space regions. These are initially independently prepared in thermal equilibrium at temperatures $T_\mathrm{L}$ and $T_\mathrm{R}$, respectively. At time $t=0$, we bring the two space regions into contact at $x=0$, which gives rise to an initial state with temperature profile 
\begin{align}
T(t=0,x)= T_\mathrm{L}\,\theta(-x)+T_\mathrm{R}\,\theta(x) \, ,
\label{introduction: initial temperature profile}
\end{align}
and let the system evolve.  Around $x=0$, a growing region with a constant energy flow $\left\langle J_E\right\rangle\neq0$, the steady state, develops. Within field theory, this was discussed by Bernard and Doyon in \cite{Bernard:2012je,Bernard:2013bqa,Bernard2015}. In their work, they showed that  for late times, the steady-state region can asymptotically be described by a thermal distribution at shifted temperature. Such a  local quench-like system can be modeled for example as in \cite{Astaneh:2014fga,Hoogeveen:2014bqa}, where two independently thermalised BCFTs are joined at $t=0$. In this work, in contrast, we will follow \cite{Bhaseen:nature} and utilise a different setup where equation \eqref{introduction: initial temperature profile} holds at $t=0$, and a steady state forms when evolving the state both forwards and backwards in time. However, we will only consider the regime $t\geq0$ as physically interesting.

\begin{figure}[h]
\begin{tabular}{ccc}
\hspace{-0.4cm}\includegraphics[width=63mm]{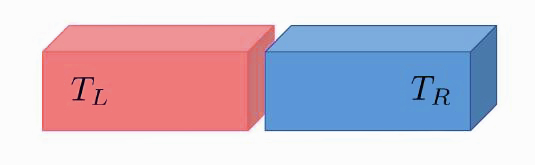} &
\hspace{-0.7cm}\includegraphics[width=20mm]{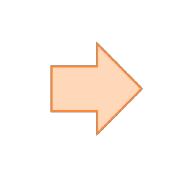} &
\hspace{-0.5cm}\includegraphics[width=63mm]{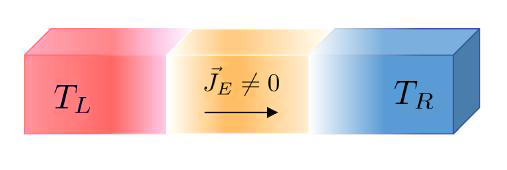} \\
\end{tabular}
\caption{At $t=0$, both systems are isolated and independently at equilibrium. Evolving forward in time from $t=0$, a spatially homogeneous non-equilibrium steady state develops in the middle region, carrying an energy current $J_E$.}
\label{fig:system}
\end{figure}

In principle, the time-dependent system described above can be set up and studied in arbitrary dimensions. However, in 1+1 dimensions the numerical analysis can be supplemented with analytical results, due to the fact that in this case conformal field theory techniques can be applied. From the holographic perspective, it is relevant that 2+1-dimensional gravity is non-dynamical. We study the 1+1-dimensional case and its gravity dual in the present paper. An important difference between the 1+1-dimensional and the higher-dimensional case is given by the following: In 1+1 dimensions, the shock waves with which the steady state region expands are dissipation-free. Both in the field theory and in the gravity dual, for all times the transition between the heat baths and the steady-state region is described by a step function. On the other hand, in the higher-dimensional case the shock waves experience diffusion and the temperature profile is no longer described by a step function. This may be referred to as a rarefaction wave.

A main focus of the present paper is the study of the time dependence of entanglement entropy and of mutual information in the steady-state system described above. In particular, our analysis describes how these quantities  change as the shock wave moves through the chosen entangling region, for instance a spacelike interval of length $\ell$ located away from $x=0$. 

To our knowledge, the time evolution of the entanglement entropy in this setup has not been studied yet. For our analysis we take the holographic perspective, which allows us to make use of the prescriptions proposed in \cite{Ryu:2006bv,Ryu:2006ef,Hubeny:2007xt}. The original Ryu-Takayanagi prescription states that in a $d$-dimensional CFT the entanglement entropy of any region $\mathcal{A}$ is proportional to the area of the minimal codimension-two surface in the $d$+1-dimensional dual static geometry anchored at the boundary of $\mathcal{A}$. Later this prescription was generalized to the time-dependent case, in which the entanglement entropy of $\mathcal{A}$ is proportional to the extremised spacelike codimension-two surface in the time-dependent bulk geometry. For our 1+1-dimensional boundary setup, the extremal surfaces we are looking for are geodesics.

AdS/CFT relates thermal states to stationary AdS black holes. Away from equilibrium, the time evolution of the field-theory system corresponds to the evolution of the spacetime dynamics subject to appropriate boundary conditions at the asymptotic AdS boundary. The holographic dual of the initial state with the temperature profile (\ref{introduction: initial temperature profile}) is thus given by a geometry consisting of two AdS black holes at two different temperatures both cut at $x=0$, and a half of each glued together at $t=0$. For this particular scenario, the asymptotic late-time   geometry is known and the steady-state region was shown to be equivalent to a boosted AdS Schwarzschild black hole at temperature $\sqrt{T_\mathrm{L}T_\mathrm{R}}$ \cite{Bhaseen:nature}. This result is in agreement with the earlier CFT result.

Taking the holographic approach, according to \cite{Bhaseen:nature}, the steady-state region in the late-time limit can be described by a boosted thermal state in the higher-dimensional case as well if the system shows time-independent asymptotic behaviour. The corresponding argument is based on black hole uniqueness theorems. A numerical analysis of the 2+1 dimensional boundary CFT \cite{Amado:2015uza} shows very good agreement with the conjecture. However, while in \cite{Bhaseen:nature} the two wavefronts in the higher-dimensional case are both shockwaves, it was later shown in \cite{Spillane:2015daa} and \cite{Lucas:2015hnv}  that the solution with one shockwave and one rarefaction wave is preferred.

Hydrodynamics provides another fruitful approach to study the time evolution of systems subject to a local quench like (\ref{introduction: initial temperature profile}). Studying the hydrodynamic expansion to first order, the authors of \cite{Chang:2013gba} conjectured the universality of the steady state regime in a sense that its emergence at late times is universal and irrespective of the dynamical details of the system or details of the initial state configuration and that the heat current can be described with a universal formula. The assumptions on which the conjecture is based are similar to the ones in \cite{Bhaseen:nature} namely that at late times the system can be described by three regions, the two heat reservoirs and a steady state regime growing in time as two shockwaves move towards spatial infinity.

A free field analysis within Klein-Gordon theory in \cite{Doyon:2014qsa} showed that in contrast to the 
1+1-dimensional case in more dimensions the emerging steady state is different from its strongly coupled analogue. A recent review \cite{Calabrese:2016xau} on quantum quenches in 1+1 dimensional conformal theories discussed global quenches at finite temperature and local quenches at zero temperature.

Much interest has also been directed towards the holographic study of the time-dependent behaviour of entanglement entropy after global quenches. For example, in \cite{AbajoArrastia:2010yt,Balasubramanian:2010ce,Balasubramanian:2011ur,Balasubramanian:2011at,Hartman:2013qma,Liu:2013iza,Li:2013sia,Liu:2013qca,Kundu:2016cgh} it was found that after a global quench, an initial quadratic growth of entanglement with time is followed by a universal linear growth regime. The special case where the final state is an AdS-Schwarzschild black hole is  referred to as {\it entanglement tsunami} \cite{Liu:2013iza,Li:2013sia,Liu:2013qca}. Noteworthy, in \cite{Liu:2013qca} the linear growth is found to be independent of the choice of entangling regions. It is also interesting to note that the cases studied in \cite{Keranen:2014zoa,Keranen:2015fqa} display a linear growth independent of the equation of state, showing more evidence that points towards a universal behavior. Related work on the evolution of entanglement entropy after a local quench also include  \cite{Caputa:2014vaa,Caputa:2014eta,Caputa:2015waa,Rangamani:2015agy,David:2016pzn,Rozali:2017bco,Nozaki:2013wia,Jahn:2017xsg,Astaneh:2014fga}. In most of these references the authors complemented the holographic analysis with results from a CFT analysis. In particular  in \cite{David:2016pzn}, the authors consider a local quench with a small time width $\epsilon$ and find universal features of the time evolution of the entanglement entropy.

A related, but distinct line of research involves the deformation of strongly coupled matter by time dependent perturbations of a relevant scalar operator.  Numerical investigations into this situation \cite{Buchel:2014gta} involve an uncharged black brane solution which is perturbed by varying the non-normalizable boundary mode of a massive bulk scalar in time. One  of  the  most  interesting  results  to  emerge  from  such a study has been the appearance of a “universal fast quench regime” in which the change in energy density after the quench scaled as a power law in the quench width.

The first focus of this paper (section \ref{sec:Holographic-Setup}) is to investigate the time evolution of the steady state itself. We analyze the causal structure of the dual geometry and find that the hypersurfaces that, in the chosen coordinate system, extend the boundary shockwaves into the bulk are spacelike. Therefore they appear to be superluminal. However, as we explain, causality is not violated.

In sections \ref{sec:numerics} and \ref{sec:Matching}, we then numerically compute the time evolution of the entanglement entropy for a spacelike interval which at $t=0$ is entirely enclosed in one of the heat baths, say the left one. As we work with a 1+1 dimensional boundary the interval is one-dimensional. The entanglement entropy of such an interval quantifies the quantum entanglement between the interval and its complement. Let us describe what happens when the shockwave passes the interval. Before the shockwave propagating outwards enters the interval on its right edge, the entanglement entropy is constant in time. The same is true once the shockwave has left the left edge of the interval behind. While the shockwave is passing through the interval, we observe a universal time evolution: The functional dependence on the interval length and the two temperatures $T_\mathrm{L,R}$ is the same for a wide range of temperatures and intervals, as long as the temperatures and their difference are small compared to the inverse of the interval length considered. In section \ref{sec:AnalyticalResults}, we present an analytic proof for the universal behaviour. For larger temperatures and temperature differences there are deviations from this universality which we see both in our numerical result and our analytical computation. For the latter we give an estimate.

In section \ref{sec:numerics}, we also study the time dependence of the mutual information for which we consider equal length intervals at equal distance from $x=0$. The mutual information quantifies the amount of information obtained about the degrees of freedom in the one interval from the degrees of freedom in the other. We find that mutual information for the geometry described above grows monotonically in time. Furthermore, we look at  an interval initially located within the smaller temperature heat bath. Its entanglement entropy increases with time. In section \ref{sec:AnalyticalResults} we show analytically that its average increase rate is bounded. This is similar to what is observed for the entanglement tsunami. A further analytically tractable case is when one of the temperatures is zero and the other temperature is large compared to the inverse of the interval length. For this case we show that the entanglement entropy grows linearly in time.
  
Our second main focus, considered in section  \ref{sec:Inequalities}, concerns entanglement inequalities for $n$ disconnected intervals. A famous example is the strong subadditivity for $n=3$ intervals. In this paper we numerically study generalized entanglement inequalities introduced in \cite{Bao:2015bfa} for  $n=5$ intervals. These authors proved these inequalities in the static case. We numerically verify by considering a large number of examples that these inequalities also hold in the time dependent geometry considered here. For obtaining this result we have developed a new algorithm counting the number of physical ways to link the boundary intervals by curves in the bulk. Details of this algorithm are given in the supplementary material joined to the hep-th submission of this paper.

In this work we use two complementary numerical methods for evaluating the entanglement entropy. 
They are described in sections \ref{sec:numerics} and \ref{sec:Matching}, respectively.
Their results are consistent and allow us to support the assumptions we make in each of the two numerical approaches. For the first method we explicitly solve the geodesic equation numerically and employ a shooting method to handle the boundary conditions. This approach requires a smooth geometry which we realize with a hyperbolic tangent ansatz. We refer to this method as the shooting method. The second method uses analytic expressions for the geodesic length, available for the pure AdS Schwarzschild or boosted AdS Schwarzschild geometries. From these we can write down the geodesic length of a piecewise defined geodesic parametrized by the point in spacetime where the two pieces meet on a specific hypersurface. Extremising the expression with respect to the meeting point gives the entanglement entropy of the interval. We refer to this method as the matching method. A corresponding Mathematica code is provided in a supplementary file together with this paper. In contrast to the first method, the second method does not require the knowledge of the geodesics themselves nor does it resort to a smoothened version of the geometry. The advantage of the matching method is that it allows us to study a wider spectrum of temperatures compared to the shooting method. Note that in this paper we only consider intervals that experience at most two of the three regions, the two heat baths and the steady state region.

This paper is organised as follows. In section \ref{sec:Holographic-Setup} we describe the holographic ansatz we work with and discuss the causal structure of the geometry considered. In section \ref{sec:numerics} and \ref{sec:Matching} we explain the two complementary numerical methods, shooting and matching, in detail and present the consistent results on the time evolution of the entanglement entropy and the mutual information. Analytical results are presented in section \ref{sec:AnalyticalResults}. In \ref{sec:universalformula} we analytically prove the universal behaviour of the time dependence of the entanglement entropy. We present analytical results for the special case in which one of the heat baths is at zero temperature and discuss bounds on the entropy increase rate in sections \ref{sec:zerotemperature} and \ref{sec:Tsunami}. Section \ref{sec:Inequalities} is devoted to the study the entanglement entropy of setups with many disconnected intervals. An algorithm for dealing with the large number of configurations is introduced and subsequently used to explore entanglement inequalities. In section \ref{sec:Higher-Dimensions} we present some analytical results on the higher dimensional case and comment on the challenges of the numerical approach in this case. We conclude in section \ref{sec:Conclusion} .


\section{Holographic Setup}
\label{sec:Holographic-Setup}
We are interested in studying a strongly coupled CFT in $d-1=1$ spatial dimensions. The energy flow in such a system can be qualitatively captured by pure gravity alone in holography, so the bulk action that we will consider is simply
\be
S=\frac{1}{16\pi G}\int d^3x\,\sqrt{-g}\,(R-2\Lambda)\,,
\ee
where $\Lambda=-1/L^2$ is the cosmological constant of $AdS$.  
A static CFT configuration at finite temperature $T$ is dual to the BTZ black hole \cite{Banados:1992wn,Banados:1992gq},
\be
ds_{T}^2=\frac{L^2}{z^2}\left[ -\left(1-(2\pi T z)^2\right)dt^2+\frac{dz^2}{1- (2\pi T z)^2} +dx^2 \right]\,,
\label{eq:btz}
\ee
where $L$ is the radius of AdS spacetime, which we will set to $L=1$  later, and the temperature is related to the horizon's position $z_H$ via $1/T=2\pi z_H$. We will always assume the spatial coordinate $x$ to be decompactified, such that $-\infty<x<+\infty$. In this geometry, the calculation for the entanglement entropy can be easily derived from the fact that the BTZ black hole is obtained from a quotient of pure $\text{AdS}_3$ \cite{Hubeny:2007xt}. Given a spatial interval with separation $\ell$ in the CFT, the holographic result for the entropy of the entanglement between this region and its complement is given by \cite{Ryu:2006bv}
\be
S_{\text{BTZ}}=\frac{L}{4G}\log\left(\frac{1}{\pi^2 \epsilon^2 T^2}\sinh^2(\pi \ell T) \right)\,,
\label{eq:sbtzren}
\ee
where $\epsilon$ is a UV cut-off. Using minimal subtraction, this result may be  regularized to read
\be
S^{\text{ren}}_{\text{BTZ}}=\frac{L}{4G}\log\left(\frac{1}{\pi^2 T^2}\sinh^2(\pi \ell T) \right)\,.
\label{eq:sbtz}
\ee

In this paper we study the particular dynamical configuration investigated in \cite{Bhaseen:nature}. We consider two thermal reservoirs, each of them initially at equilibrium but at different temperatures, $T_L$ and $T_R$. After bringing the two systems into thermal contact at $t=0$, a spatially homogeneous steady state develops, carrying a heat flow $J_E$ which transfers energy from the heat bath at higher temperature to the other. This physical situation is presented in figure \ref{fig:system}. The steady state configuration in the CFT is described by a Lorentz-boosted stress-energy tensor, which is dual to a boosted black hole geometry,
\begin{align}
ds_{\text{boost}}^2=\frac{L^2}{z^2}\left[ -\left(1-\frac{z^2}{z_H^2}\right)(dt\,\cosh\theta-dx\,\sinh\theta)^2+\frac{dz^2}{1-z^2/z_H^2} +(dx\,\cosh\theta-dt\,\sinh\theta)^2 \right].
\label{eq:boosted}
\end{align}
This is dual to a boosted thermal state with boost parameter $\theta$, temperature $T$ and velocity $\beta$, which are determined by
\be
T = \sqrt{T_L T_R}  \,,  \quad \chi =  \frac{T_L}{T_R}  \,,  \quad \beta =  \frac{\chi-1}{\chi+1} \,, \quad \theta = \text{arctanh}\, \beta\,. 
\label{eq:boostparam}
\ee
This is a particular case of equations \eqref{theta} for $d=2$. Its associated entanglement entropy is given by\footnote{By carrying out the boost, this can be obtained from the entanglement entropy for a static black hole for boundary intervals with endpoints at different (boundary) times $t_1\neq t_2$.  }
\be
S_{\text{boost}}=\frac{L}{4G}\log\left(\frac{1}{\pi^2 T_R T_L \epsilon^2}\sinh(\pi \ell T_L) \sinh(\pi \ell T_R) \right)\,.
\label{eq:boostedEE}
\ee
This result also gives the late-time limit of our case, since the central region expands progressively as the shockwaves advance towards the heat reservoirs located at spatial infinity. As with \eqref{eq:sbtzren}, it must be renormalized by subtracting $(L/4G) \log \epsilon^{-2}$, according to our scheme of minimal subtraction.

For the case $d=2$, the shockwaves move with the same speed $u=1$ in both directions, so generically, at a time $t>0$, the geometry is divided into three regions. Schematically,
   \begin{equation}
     \label{eq:discontscheme}
     ds^2 = \left\{
	       \begin{array}{ll}
		 ds_{T_L}^2      & \mathrm{if\ } x < -t \\
		 ds_{\text{boost}}^2 & \mathrm{if\ } -t < x < t \\
		 ds_{T_R}^2     & \mathrm{if\ } x > t
	       \end{array}
	     \right.
   \end{equation}
Such a dynamical solution corresponds to the idealized limit in which the initial temperature profile of the system includes a step function of zero width, leading to sharp shockwaves in the CFT. In this limit, there are three different regions, the central one corresponding to the steady state, which is formed by the propagating shockwaves.
Note that this simple solution only applies to the (1+1)-dimensional case. In a generic number of dimensions, the dynamics is non-linear and the nature of the right and left
shockwaves is very different. See section \ref{sec:Higher-Dimensions} and \cite{Bhaseen:nature} for a discussion of the higher-dimensional case.

Given a generic smooth temperature profile of finite width, it is convenient to work in Fefferman-Graham coordinates $(\tilde z, \tilde t, \tilde x)$.
The dynamical solution in these coordinates can be found in references \cite{Banados:1998gg,Bhaseen:nature}. It may be written as
\be
ds^2=\frac{L^2}{\tilde z^2}\left( d\tilde z^2 +  g_{\mu\nu}(\tilde z, \tilde x,\tilde t) d\tilde x^\mu d\tilde x^\nu  \right) \,, \label{eq:FG}
\ee
where 
\begin{subequations}
\begin{align}
g_{tt}(z,x,t) &= -\left[1-\frac{z^2}{L^2}\left( f_R(x-t) + f_L(x+t) \right)\right]^2  +  \left[ \frac{z^2}{L^2}\left( f_R(x-t) - f_L(x+t) \right) \right]^2  \,,  \\
g_{tx}(z,x,t) &=   -2 \frac{z^2}{L^2}\left( f_R(x-t) - f_L(x+t) \right)  \,,  \\
g_{xx}(z,x,t) &= \left[1+\frac{z^2}{L^2}\left( f_R(x-t) + f_L(x+t) \right)\right]^2  -  \left[ \frac{z^2}{L^2}\left( f_R(x-t) - f_L(x+t) \right) \right]^2  \,. 
\end{align}
\label{eq:gxx}
\end{subequations}
The functions $f_L(x+t)$ and $f_R(x-t)$ are to be determined by the boundary conditions. The way to do this is to calculate the boundary stress-energy tensor.
Its vacuum expectation value is given by\footnote{These are expectation values of the boundary stress-energy tensor. The gravitational solution in the bulk is a vacuum solution.}
\begin{subequations}
\begin{align}
\langle T^{tt} \rangle &= \langle T^{xx}\rangle = \frac{c}{6\pi^2 L^2} \left( f_R(x-t) + f_L(x+t) \right)   \,, \\
\langle T^{tx} \rangle &= \frac{c}{6\pi^2 L^2} \left( f_R(x-t) - f_L(x+t) \right) \,,  
\end{align}
\label{eq:set}
\end{subequations}
where $c$ is the central charge of the CFT. The initial condition $\langle T^{tx}\rangle = 0$ (i.e.~the absence of a heat current at $t=0$) demands that $f_L(v) = f_R(v)$.
In the following we will consider a profile
\begin{equation}
f_L(v) = f_R(v) = \frac{\pi^2 L^2}{4} \Big( (T_L^2 + T_R^2) + (T_R^2 - T_L^2) \tanh(\alpha v) \Big) \,, \label{eq:fLfR}
\end{equation}
which corresponds to a step of width $w \approx 1/(2\alpha)$. In the limit $\alpha \to \infty$, it asymptotes to a sharp step function,
\begin{equation}
f_{L/R}(v) \to \frac{\pi^2 L^2}{2}\left( T_L^2 + \left( T_R^2 - T_L^2 \right)\theta(v) \right) \,. \label{eq:fLfRlimit}
\end{equation}
The discontinuous metric in this case is given by
\be
g_{tt} = - \left( 1-\pi^2 z^2 T_{L/R}^2 \right)^2\,,\quad g_{tx}=0\,,\quad g_{xx} =\left( 1+\pi^2 z^2 T_{L/R}^2 \right)^2
\label{eq:discont1}
\ee
on the  left ($L$) and right ($R$) sides respectively, and
\begin{subequations}
\begin{align}
g_{tt} &= -1 + \pi^2 \left( T_R^2 + T_L^2 \right) z^2 - \pi^4 T_L^2 T_R^2 z^4  \,,  \\
g_{tx} &=  \pi^2 \left( T_R^2 - T_L^2 \right) z^2 \,,  \\
g_{xx} &= 1 + \pi^2 \left( T_R^2 + T_L^2 \right) z^2 + \pi^4 T_L^2 T_R^2 z^4 
\end{align}
\label{eq:discont2}
\end{subequations}
in the central region.\footnote{This shows that when setting $T_L=T_R$, the bulk metric will just be a static BTZ black hole, and there will be no non-trivial time evolution of entanglement entropy. This distinguishes our setup from the one studied in \cite{Astaneh:2014fga,Hoogeveen:2014bqa}, where two BCFTs are joined at $t=0$, and non-trivial time evolution takes place even when the temperatures of both sides where equal.} The relation between this solution and the equivalent in Schwarzschild-type coordinates
is discussed in section \ref{sec:Matching}.

It is now worth to point out that this discontinuous geometry is formed by gluing different spacetimes together along co-dimension one hypersurfaces which represent the extension of the shockwaves from the boundary into the bulk. The procedure by which spacetimes are matched to one another in GR involves the use of Israel junction conditions \cite{Israel1966}. Generically, each chunk of spacetime ends in a codimension one hypersurface, and when two of these hypersurfaces $\Sigma_1, \Sigma_2$ are identified, they must have the same topology and induced metric $\gamma_{ij}$. The identification is generally only possible provided that the energy-momentum tensor $S_{ij}$, defined on the hypersurface $\Sigma_1\equiv\Sigma_2$ which glues regions of spacetime together, satisfies
\be
\left( K_{ij}^+-\gamma_{ij} K^+ \right)-\left( K_{ij}^- -\gamma_{ij}K^- \right)=-\kappa S_{ij}\,.
\label{eq:israel}
\ee
Note, however, that $S_{ij}$ vanishes for our case, as the bulk solution is supposed to be a vacuum solution everywhere. In these equations, $K^+, K^-$ are extrinsic curvatures of the hypersurface, computed from the induced metric on the right and left sides respectively. They correspond to different embeddings, since this hypersurface is embedded from both sides. We checked that this condition is satisfied for the present geometry. There is, however, a non-trivial conceptual question, since the spacetimes that are being glued include a horizon which, apparently, is cut into three pieces. In order to visualize this, it is useful to employ a causal Kruskal diagram of the spacetime \cite{Banados:1992gq}. This will also allow us to understand and interpret the fact that in the bulk, the ``shockwaves'' form \textit{spacelike} hypersurfaces. Of course, this means that while on the bounday, via the AdS/CFT correspondence, we describe actual shockwaves in the CFT, the spacelike hypersurfaces that extend these shockwaves from the boundary into the bulk in our dual AdS picture should not be referred to as genuine shockwaves from a bulk perspective. Nevertheless, for the sake of brevity we will from now on leave away inverted commas and also refer to the bulk hypersurfaces as shockwaves. 

\begin{figure}[h]
\centering
\includegraphics[width=0.8\textwidth]{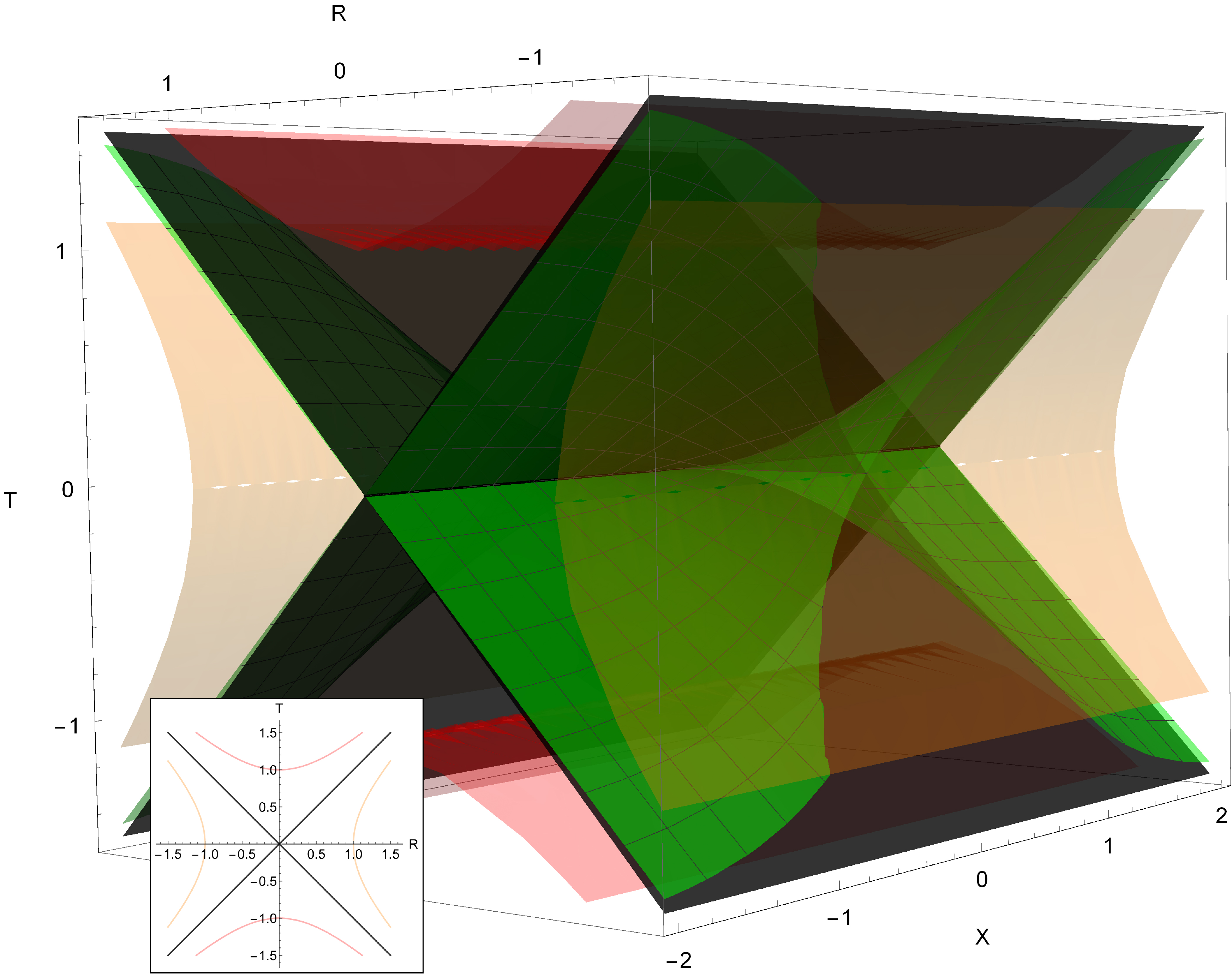}
\caption{Kruskal diagram of the bulk spacetime. The black diagonal sheets correspond to the location of the black hole's horizon. The red surfaces show the location of the singularity, and the orange surfaces show the location of the boundary at $z=0$. The green surface is the worldvolume of the bulk shockwaves along which the three regions of spacetime are glued together. A steady state region forms both for $t>0$ and for $t<0$. However, note that the physically relevant part of this spacetime for our investigation is assumed to be $t \geq 0$ only.}
\label{graf:Kruskal}
\end{figure}

The bulk spacetime has two spatial coordinates $z, x$. In order to obtain a Kruskal diagram\footnote{The global extensions of metrics of the form \eqref{eq:FG} have been studied in \cite{Mandal:2014wfa} in more generality. }, we compactify $z$ and the time $t$ for each slice of the $x$ direction, thus obtaining the Kruskal coordinates $(R,T)$, defined by
\be
T=\left|\frac{z-z_H}{z+z_H}\right|^{1/2}\sinh\left(\frac{t}{z_H}\right),\quad R=\left|\frac{z-z_H}{z+z_H}\right|^{1/2}\cosh\left(\frac{t}{z_H}\right).
\ee
Let us now analyze the resulting diagram of figure \ref{graf:Kruskal}, in order to understand how the spacelike bulk shockwaves are still in agreement with causality of the bulk. A two-dimensional Kruskal diagram is included in the bottom left corner, it shows a two-dimensional spacetime divided into four quadrants: the external universe is captured by the right quadrant, the interior of the black hole by the quadrant at the top, and the other two correspond to the respective analytical continuations. The lines of constant $z$ correspond to hyperbolae, which get closer to the horizon with increasing value of $z$, to the extreme that the line corresponding to $z=z_H$ degenerates to the diagonals that separate the interior and the exterior of the black hole. The lines of constant $t$ correspond to  straight lines emanating from the center, which at $t=0$ appear as horizontal in the right quadrant and as vertical in the top quadrant.
They get closer to the future horizon with increasing value of $t$.

Focusing on the exterior of the black hole in figure \ref{graf:Kruskal}, the shockwaves leave the central point $x=0$ at $t=0$. Therefore the initial location of the shockwave is marked by the horizontal ray $t=x=0,0\leq z\leq z_H$. Any other boundary point $x=x_0$ experiences the shockwave crossing at $t=|x_0|$ (and this extends radially all along $z$), so the location of the shockwave is marked by a straight line that increasingly separates from the initial horizontal line as $|x_0|$ increases. Gathering the locations corresponding to a shockwave for all values of $x$, we obtain the green surface in figure \ref{graf:Kruskal}. This figure displays that in this causal diagram, the shockwave worldvolume  does not touch the horizon surface, except at the line corresponding to $T=R=0$, which is the bifurcation surface. Consequently, the only part of the event horizon of the static regions on the left and on the right that is retained in this construction is a half of the bifurcation surface for each side. Apart from that, only the event horizon of the steady-state region appears, it remains untouched by the shockwaves on which the gluing of spacetimes takes place.

As mentioned above, the analysis of the causal diagram in figure \ref{graf:Kruskal} reveals another important aspect of the shockwaves: their spacelike nature, i.e.~the fact that their induced metric will have positive determinant. Intuitively, this means that in the bulk, they would be perceived as being superluminal. Of course, this raises puzzling questions concerning whether this system should be considered to be physical or not. However, we note that the present geometry is a solution of vacuum in three dimensions, in which general relativity does not have propagating degrees of freedom. Therefore,  these shockwaves do not transport information in the bulk, and every bulk observer will locally observe AdS space everywhere. However as we will see from the time dependence of entanglement entropy later on, from the dual CFT perspective the shockwaves, which travel at the speed of light on the boundary, \textit{do} transport information. Considerations of energy conditions in the bulk are also unnecessary, given that it is a vacuum solution (including the fact that $S_{ij}=0$ in \eqref{eq:israel}), so most common energy conditions are trivially satisfied.

\begin{figure}[hbtp]
\centering
\includegraphics[width=0.6\textwidth]{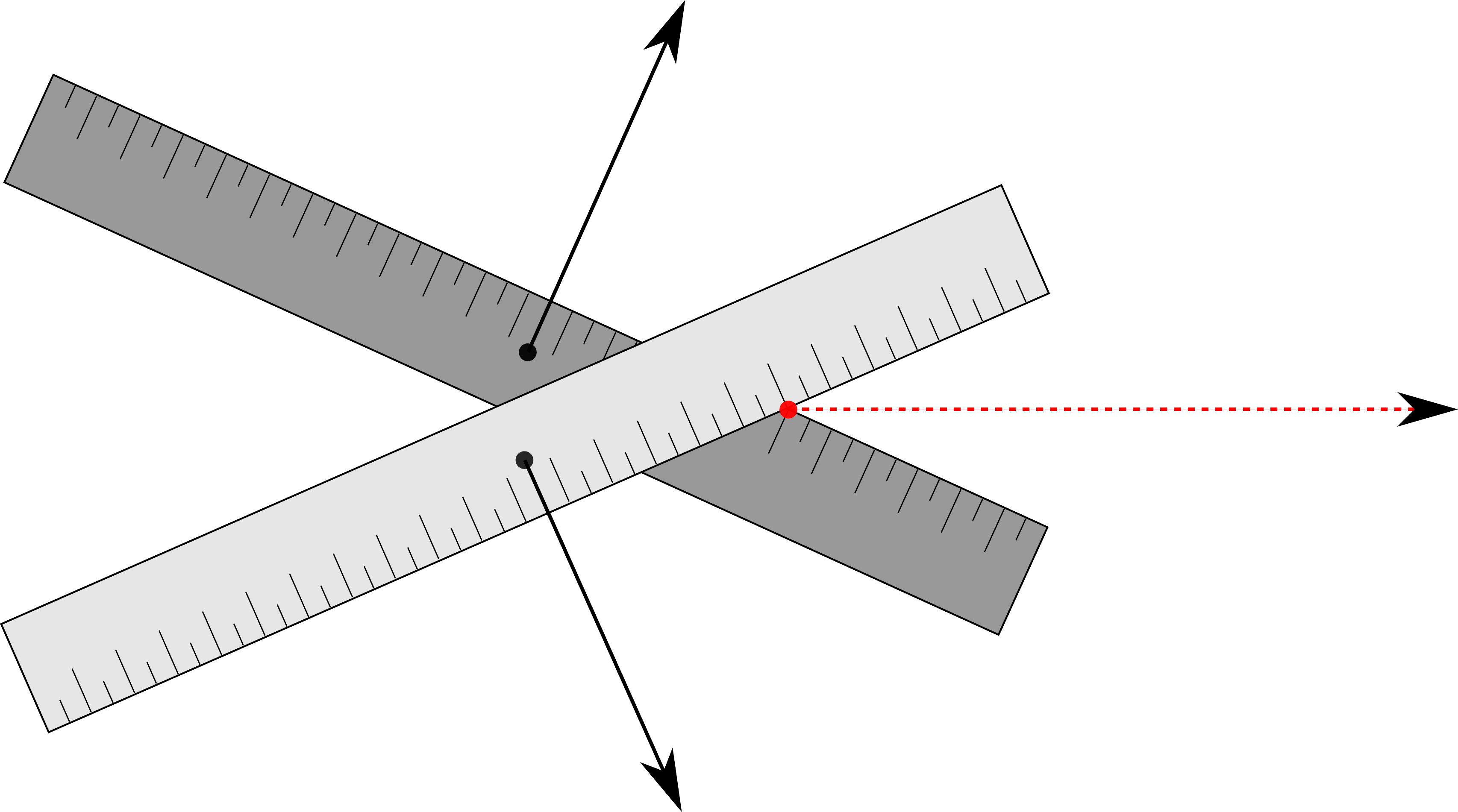}
\caption{Two rulers moving at an angle. The red dot to the right indicates the point of intersection of the rulers. As the rulers move away in diverging directions, their point of intersection may move at superluminal speed, but without transport of information.}
\label{fig:rulers}
\end{figure}

The intuitive picture of why  information is not transported by these shockwaves in the bulk can be understood by taking into account that apparent faster than light propagation is present in many physical situations.
In order to illustrate this point, we look at the example of two rulers, as in figure \ref{fig:rulers}. The red dot corresponds to their point of intersection. As the rulers move in diverging directions (at speeds slower than light), as indicated by the arrows of the figure, their point of intersection moves forward at a higher speed. The velocity of this point depends on the angle between the two rulers, and it can move superluminally if the initial conditions conspire accordingly.

However, this point of intersection is an emergent object and not a physical one, so it does not carry information. As a consequence, causality is not violated. In other words, consecutive positions of that point are not causally related, even though the information about these events is encoded in the initial conditions. Similarly, the shockwaves of our system have dynamics encoded in the initial conditions and can develop apparent superluminal speeds, but since they do not carry information, causality is preserved.

Another particular feature present in the $2+1$-dimensional case is that spacetime is always locally $AdS$, even at the matching surface. This means that a local observer traveling with the shockwave still sees $AdS$ everywhere.  This is why the velocity and features of the shockwave cannot be physically relevant in the bulk.

In the following we holographically study the evolution of entanglement entropy in this setup. We find that it has physical behaviour in agreement with field-theory expectations.
For instance, there is a well-defined  velocity of propagation for entanglement entropy. This is in agreement with arguments using a quasiparticle picture \cite{Calabrese:2005in}, according to which the initial condition acts as a source of pairs of quasiparticle excitations. As they propagate causally throughout the system, larger regions get entangled. In this picture, if a maximum quasiparticle velocity exists, then the entanglement entropy grows linearly in time for certain boundary regions. We will see that also holographically, there is a velocity associated to entanglement, independently of the gluing of the spacetimes. In fact, in the following section we will see that entanglement entropy does evolve in a causal manner, and obeys the velocity bound known from the study of entanglement tsunamis.


\section{Numerical results I: Shooting method}
\label{sec:numerics}

We now turn to the numerical computation of entanglement entropy in the background introduced in the preceding section. For this purpose, we study minimal surfaces whose boundary at $z=0$ is in $x = x_L$ and $x = x_R$, and consider space-like intervals with $t(x_L) = t(x_R)$.~\footnote{In this section we are working in Fefferman-Graham coordinates, and we denote them by $(z,t,x)$.} By the HRT prescription in $2+1$ dimensions \cite{Hubeny:2007xt}, the minimal surface compatible with these boundary conditions corresponds to a geodesic in the bulk. This follows from a solution of the geodesic equations, which read
\begin{equation}
\frac{d^2 x^P}{ds^2}  +\Gamma^P_{MN} \frac{dx^M}{ds} \frac{dx^N}{ds} = 0 \,, \qquad P= t,x,z \,.  \label{eq:geodesics}
\end{equation}
We assume an affine parametrization of the geodesic, which implies
\begin{equation}
\frac{\partial x^M}{\partial s}\frac{\partial x^N}{\partial s} g_{MN}   = 1 \,. \label{eq:affine}
\end{equation}
The induced metric on the minimal surface reads
\begin{equation}
h_{ab} = \frac{\partial x^M}{\partial x^a} \frac{\partial x^N}{\partial x^b} g_{MN} = h_{ss}  \,, \label{eq:hab}
\end{equation} 
where $s$ is the coordinate of the surface. The entanglement entropy then follows from the area of the manifold $\gamma_A$, which can be computed from the induced metric as
\begin{equation}
S_A = \frac{1}{4 G} \int_{s(x_L)}^{s(x_R)} ds \, {\cal L} \,, \qquad \textrm{with} \qquad {\cal L} = \sqrt{h_{ss}} \,. \label{eq:S1}
\end{equation}
From (\ref{eq:affine}) and (\ref{eq:hab}) we find that the entanglement entropy of  (\ref{eq:S1}) reduces to the trivial integral $S_A = \frac{1}{4 G} \int_{s(x_L)}^{s(x_R)} ds$. The solution of the geodesic equations leads to the behavior $z \sim e^{-|s|}$ in the regime $s \to \pm \infty$. Consequently, the entanglement entropy is divergent. In the present case the divergence behaves as~${\textrm{Area}}(\gamma_A^{div}) \sim -2L \log \epsilon + \cdots$, and a renormalization scheme is required. We use a minimal subtraction scheme, so that the renormalized entanglement entropy is defined as
\begin{equation}
S_A^{\textrm{ren}} = \frac{1}{4 G} \left( \textrm{Area}(\gamma_A) -  \textrm{Area}(\gamma_A^{div}) \right)  \qquad \textrm{with} \qquad   \textrm{Area}(\gamma_A^{div}) = -2 L \log \epsilon \,. \label{eq:Sreg}
\end{equation}
In the following we will compute renormalized entropies according to this formula.

\subsection{Numerical solution of the geodesic equations}
\label{subsec:entanglement_entropy}

The geodesic equations of (\ref{eq:geodesics}) consist of three coupled differential equations of second order, whose solution can be expressed in the parametric form
\begin{equation}
t= t(s) \,, \qquad x = x(s) \,, \qquad z=z(s) \,. 
\end{equation}
These equations can be solved by imposing six boundary conditions, which are
\begin{equation}
\begin{cases}
t(s_L) = t(s_R) = t_0       &   \\
x(s_L) = x_L \,, \qquad x(s_R) = x_R    &   \\
z(s_L) = z(s_R) = \epsilon  \, .     &  
\end{cases} \label{eq:bc}
\end{equation}
We use the shooting method for the numerical computation of the geodesic equations: We shoot from $s=0$ with given values of $\{t(0),x(0),z(0)\}$ and $\{t^\prime(0),x^\prime(0),z^\prime(0)\}$, and then find the values of these initial conditions that lead to the desired boundary values at $s \to s_{L,R}$ given  in  (\ref{eq:bc}).~\footnote{There are in the literature other numerical methods for the solution of this two-point boundary value problem. An example is given by the relaxation methods, in which the solution is determined by starting with an initial guess and improving it iteratively, see e.g.~\cite{Ecker:2015kna}.}

\begin{figure*}[h]
\includegraphics[width=6.65cm]{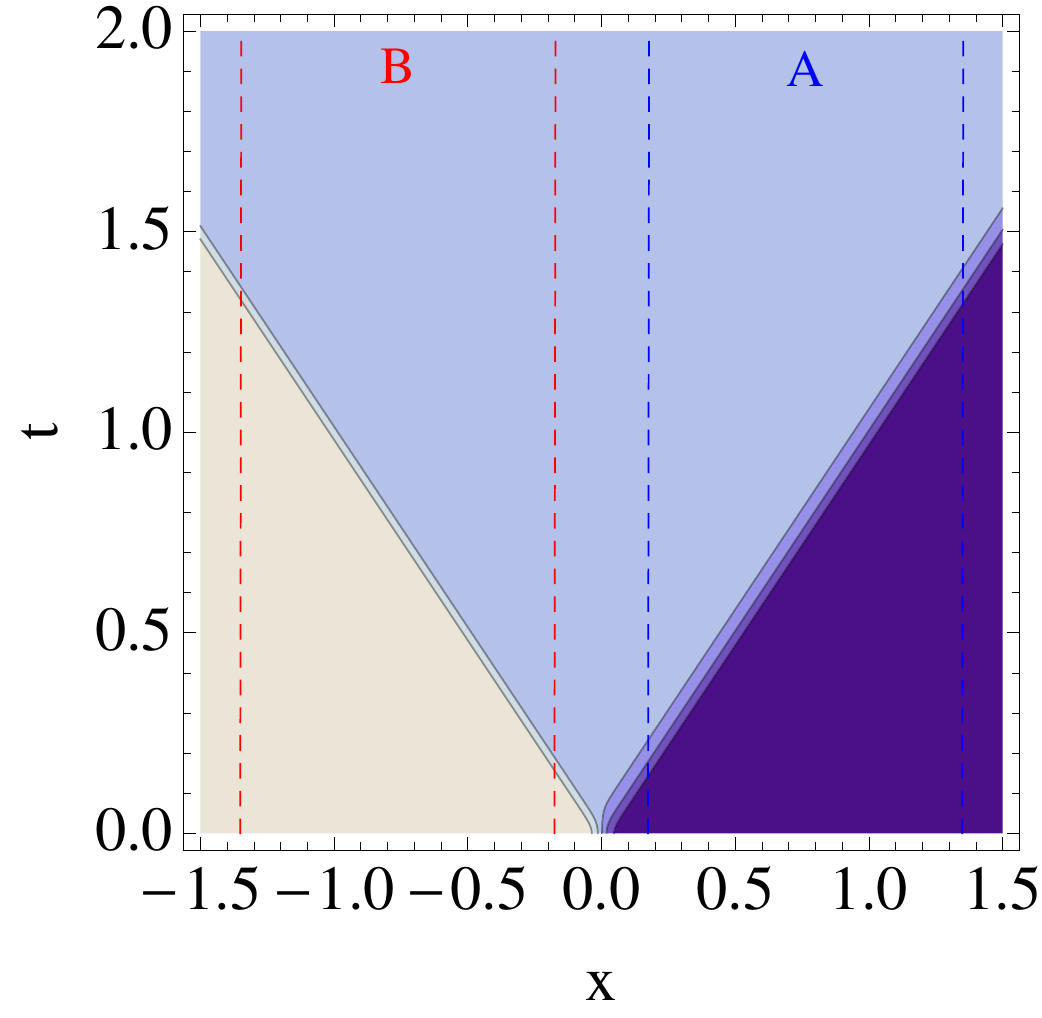} \hspace{2pc}%
\begin{minipage}{18.8pc}
\vspace{-6cm}\caption{ Contour plot of energy density $\langle T^{tt}(t,x)\rangle$ with the model in $d=2$, see section \ref{sec:Holographic-Setup}. Dashed lines show the time evolution of the endpoints of the intervals $A$ and $B$, in the positive and negative semiplane respectively.  We consider the intervals $x^A \in [0.175,1.35]$ (blue) and $x^B \in [-1.35,-0.175]$ (red), temperatures $T_L = 0.2$, $T_R=0.195$ and $\alpha = 25$ (in equation \eqref{eq:FG}), in units in which $L=1$.}
\label{fig:Tttd1}
\end{minipage}
\end{figure*}

\begin{figure*}[h]
\begin{tabular}{cc}
\includegraphics[width=7.0cm]{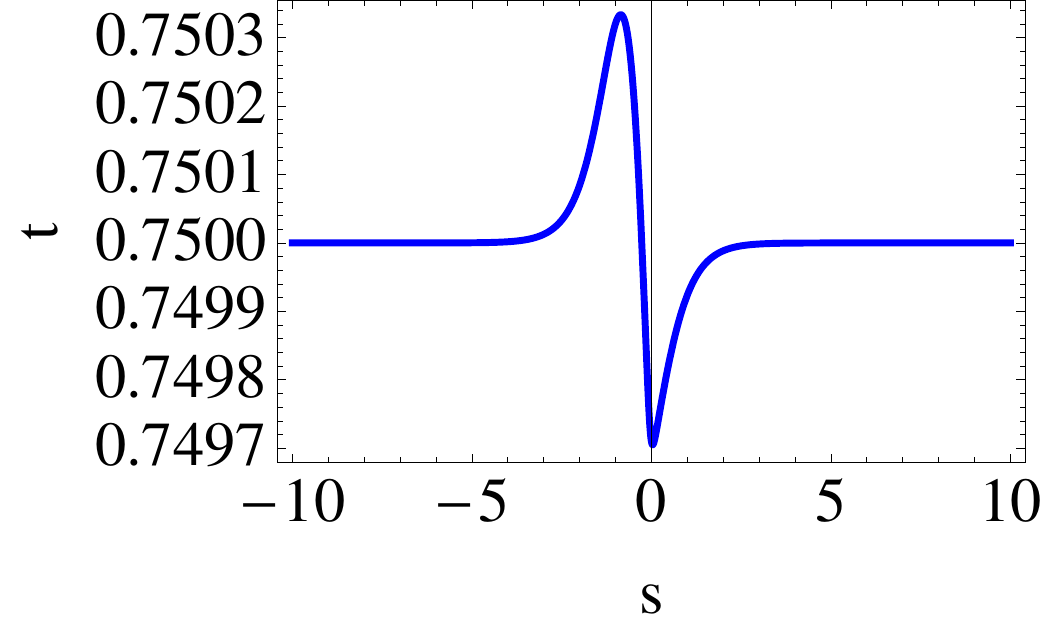} & 
\includegraphics[width=6.35cm]{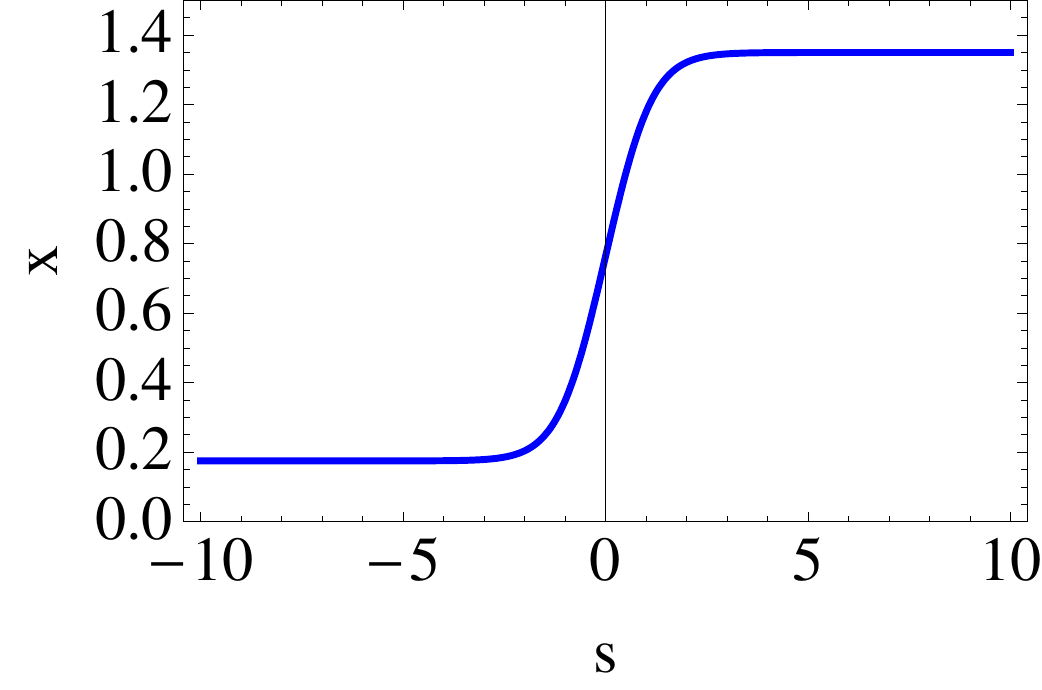} \\
\end{tabular}
\caption{ Parametric dependence of the geodesic as a function of the affine parameter~$s$. We show $t = t(s)$ (left) and $x = x(s)$ (right). We have considered the interval $x^A \in [0.175,1.35]$ as shown in figure \ref{fig:Tttd1}, and $t_0 = 0.75$, see equation (\ref{eq:bc}).}
\label{fig:geodesic1}
\end{figure*}

\begin{figure*}[h]
\begin{tabular}{cc}
\includegraphics[width=6.65cm]{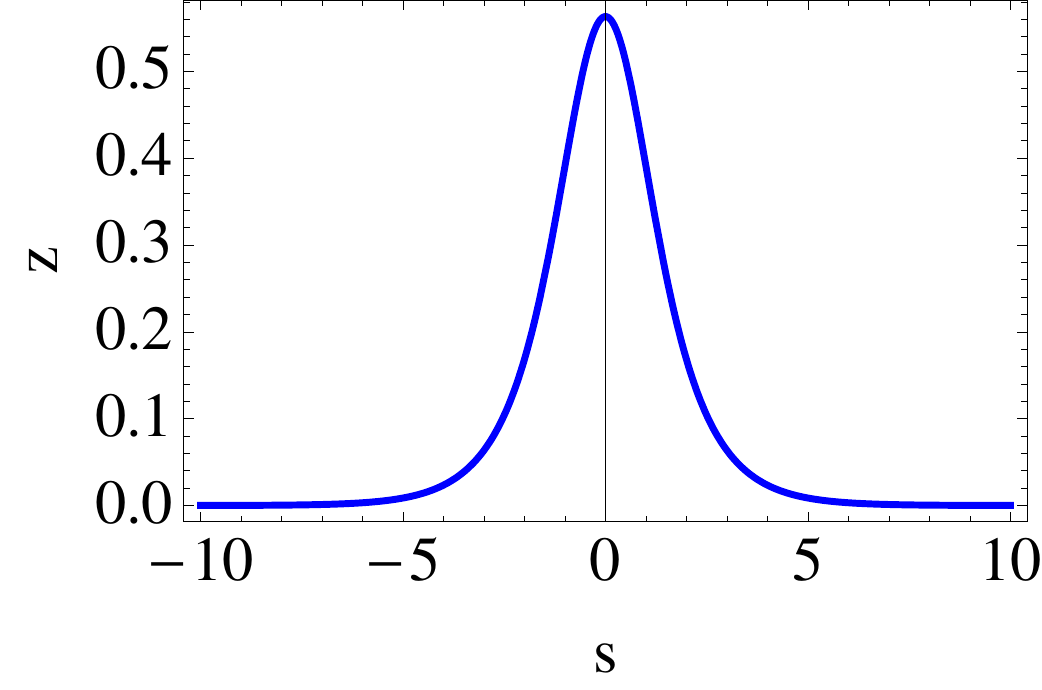} & 
\includegraphics[width=5.65cm]{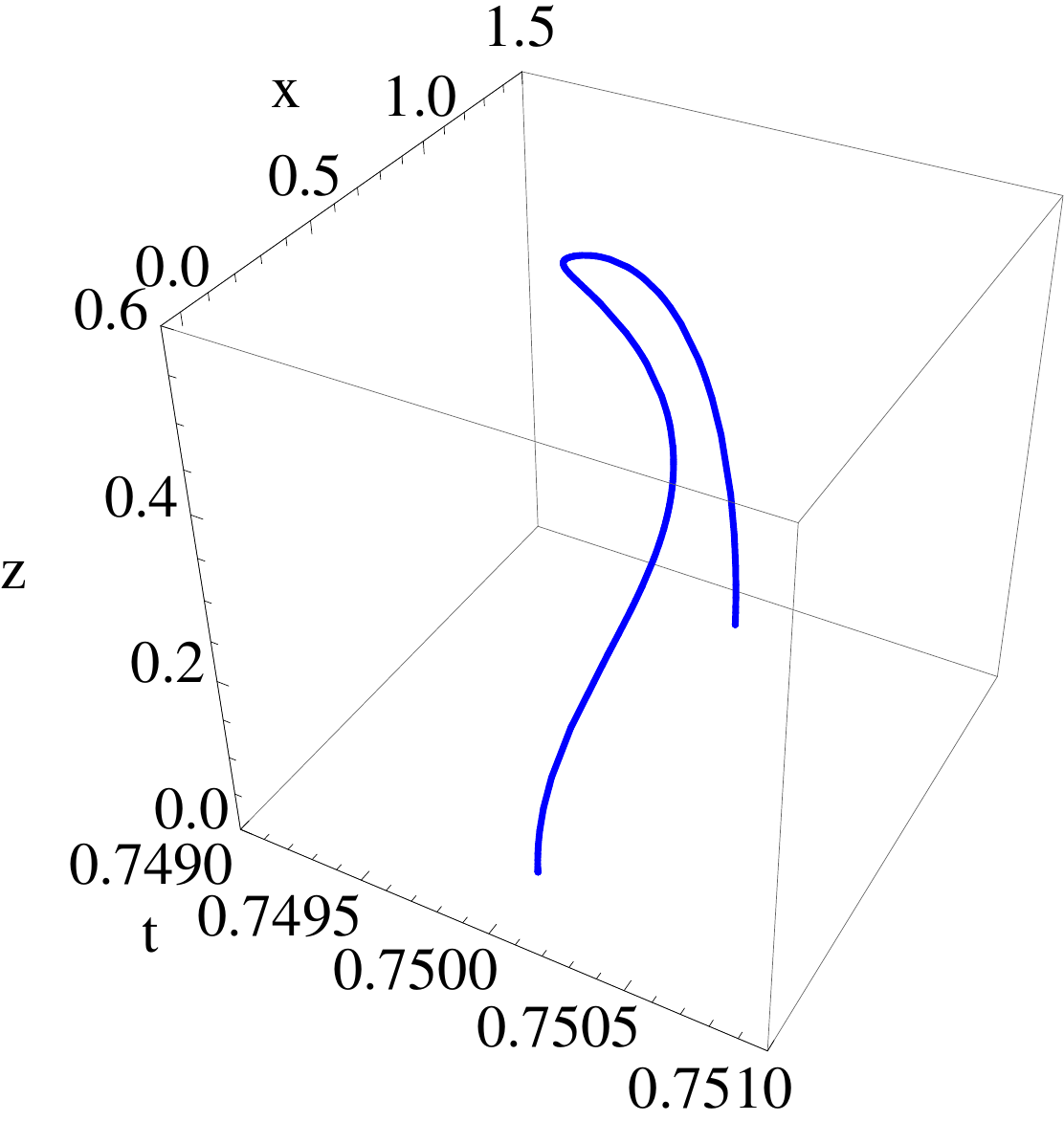}
\end{tabular}
\caption{  Parametric dependence of the geodesic. (Left) We show $z = z(s)$. (Right) Geodesic in the space $(t,x,z)$. See figure \ref{fig:geodesic1} for further details.}
\label{fig:geodesic2}
\end{figure*}

We introduce a cutoff $\epsilon \ll 1$ for regularization. This also induces a cutoff in the affine parameter, i.e. $s_L \sim -|\log \epsilon|$ and $s_R \sim |\log \epsilon|$. In the following we will consider space-like intervals $A$ and $B$ as shown in figure \ref{fig:Tttd1}.  Figures \ref{fig:geodesic1} and~\ref{fig:geodesic2} display a typical solution of the geodesic equations, which satisfies the boundary conditions of  (\ref{eq:bc}). Once the geodesics are obtained, the next step is to compute the area of these curves and then the entanglement entropy from  (\ref{eq:Sreg}). We now present results for the time evolution of the entanglement entropy in the system of section \ref{sec:Holographic-Setup}.

\subsection{Entanglement entropy and universal time evolution}
\label{subsec:entanglement_entropy2}

\begin{figure*}[htb]
\begin{tabular}{cc}
\includegraphics[width=6.85cm]{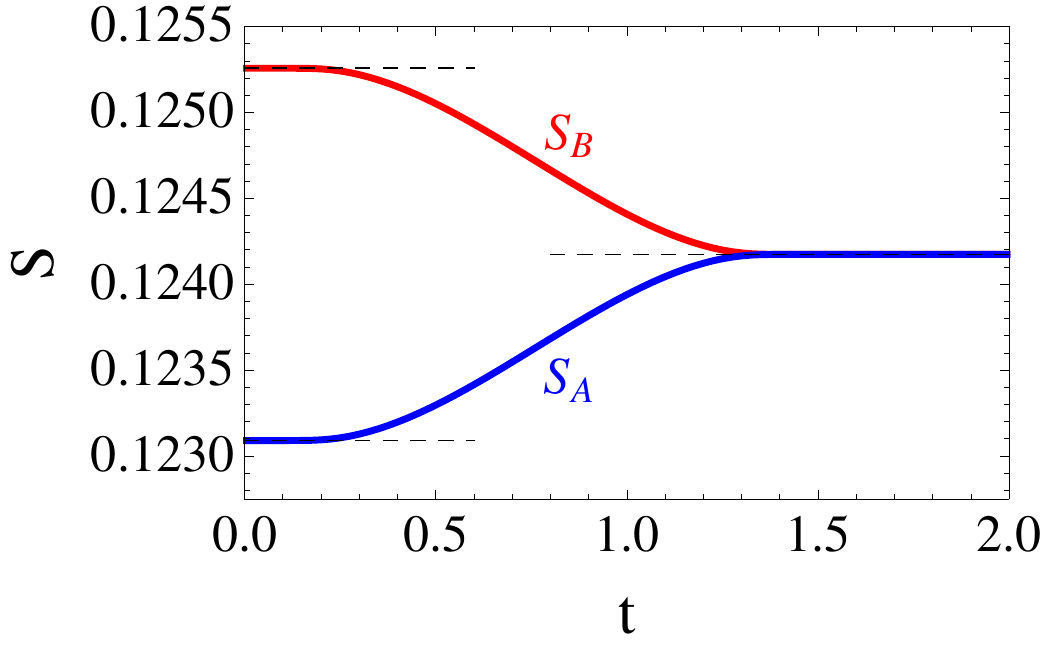} & 
\includegraphics[width=6.45cm]{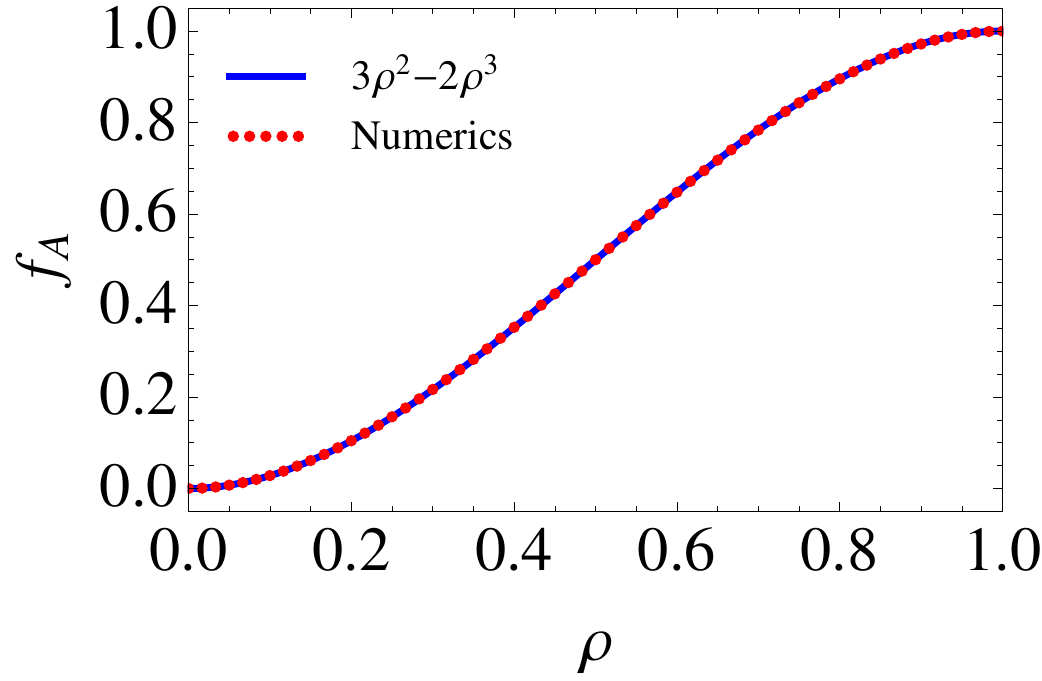}
\end{tabular}
\caption{ (Left) Renormalized entanglement entropies of intervals $A$ and $B$ as a function of time, see figure \ref{fig:Tttd1}. The intervals correspond to $x^A \in [0.175,1.35]$ and $x^B \in [-1.35,-0.175]$, and temperatures $T_L = 0.2$ and $T_R=0.195$. The (dashed) horizontal lines correspond to the results by using the analytical formulae (\ref{eq:St0}) and (\ref{eq:Sboost}). We have set $G=1$ and $L=1$. (Right) Renormalized entanglement entropy $S_A$ as a function of time, normalized to $[0,1]$ in both horizontal and vertical axes, see  (\ref{eq:fA}). The dots correspond to the numerical result with the interval $A$ in figure \ref{fig:Tttd1}, while the continuous line is the universal behavior~$f_A(\rho) = 3\rho^2 - 2\rho^3$. }
\label{fig:Suniversal}
\end{figure*}

For the moment we consider a single interval $x \in [x_L,x_R]$ denoted by $A$, placed in the positive semiplane, i.e. $x_{L,R}>0$. Let us study the time evolution of the entanglement entropy $S_A$ during the formation of the steady state. We are considering the model in $d=2$, so that the shockwaves are at $t = |x|$. This means that the shockwaves touch the two ends of the interval at times $t=|x_L|$ and $t=|x_R|$, see figure \ref{fig:Tttd1}. We will denote these values by $t_1$ and $t_2$, respectively. In the limit $\alpha \to \infty$ in  (\ref{eq:fLfR}), the entanglement entropy turns out to be constant in the regimes $0 \le t \le t_1$ and $t_2 \le t$, and there is a non-trivial time evolution only in the interval $t_1 \le t \le t_2$, i.e.
\begin{equation}
S_A(t) =
\begin{cases}
S_A(t=0)       &   0 \le t \le t_1 \\
S_A(t)         &   t_1  \le   t  \le t_2  \\
S_A(t=\infty)  &   t_2 \le t  
\end{cases}  \,.  \label{eq:SAt}
\end{equation} 
We display in figure \ref{fig:Suniversal} (left) the time evolution of the entanglement entropy of interval $A$ of figure \ref{fig:Tttd1}, from a numerical computation of the geodesic equations. In this and subsequent figures we display the entanglement entropy renormalized as shown in equation (\ref{eq:Sreg}). Let us focus on the regime $t_1  \le   t  \le t_2 $. It is convenient to define the normalized entanglement entropy~$f_A(\rho)$ as
\begin{equation}
f_A(\rho ) \equiv \frac{S_A(t) - S_A(t=0)}{S_A(t=\infty) - S_A(t=0)} \qquad \textrm{with} \qquad \rho \equiv (t-t_1)/\Delta t  \,, \label{eq:fA} 
\end{equation}
where $\Delta t = \ell = |x_R - x_L|$. This corresponds to the function $S_A(t)$ normalized to the interval [0,1] in both horizontal and vertical axes for $t_1\leq t\leq t_2$. It is clear from equations(\ref{eq:SAt}) and (\ref{eq:fA}) that $f_A(\rho)$ has the values $f_A(0)=0$ and $f_A(1)=1$. We have computed numerically the entanglement entropy $S_A(t)$ in number of configurations with different temperatures $T_L$, $T_R$ and lengths~$\ell$, and find that for a large range of temperature differences, the behavior of $f_A(\rho)$ may be approximated by
\begin{equation}
f_A(\rho ) \simeq 3\rho^2-2\rho^3  \,, \qquad 0 \le \rho \le 1 \,. \label{eq:fAuniversal}
\end{equation}
Specifically, this function fits extremely well the numerical results of the entanglement entropies as long as $\ell T_L < 1$ and $\ell T_R < 1$. This is illustrated in figure \ref{fig:Suniversal} (right) for a particular case. The result of equation (\ref{eq:fAuniversal}) is independent of the values of the parameters $T_L$, $T_R$ and $\ell$, and so it implies the existence of an {\it 'almost'} universal time-evolution of entanglement entropy in the theory with $d=2$ at small temperatures. Eq.~(\ref{eq:fAuniversal}) will be proven analytically in section~\ref{sec:universalformula} within a small temperature expansion.

The analysis presented above applies also to intervals in the negative semiplane. We show in figure \ref{fig:Suniversal}~(left) the entanglement entropy of interval $B$ of figure \ref{fig:Tttd1}.  Note that both functions, $S_A(t)$ and $S_B(t)$, tend to the same value when the intervals reach the steady-state regime.

\subsection{Time evolution of mutual information}
\label{subsec:mutual_information}

A quantity of interest related to the entanglement entropy is the {\it mutual information}. It measures which information of subsystem $A$ is contained in subsystem $B$, or in other words the amount of information that can be obtained from one of the subsystems by looking at the other one.  An advantage of this quantity is that it is finite, so that it does not need to be regularized. It is defined as
\begin{equation}
I(A,B) = S_A + S_B - S(A \, \cup B) \,,
\end{equation}
where, holographically, 
\begin{equation}
S(A \, \cup B) = \textrm{min} \bigg\{ S_A + S_B , S_1 + S_2 \bigg\} \,,
\end{equation}
and $S_1$ and $S_2$ are defined as the entanglement entropy of the intervals $[x_R^B,x^A_L]$ and $[x_L^B,x^A_R]$ respectively, see figure \ref{fig:Tttd1}. Note that the mutual information satisfies $I(A,B) \ge 0$. This corresponds to the simplest example of inequalities of entanglement entropies in a system involving a number of subsystems. See also section \ref{sec:Inequalities} for a further discussion. 

\begin{figure*}[htb]
\begin{tabular}{cc}
\includegraphics[width=7.2cm]{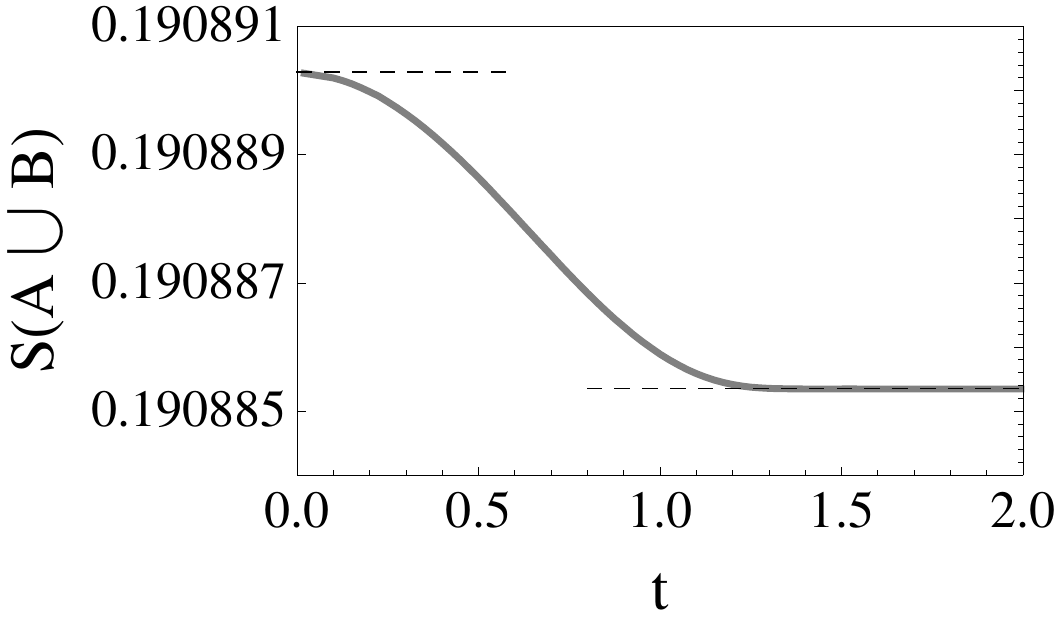} & 
\includegraphics[width=7.2cm]{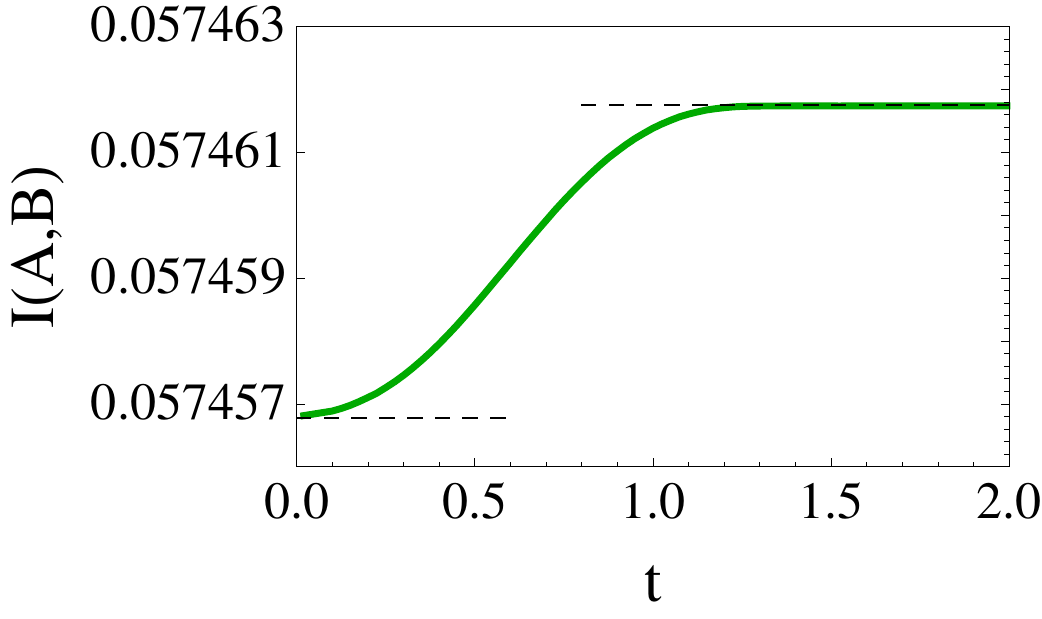}
\end{tabular}
\caption{ (Left) Renormalized entanglement entropy of $A\, \cup B$ as a function of time, see figure \ref{fig:Tttd1}. (Right) Mutual information of $A$ and $B$ as a function of time. In these figures the (dashed) horizontal lines correspond to the results by using the analytical formulas, equations(\ref{eq:Sstepwise}), (\ref{eq:St0}) and (\ref{eq:Sboost}). The intervals correspond to $x^1 \in [-0.175,0.175]$ and $x^2 \in [-1.35,1.35]$, and temperatures $T_L = 0.2$ and $T_R=0.195$. We have set $G=1$ and $L=1$.}
\label{fig:mutual_information}
\end{figure*}

We have numerically studied the time evolution of $S(A\, \cup B)$ and the mutual information $I(A,B)$. The results are shown in figure \ref{fig:mutual_information}. An important property that we can infer from this result is that, contrary to the entanglement entropy $S_A$ or $S_B$, the mutual information always grows with time, i.e.
\begin{equation}
\partial_t I(A,B) \ge  0 \,. \label{eq:dtI}
\end{equation}
We have checked this property for a large number of configurations, with different temperatures and intervals, and it always remains valid\footnote{An analytical guess for the time evolution of the mutual information in analogy with the universal formula of equation (\ref{eq:fAuniversal}) turns out to be more complicated than in previous section, due to the structure of the term $S(A\, \cup B)$.}. This seems to imply that in the boundary picture, the shockwaves transport information about the presence of the other heat bath throughout the system. Note that while the hypersurfaces describing the shockwave in the bulk are spacelike and can hence not carry information in the bulk picture (as explained in section \ref{sec:Holographic-Setup}), the shockwaves are null on the boundary, and hence they can be interpreted to transport information from the boundary perspective.  

\subsection{Conservation of entanglement entropy}
\label{sec:conservation_entanglement_entropy}

Let us consider the two extreme regimes $t=0$ and $t \to \infty$. It is possible to obtain analytical results for the entanglement entropies in these cases for the model with $d=2$ presented in section \ref{sec:Holographic-Setup}. On the spacelike slice defined by $t=0$, the metric corresponds to two black holes of different temperature located to the left and right of $x=0$ each, i.e.
\begin{equation}
ds^2 = ds_L^2 \theta(-x) + ds_R^2 \theta(x) \,,
\end{equation}
and in this case the entanglement entropy for an interval $[x_L,x_R]$ with $x_L< 0$ and $x_R>0$ becomes, using minimal subtraction, 
\begin{eqnarray}
&&S(T_L, x_L;T_R, x_R) =  \label{eq:Sstepwise}\\
&&\qquad\qquad \frac{L}{2G}\log\left[\frac{T_\mathrm{L}\cosh\left(\pi T_\mathrm{L}x_\mathrm{L}\right)\sinh\left(\pi T_\mathrm{R}x_\mathrm{R}\right)-T_\mathrm{R}\sinh\left(\pi T_\mathrm{L}x_\mathrm{L}\right)\cosh\left(\pi T_\mathrm{R}x_\mathrm{R}\right)}{\pi T_\mathrm{L}T_\mathrm{R}}\right]  \,. \nonumber 
\end{eqnarray}
To obtain this expression we considered the length of a curve piecewise defined in the two heat baths, consisting of two pieces at $x<0$ and $x>0$ glued together at $x=0$. They are parts of two geodesics, defined in a geometry with a black hole at temperature $T_\mathrm{L}$ and $T_\mathrm{R}$, respectively. The curve is chosen such that its length is minimal with respect to the value of the radial coordinate at which the two geodesics meet at $x=0$. For a similar matching-based method see section \ref{sec:Matching}.

If we place the interval in just one semiplane, i.e. $x_{L,R} > 0$ (or $x_{L,R} < 0$), the entanglement entropy at $t=0$ corresponds to the one for a stationary black hole at temperature~$T$, which reads
\begin{equation}
S(T,\ell;t=0) = \frac{L}{2G} \log \left( \frac{1}{\pi T}  \sinh( \pi \ell T) \right) \,, \qquad \ell := |x_R - x_L| \,.  \label{eq:St0}
\end{equation}
In this equation $T=T_{L}$ (or $T_R$) when $x_{L,R}<0$ (or $x_{L,R}>0$).  In the other extreme, $t \to \infty$, the system is in the steady-state regime, and the entanglement entropy is the one for a boosted black hole, \footnote{Note that equation (\ref{eq:Sboost}) is valid when $t \ge \max(|x_L|,|x_R|)$ if the initial profile $F(v)$ in equation (\ref{eq:fLfR}) is a stepwise function, i.e. in the limit $\alpha \to \infty$. When~$F(v)$ is a smooth function, the right-hand side of equation (\ref{eq:Sboost}) corresponds to the asymptotic value of the entanglement entropy at very late times, i.e. for $t \gg \max(|x_L|,|x_R|)$.}
\begin{equation}
S(T_L,T_R,\ell;t=\infty) =  \frac{L}{4G} \log \left( \frac{1}{\pi^2 T_L T_R} \sinh\left( \pi \ell T_L \right)  \sinh\left( \pi \ell  T_R \right) \right) \,. \label{eq:Sboost}
\end{equation}
These analytical results, equations(\ref{eq:St0}) and (\ref{eq:Sboost}), correspond to $S_A(t=0)$ and $S_A(t=\infty)$ in equation (\ref{eq:SAt}), respectively. From these formulae we easily obtain the property
\begin{equation}
S_A(t=0) + S_B(t=0) = S_A(t=\infty) + S_B(t=\infty) \,, \label{eq:SASBconservation}
\end{equation}
where we  consider intervals $A$ with $x^A_{L,R}>0$, and $B$ with $x^B_{L,R}<0$, and lengths $\ell = \ell_A = \ell_B$. This property is non-trivial, as in the left-hand side of equation (\ref{eq:SASBconservation}) there is the contribution of stationary black hole solutions at temperatures $T_L$ and $T_R$, while in the right-hand side there is a boosted black hole and the corresponding energy flow contributes as well to the entanglement entropy. This relation is significant as it implies the {\it 'conservation'} of entanglement entropies between $t=0$ and $t=\infty$. However, there is a non-trivial time evolution at intermediate times, as we discuss below. Interestingly, \eqref{eq:SASBconservation} has also been obtained in a slightly different setup in \cite{Hoogeveen:2014bqa}.

\begin{figure*}[t]
\begin{tabular}{cc}
\includegraphics[width=7.15cm]{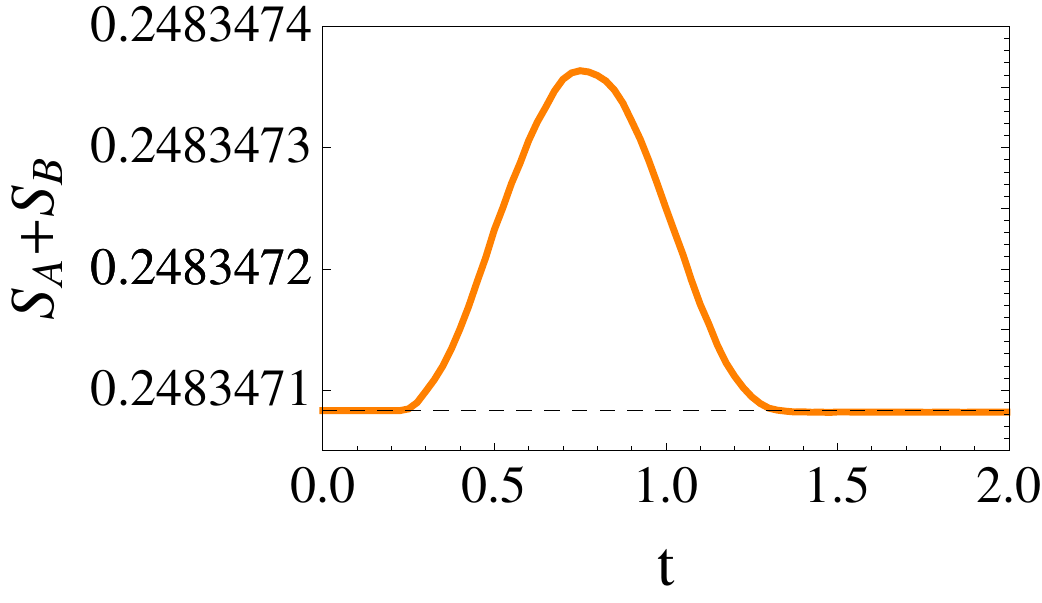} & 
\includegraphics[width=6.15cm]{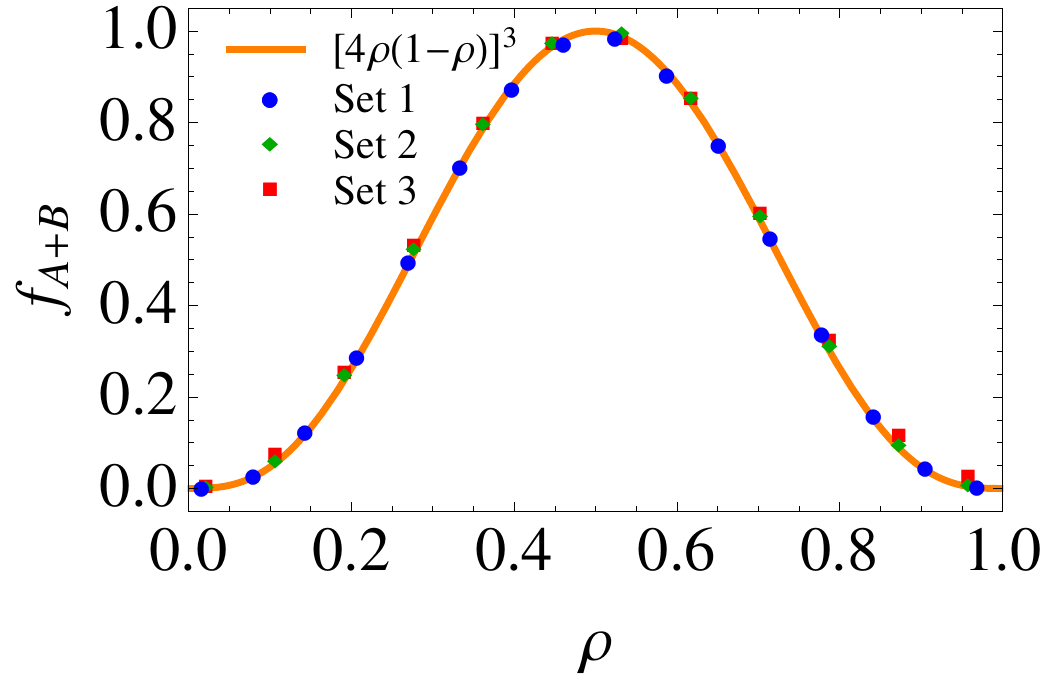}
\end{tabular}
\caption{ (Left) Renormalized entanglement entropy $S_A + S_B$ as a function of time, see figure \ref{fig:Tttd1}.  The (dashed) horizontal line corresponds to the result by using the analytical formulae (\ref{eq:St0}) and (\ref{eq:Sboost}). We have set $G=1$ and $L=1$. (Right) Renormalized entanglement entropy $S_{A+B}$ as a function of time, normalized to $[0,1]$ in both horizontal and vertical axes, see (\ref{eq:fAB}). The dots correspond to the numerical result with intervals $A$ and $B$, placed symmetrically with respect to $x=0$ as shown in figure \ref{fig:Tttd1}, in different configurations: Set 1 is $(T_L=0.2, T_R=0.195, \ell_A = \ell_B =  1.175)$, Set 2 is $(T_L=0.2, T_R=0.175,  \ell_A = \ell_B  = 1.175)$ and Set 3 is $(T_L=0.2, T_R=0.175,  \ell_A = \ell_B = 1.475)$. The continuous line is the universal behavior~$f_{A+B}(\rho) = \left[ 4\rho(1-\rho)\right]^3 $.}
\label{fig:SASBpeak}
\end{figure*}

In figure \ref{fig:SASBpeak} (left), the time evolution of $S_{A+B} \equiv S_A + S_B$ is displayed. We see that our numerics confirm the conservation law of equation (\ref{eq:SASBconservation}). In the next subsection we will study this system in the quenching regime, i.e. $t_1 \le t \le t_2$ in equation (\ref{eq:SAt}), and characterize the violations of the entanglement entropy conservation in this case. 

\subsection{Non-universal effects in time evolution}
\label{subsec:non_universal}

As it is shown in figure \ref{fig:SASBpeak} (left), we find from our numerics that $S_{A+B}(t) \ne const$ in the quenching regime. This implies that the entanglement entropy is not conserved at intermediate times. A straightforward computation shows that these non-conservation effects are only possible if there are non-universal contributions in equation (\ref{eq:fAuniversal}), otherwise this equation would predict $S_{A+B}(t) = const$.

In the following we restrict to intervals $A$ and $B$ with the same length and placed symmetrically with respect to $x=0$, i.e. $\ell_A = \ell_B$ and $x_{L,R}^A = -x_{R,L}^B$. While the function $S_{A+B}(t)$ has the same value at $t=0$ and $t=\infty$ (see  (\ref{eq:SASBconservation})), we find from our numerics that it has a maximum at $t_{\textrm{max}} \approx (t_1 + t_2)/2$. In order to characterize the time evolution of $S_{A+B}(t)$, let us define the normalized entanglement entropy
\begin{equation}
f_{A+B}(\rho) \equiv \frac{S_{A+B}\left( t \right) - S_{A+B}(t=0)}{S_{A+B}(t_{\textrm{max}})-S_{A+B}(t=0)} \qquad \textrm{with} \qquad \rho \equiv (t-t_1)/\Delta t  \,,  \label{eq:fAB}
\end{equation}
where $t_1$ and $\Delta t$ are defined as in equation (\ref{eq:fA}). Finally, from a numerical computation of~$f_{A+B}(\rho)$ in a variety of intervals, we find that its behaviour is  well-approximated by
\begin{equation}
f_{A+B}(\rho) \simeq \left[ 4\rho(1-\rho)\right]^3 \,, \qquad 0 \le \rho \le 1 \, . \label{eq:fABuniversal}
\end{equation}
This is illustrated in figure \ref{fig:SASBpeak} (right) for several configurations. From a combination of the results in (\ref{eq:fAuniversal}) and (\ref{eq:fABuniversal}), we conclude that for small temperatures, the normalized entanglement entropy defined in equation (\ref{eq:fA}) can be approximated by
\begin{equation}
f_A(\rho) = 3 \rho^2 - 2 \rho^3 + \Delta_A(\rho)   \,, \quad  \textrm{with} \quad \Delta_A(\rho) \simeq   C(T_L, T_R, \ell)  \cdot [4\rho(1-\rho)]^3 \,. \label{eq:fA2}
\end{equation}
The factor $C(T_L, T_R, \ell)$ has a non-universal dependence on the parameters of the interval, so that $\Delta_A(\rho)$ is a non-universal contribution to $f_A(\rho)$. Note, however, that $C(T_L,T_R,\ell)$ does not appear to affect the universal behavior of $f_{A+B}(\rho)$, see  (\ref{eq:fABuniversal}). Some remarks deserve to be mentioned: On the one hand, $\Delta_A(\rho)$ is a correction of order ${\cal O}(\rho^3)$, so that it does not jeopardize the early-time behavior $S_A(t) \sim t^2$ which is present in a wide variety of systems, see e.g.~\cite{Balasubramanian:2010ce,Balasubramanian:2011ur,Balasubramanian:2011at,Hartman:2013qma,Liu:2013iza,Li:2013sia,Liu:2013qca,Kundu:2016cgh,OBannon:2016exv,Lokhande:2017jik}. On the other hand, the effect of $\Delta_A(\rho)$ is extremely small in the configurations we  studied numerically.\footnote{One can see from figure \ref{fig:SASBpeak} (left) that in this case the peak in $S_{A+B}(t_{\textrm{max}})$ is a correction of order ${\cal O}(10^{-6})$ with respect to $S_{A+B}(0)$, so that the order of magnitude of the non-universal contribution in equation (\ref{eq:fA2}) is 
\begin{equation}
C(T_L,T_R,\ell) \simeq \Delta_A\left(\rho=\frac{1}{2}\right)  \simeq \frac{1}{2} \frac{ S_{A+B}(t_{\textrm{max}}) - S_{A+B}(0)  }{ S_A(\infty) - S_A(0) } \sim {\cal O}(10^{-4}) \,.
\end{equation}
} The range of validity of equation \eqref{eq:fAuniversal} will be further discussed in sections \ref{sec:Matching} and \ref{sec:AnalyticalResults}.

\newcommand{\rdm}{(\ref{eq:discont1}, \ref{eq:discont2})}
\newcommand{\rcm}{(\ref{eq:gxx}, \ref{eq:fLfR})}
\newcommand{\CSP}{{\b (CSP)}}
\newcommand{\CSE}[1]{{\b (ref: #1)}}
\newcommand{\Sss}{S_\text{\ss}}
\newcommand{\ttt}{\tilde{t}}
\newcommand{\xtt}{\tilde{x}}
\newcommand{\zt}{\tilde{z}}

\section{Numerical results II: Matching of  geodesics}
\label{sec:Matching}
This section is devoted to an approach complementary to solving the geodesic equation as above -- a method that we refer to as the matching approach. The idea is rather simple and based on the same principle as the variational derivation of the light refraction law. In short: we take the discontinuous shockwave geometry, where the metric is piecewise constant and coincides  either with the standard or the boosted Schwarzschild metric. We calculate geodesics in each of these two spacetime regions and parametrize them by the positions of two points: One of these is located where the geodesic meets the conformal boundary of AdS, and the other where the geodesic meets the shockwave. We take two geodesics that reach the shockwave at the same point, each of them being located in one of the two regions of spacetime. We  add their (renormalised) lengths and extremise the sum with respect to the position of the 'meeting point' at the shockwave. The value of the length at the extremum yields the desired entanglement entropy of an interval enclosed by the 'boundary' endpoints of our geodesics. Having painted the procedure by a broad brush, we shall now describe some technical details and assumptions made to carry out this procedure.

\subsection{Setup and assumptions}
Let us take the metric in its piecewise form \rdm{}. The metric is a piecewise smooth function of Fefferman-Graham coordinates, denoted by $\zt, \ttt, \xtt$ in this section. We use the name \emph{region} to refer to the whole subset of our space on which the metric coefficients are smooth, e.g \emph{steady state region} (denoted $S_\text{\ss}$) is given by $\ttt>0,~ |\xtt|<\ttt,~\zt\in \left(0, (\pi{}\sqrt{T_L T_R})^{-1}\right)$. In the same manner, the left and right thermal regions will be denoted by $L$ and $R$, respectively.  The dimension two surface along which the metric is discontinuous will be referred to as \emph{the shockwave}. Our aim is to calculate (regularised) geodesics length between two points lying on the conformal boundary of the space-time. If both endpoints belong to the same region, the answer is already known to be \eqref{eq:sbtz} in a thermal region and \eqref{eq:boostedEE} in the steady state region. If, however, the boundary points belong to different regions, finding the solution is a more complicated task. We therefore make some assumptions about the spacetime. \\
Most importantly, we assume that the coordinates $\zt, \ttt, \xtt$ cover all the regions in a smooth way, and it is actually \emph{the metric components} as functions of $\zt, \ttt, \xtt$ that are discontinuous\footnote{Note, however, that as the Israel junction conditions \eqref{eq:israel} are satisfied, the metric is continuous in a strict mathematical sense. Especially, the induced metrics on the shockwave both from the static side and from the steady state side agree.}. This assumption can be motivated by the fact that our metric \rdm{}~can be obtained as a limit of a continuous metric \rcm{}~where the shockwave width ($w$) tends to zero ($\alpha$ goes to infinity). It is reasonable to assume that taking the limit described does not influence the domain of our coordinates. The agreement between numerical results from the continuous model at large $\alpha$ and the  results of this section shall confirm that the assumption made  yields correct results. From the assumption follows in particular that curves which are continuous in our coordinates are also continuous on the manifold itself. This will be essential in our calculation, since it is based on joining two smooth curves in a way that it is still continuous.
\begin{figure}[hbtp]
\centering
\includegraphics[scale=0.5]{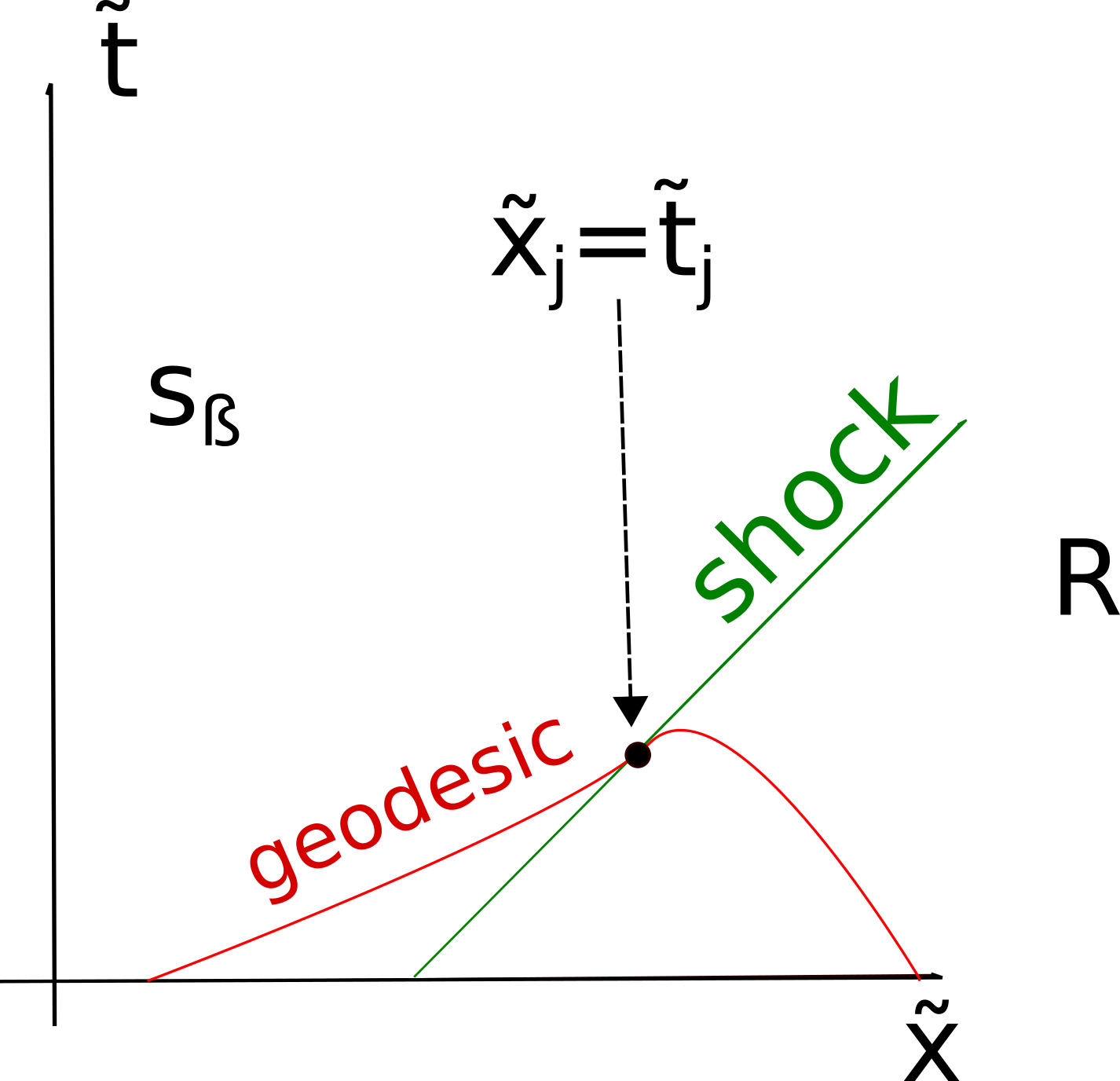}
\caption{A cartoon of a curve used in our procedure, projected onto a $\ttt,x$ hypersurface. $\Sss$ and $R$ denote regions of spacetime, the steady state and right thermal region, respectively. The red line is a piecewise geodesic  (it is a geodesic in any of both regions) connecting boundary endpoints and a point on a shockwave ($\xtt=\ttt$ in our coordinates). Then the (renormalised) length of that geodesic is extremised with respect to the coordinates of the joining point,  which yields the entanglement entropy.}
\label{graf:JoiningCartoon}
\end{figure}
\subsection{Geodesics and distance}

Given our setup, we need to calculate a spacelike distance between a given point on the boundary and an arbitrary point on the shockwave\footnote{Of course on a Lorentzian manifold, not every point on the shockwave will be spatially separated from a fixed point on the boundary, as we shall directly see later. It is enough that there will always be a set of points that satisfy this condition -- a situation that indeed occurs in our case.}. To achieve this, we shall follow the logic of \cite{Shenker:2013pqa} and utilise the fact that every three-dimensional asymptotically AdS manifold that is a solution to Einstein's equations is locally isometric to AdS$_3$. So, if we identify the isometry, we may use the ready formul{\ae} for geodesic distance in AdS$_3$ space. It is worth noting that thanks to this property special to three dimensions, we do not need to calculate the geodesics explicitly, we just express their length as a function of coordinates of the two points. This considerably simplifies the calculation.  However, we still  have to find formul{\ae} for the geodesic distance in our problem. The metric \rdm{}~is given in  Fefferman-Graham (FG) coordinates. To use the results of \cite{Shenker:2013pqa}, we need to change to Schwarzschild-type coordinates. There is a technical difficulty in that step: in every region the change of coordinates takes a different form. We denote FG-type coordinates as $(\zt,~\ttt,~\xtt)$ and the Schwarzschild coordinates as $(z,~t,~x)$. In Schwarzschild coordinates,  the metric takes the form \eqref{eq:btz}
\begin{equation}
\dd{s}^2=\frac{L^2}{z^2}\left[ -\left(1-(2\pi T z)^2\right)\dd{t}^2+\frac{\dd{z}^2}{1- (2\pi T z)^2} +\dd{x}^2 \right]\, ,
\label{eq:ThermalSchw} 
\end{equation}
where $T$ is a real, positive constant -- an (effective) temperature. Then, the coordinate transformations are obtained as follows.
For the steady state they read
\begin{align}
z&=\frac{\zt}{1+\pi^2T_LT_R\zt^2},\label{eq:ztST0}\\
\zt&=\frac{1-\sqrt{1-4\pi^2 T_LT_Rz^2}}{2\pi^2 T_LT_R z}\label{eq:ztST},\\
\left(\begin{array}{c}
t\\
x
\end{array} 
\right)&=%
\left(\begin{array}{c c}
\ch{\theta}&-\sh{\theta}\\
-\sh{\theta}&\ch{\theta}
\end{array}\right) 
\left(\begin{array}{c}
\ttt\\
\xtt
\end{array} 
\right),\label{eq:BoostST}\\
\sh{\theta}&=\frac{T_L-T_R}{2\sqrt{T_LT_R}},~\ch{\theta}=\frac{T_L+T_R}{2\sqrt{T_LT_R}}. \nonumber
\end{align}
\eqref{eq:ztST0} is the inverse relation to \eqref{eq:ztST}. We quote both since there is a sign to be fixed.  From \eqref{eq:BoostST} it follows that the effective temperature from equation \eqref{eq:ThermalSchw} for the steady state can be expressed in terms of the reservoirs' temperatures as 
\begin{equation}
T = \sqrt{T_L T_R}.
\end{equation}
For the thermal regions it is sufficient to take \eqref{eq:ztST0} and set $T_L=T_R$, $\theta=0$ to obtain
\begin{equation}
\zt = \frac{1-\sqrt{1-4\pi^2z^2T_{L/R}^2}}{2\pi^2T_{L/R}^2z}. \label{eq:ztT}
\end{equation}
The distance can be expressed, following \cite{Shenker:2013pqa}, as\footnote{Note that, on contrary to the setup therein, we do not use an analytical continuation of coefficients since we are interested in a single-sided black hole, not a double sided one which is the situation there.}
\begin{equation}
\label{eq:geoDAdS}
\ch{\frac{d}{L}} = T_1 T_1^\prime + T_2 T_2^\prime -X_1X_1^\prime - X_2X_2^\prime
\end{equation}
with $d$ the geodesic distance.  In terms of the Schwarzschild coordinates, the functions $X_1,~X_2,~T_1,$ and $T_2$ read
\begin{align}
T_1&=\frac{z_H \sqrt{1-\frac{z^2}{z_H^2}} \sinh \left(\frac{t}{z_H}\right)}{z}, \\
T_2&=\frac{z_H \cosh \left(\frac{x}{z_H}\right)}{z}, \\
X_1&=\frac{z_H \sqrt{1-\frac{z^2}{z_H^2}} \cosh \left(\frac{t}{z_H}\right)}{z}, \\
X_2&=\frac{z_H \sinh \left(\frac{x}{z_H}\right)}{z},
\end{align} 
with Schwarzschild horizon radius $z_H=(2\pi T)^{-1}$. From the formulae above it is clear that if one of the points is taken to the boundary ($z=0$) while the other is being kept fixed, the distance must diverge. Therefore, a regularisation is needed for small $z$. Since from \eqref{eq:ztST0} and \eqref{eq:ztT} it follows that in every region
$$
z=\tilde{z}+\mathcal{O}(\tilde{z}^3),
$$
there is no difference in which coordinates we regularise. For concreteness, let us sketch the procedure of computing the regularised length for a case in which one end of the geodesic reaches the boundary in the steady state region and the other in the right thermal region with temperature $T_R$ (that is the case of figure \eqref{graf:JoiningCartoon}). So, we are interested in two lengths: one for the curve connecting the `starting point' on the boundary ($\tilde{t}_b,~\tilde{x}_{min},~z=\epsilon{}$) with a point on the shockwave $\tilde{x_j},~\tilde{x_j},~\tilde{z}_j$ and another joining the same point on a shockwave with the endpoint on a boundary on another side ($\tilde{t}_b,~\tilde{x}_{max}, z=\epsilon{}$), and then take a 'regularised limit' $\epsilon \rightarrow 0$ (i.e.~subtracting the divergent part and then taking the limit). To apply \eqref{eq:geoDAdS}, we need to connect these to the Schwarzschild coordinates of the respective patches. Let us note that the condition for the position of the shockwave is identical in any of the used coordinates,
$$
x_j=t_j \Leftrightarrow \tilde{x}_j=\tilde{t}_j.
$$
For simplicity, we regularise in Schwarzschild coordinates. Using the asymptotic approximation of hyperbolic cosine by an exponential, we arrive at the conclusion that the minimal counter-term used in \eqref{eq:Sreg} is indeed the proper one to regularise our length. At this point it is convenient to set the \AdS{}~radius to unity, $L=1$.\\
Now, we are ready to write the full, renormalised distance as a function of the joining point on the shockwave,
\begin{align}
d_R(z_j, x) =& \log \left[\left(1+\pi ^2 T_R^2 \zt_j^2\right) \cosh \left({2 \pi  T_R (x-\ell)}\right)-(1-(\pi  T_R \zt_j)^2) \cosh \left({2 \pi  T_R (t-x)}\right)\right]+
\nonumber \\
&\log\Big[\left(1+\pi ^2 T_L T_R \zt_j^2\right) \cosh \left({\pi  (t T_L-t T_R+2 T_R x)}\right)
\label{eq:RegLenRight}
\\
&+\left(\pi ^2 T_L T_R \zt_j^2-1\right) \cosh \left({\pi  (t (T_L+T_R)-2 T_R x)}\right)\Big]
 -\frac{1}{2} \log \left(16 \pi ^8  T_L^2 T_R^6 \zt_j^4\right).
\nonumber
\end{align}
In the above, we used a shortened notation: $\ell=\xtt_{max}-\xtt_{min}$, $x=\xtt_j-\xtt_{min}$, $t=\ttt_b-\xtt_{min}$ -- the time that has passed since the shockwave entered the boundary interval.
Now, this quantity is to be extremised with respect to, $\zt_j$ and $x$ (which is the same as extremising w.r.t $\xtt_j$). Extremal points are solutions to
\begin{equation}
\partial_{\zt_j} d_R = 0 \, , \qquad
\partial_{x}d_R = 0 \, .
 \label{eq:extremise}
\end{equation}
These equations turn out to be fourth order polynomial equations in $z_j$ and non-polynomial\footnote{Even upon expressing hyperbolic functions in terms of exponentials and changing variable from $x$ to $\xi=\exp(A x)$, the exponents of new the variable are non-integer for any choice of $A$.} equations in $x$. Therefore, we turn to numerical methods for solving non-linear algebraic equations. Note however that the above-mentioned system can be solved analytically in certain simplifying cases, see section \ref{sec:AnalyticalResults}. Upon solving the system \eqref{eq:extremise}, we obtain the coordinates of the extremum of $d_R$, namely ($z_0,~x_0$). Then the desired entanglement entropy is given by\footnote{In that place we have already set Newton's constant of supergravity theory to 1.}
\begin{equation}
S=\frac{1}{4}d_R(z_0, x_0).
\end{equation}
\subsection{Numerical algorithm}
To solve the non-linear algebraic system \eqref{eq:extremise}, we need to involve numerical analysis. Our solution was developed in Wolfram Mathematica. All the codes used in this section are available online as supplementary material to the arXiv submission of the paper. The algorithm consists of two steps: First, following the idea of \cite{wagon2010mathematica} in a slightly modified form (see \cite{GraphAnalysi}),
we find a rough approximation of the solution by plotting curves satisfying each of the two equations in \eqref{eq:extremise}. Then we use coordinates of crossing points of those as a starting point for standard Newton's solver built-in Mathematica's \verb+FindRoot+ function. To understand why such a two-step procedure is necessary, let us briefly discuss the function \eqref{eq:RegLenRight} and equations \eqref{eq:extremise}. \\
 \begin{figure}[hbtp]
 \centering
 \includegraphics[width=.5\textwidth]{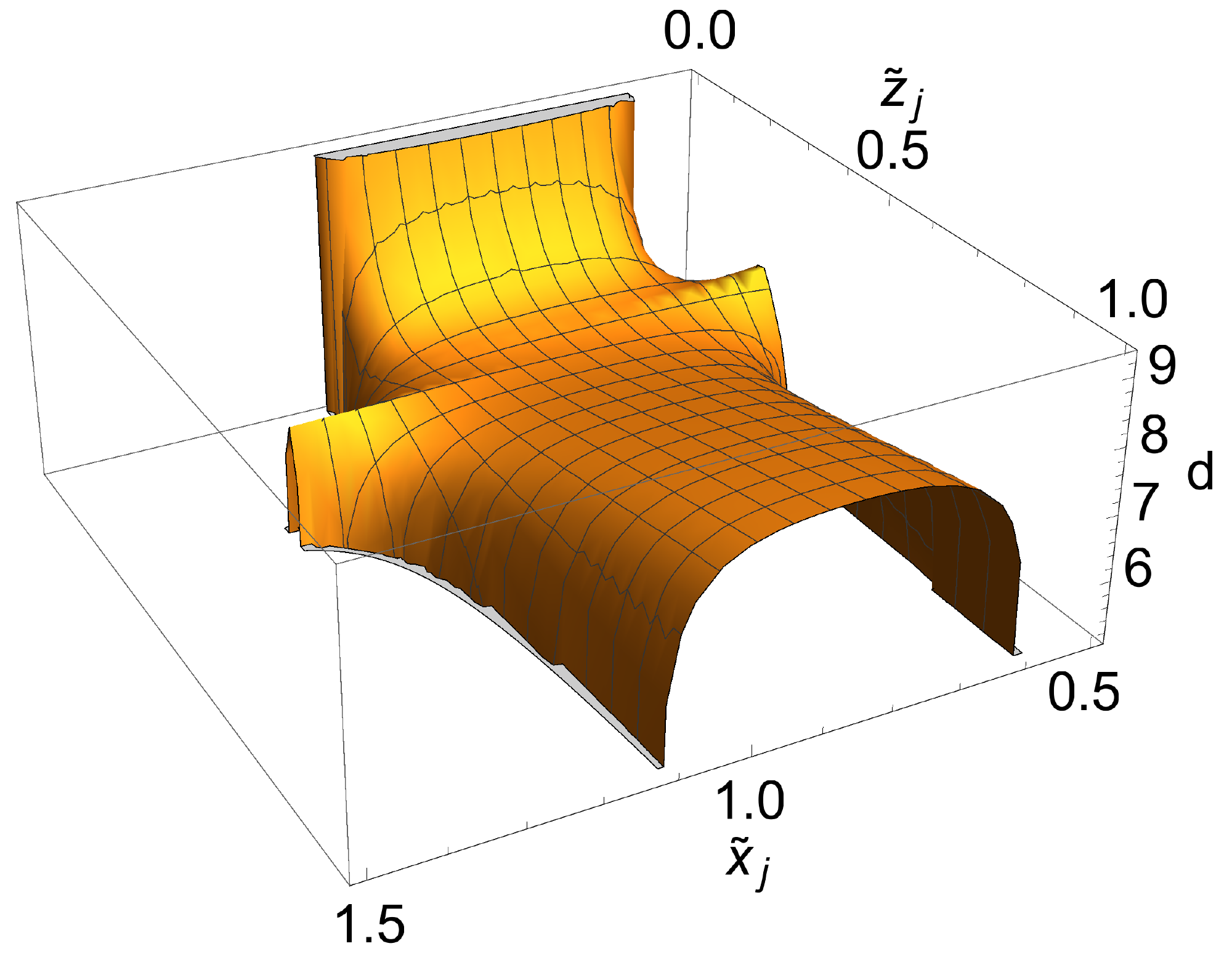}
 \caption{The distance function $d_R(z_j,x_j)$ for parameters: $t_b=0.7,~x_{min}=0.5,~x_{max}=1.5,~T_L=2,T_R=1$ in units where $L=1$. The function turns imaginary outside of some region (where the argument of the logarithm in \eqref{eq:RegLenRight} turns negative).}\label{fig:Distance}
 \end{figure}
 As figure \ref{fig:Distance} shows, the domain of the distance function and the the full domain of our coordinates are not the same. This is in agreement with the fact that on a Lorentzian manifold, not every point is spacelike-separated from a given point. Then, the function going to zero and ceasing to be real signals that one of the boundary endpoints becomes null or time-like separated from the joining point. However, since $d_R= \log(...)$, both equations \eqref{eq:extremise} have the form $\partial \log (f(...)) = 0$, so looking for their solution is equivalent to solving $\partial f(...) =0$ if $f$ does not vanish on the solution. This equivalent form is strongly preferable, given the nature of the numerical  computations, in which unnecessary divisions decrease numerical precision. On the other hand, the modified system of equations consists of two functions that are well-defined for the whole domain of our coordinates. Therefore, we begin our numerical approach with an algorithm capable of finding rough approximations of \emph{all} solutions to the system of equations in a given domain. From those solutions we choose the ones satisfying our requirement that the length evaluated on solution is positive. Then, these solutions are  refined by using Newton's method that yields the solution with the desired accuracy, in our case fifteen digits. If more than one solution is regular in the sense that the length is positive, the final answer is taken to be the one for which the value of length is smallest, according to the HRT proposal. The domain in which we search for solutions is (in terms of variables defined in \eqref{eq:RegLenRight})
 \begin{equation}
 x\in [0; L],~ \zt_j\in [0; 1.0001*(\pi \sqrt{T_L T_R})^{-1}] 
 \end{equation}
 on the left side, or
  \begin{equation}
 x\in [0; L],~ \zt_j\in [0; 1.0001*(\pi T_L)^{-1}] 
 \end{equation}
on the right\footnote{Note that we always assumed $T_L>T_R$.}, which ensures that solutions lying far below the horizon are excluded.  This however allows the algorithm to look for the solution arbitrarily close to the horizon, and  to boost the data generation.

To justify the exclusion of solutions with $\zt_j>\zt_H$, we have  numerically tested  that the solutions lying below the horizon, should they appear, are not the physical ones. The argument behind this is based on the analysis of the Kruskal diagram of our space-time (see figure \ref{graf:Kruskal}). In short, we see that the shockwave does not cross the horizon except for the bifurcation surface -- it stays entirely in the outer region of the black hole. {This means that the point where the geodesic crosses the shockwave will generically be outside of the horizon ($\zt<\zt_H$), and since this is a causality argument, this occurs both from the point of view of the static region and from the point of view of the steady state region. Indeed, we find that a solution to our matching equations with these properties always seems to exist.} Therefore, the fact that we can find a solution beneath the horizon ($\zt_j>\zt_H$) is only an artefact of our choice of coordinates. On the numerical side we allow, for test purposes, $\zt_j$ to exceed the above mentioned bounds by large values ($2-3$ times larger), and the solutions found in those regions were never chosen by the algorithm. With the restricted domain of interest and given accuracy, our algorithm has an acceptable speed: computing the entanglement entropy for a given interval and a given boundary time takes roughly $0.2$ seconds. 
\subsection{Results}
\begin{figure}[hbtp]
\centering
\begin{tabular}{cc}
\includegraphics[width=0.49\textwidth]{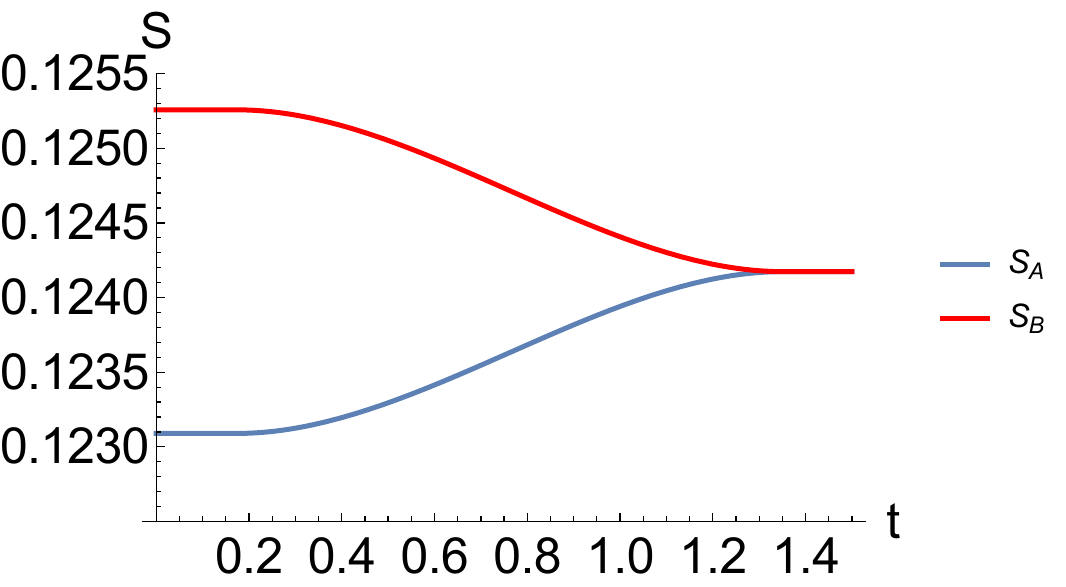}&\includegraphics[width=0.49\textwidth]{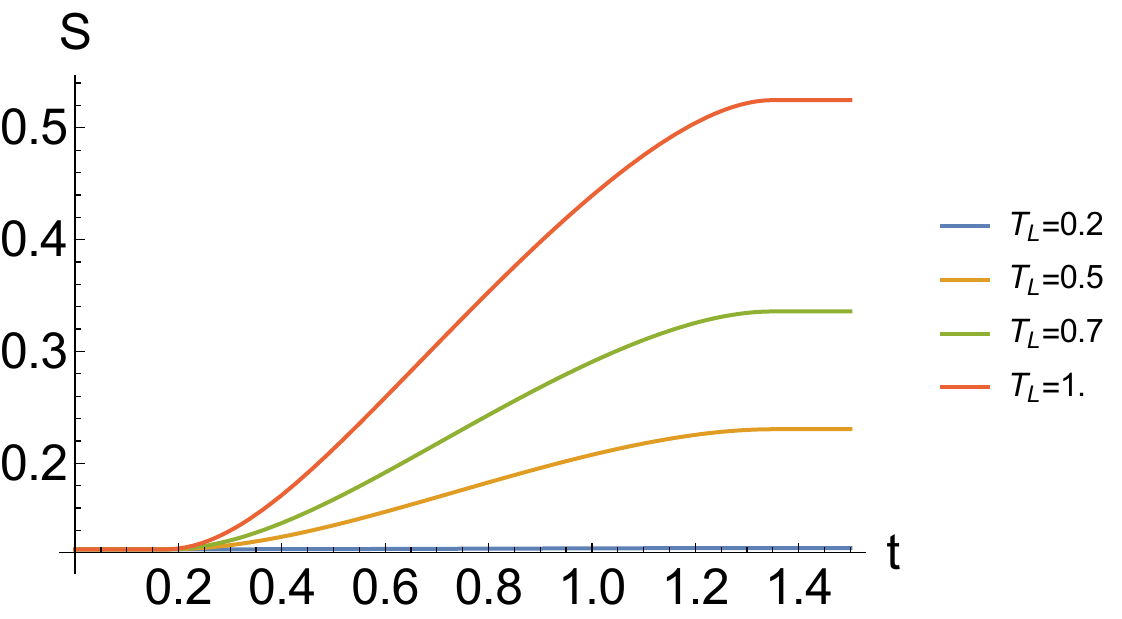} 
\end{tabular}
\caption{Comparison of entanglement entropies for intervals $A$ and $B$ at temperatures $T_L=0.2$ $T_R=0.195$ ({\bf left panel}), and entanglement entropies for the interval $A$ in different temperatures -- with $T_R=0.195$ and $T_L$ ranging from $0.2$ (blue curve) to $1.$ (orange) as functions of boundary time $t$. The left panel shows exact matching with previous results from figure \ref{fig:Suniversal}, which is a consistency check. The {\bf right panel} shows how the evolution changes when one gradually increases the temperature of one of the heat baths. All lengths in units of \AdS{}~radius ($L=1$).}\label{fig:EntanglementEntropies}
\end{figure}

\begin{figure}[hbtp]
\centering
\begin{tabular}{cc}
\includegraphics[width=.49\textwidth]{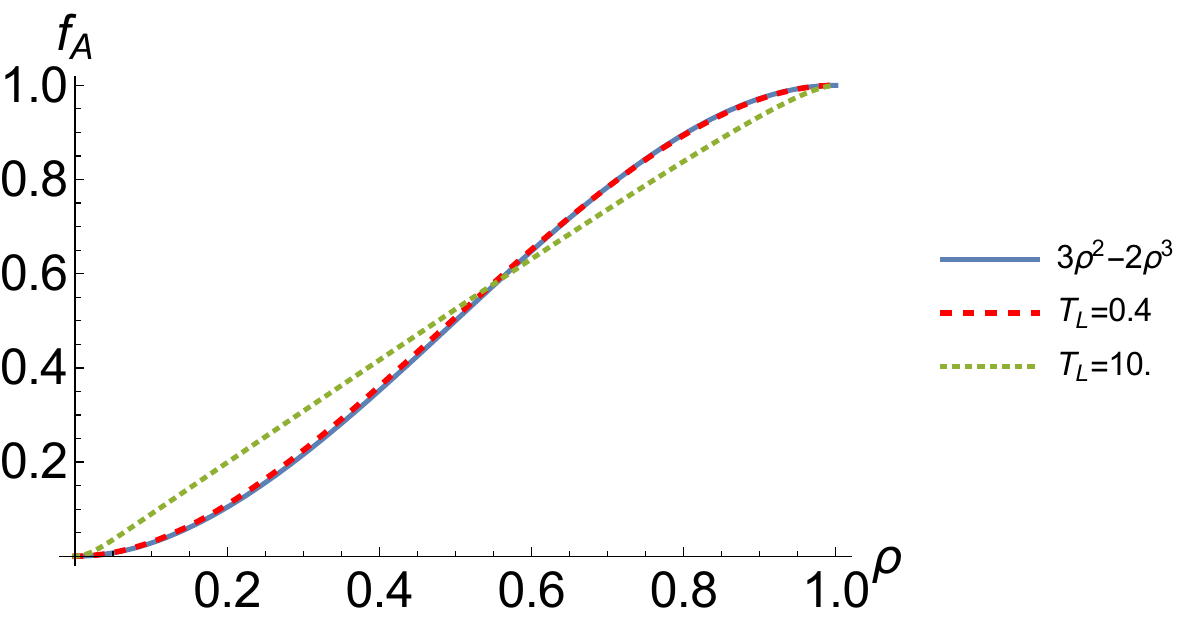} &\includegraphics[width=.49\textwidth]{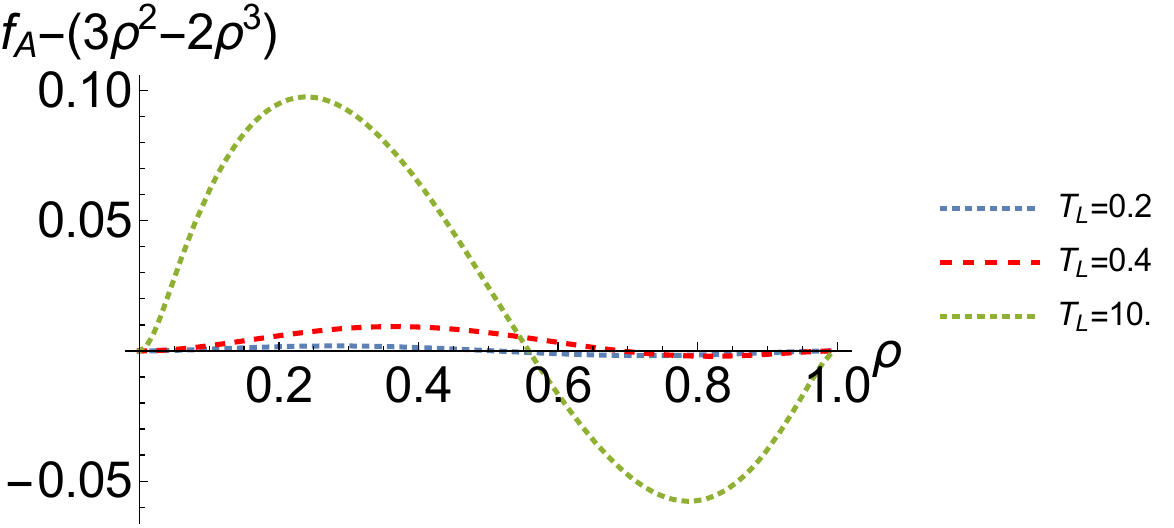} 
\end{tabular}
\caption{Normalised entanglement entropy $f_A$ for the interval $A$ in temperatures $T_L=0.4$ and $T_L=10.$ compared to the universal formula $3\rho^2-2\rho^3$ ({\bf left panel}) and deviations from universal formula for $T_L=0.2,~0.4,~10.$ ({\bf right panel}) The $T_L=0.2$ case was already shown on figure \ref{fig:UniversalFormula}. In that case the deviation is approximately $0.002$. Upon increasing the temperature of the left bath, the deviation from the universal formula grows and the time evolution resembles more a straight line -- however the difference $f_A(t)-t$ still reaches values around $0.04$ for $T_L=10.$. See section \ref{sec:Tsunami} for a discussion of the high temperature linear behaviour. All lengths in units of \AdS{} radius ($L=1$).}
\label{fig:UniversalFormula}
\end{figure}

\begin{figure}[hbtp]
\centering
\begin{tabular}{cc}
\includegraphics[width=.49\textwidth]{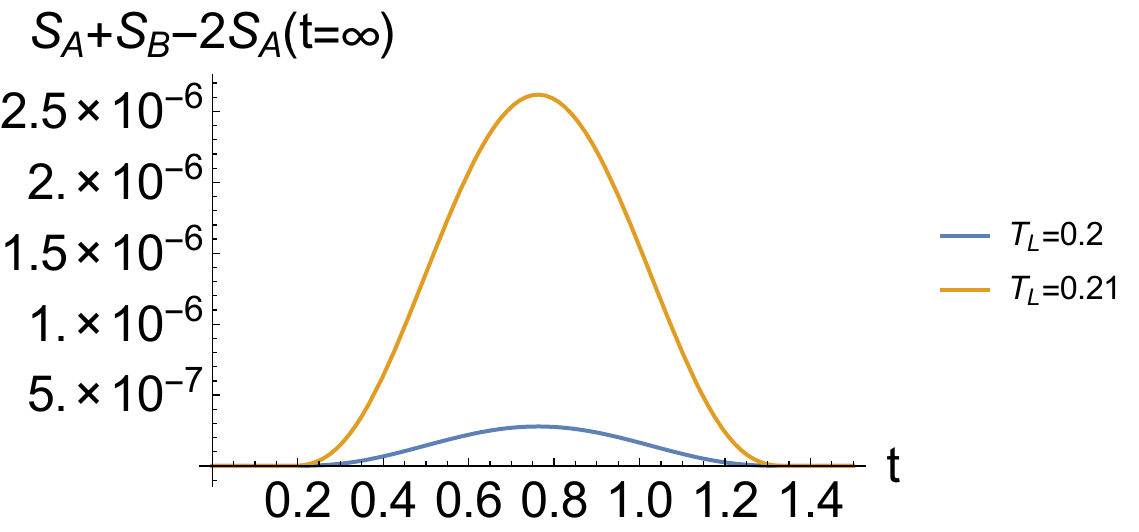} & \includegraphics[width=.49\textwidth]{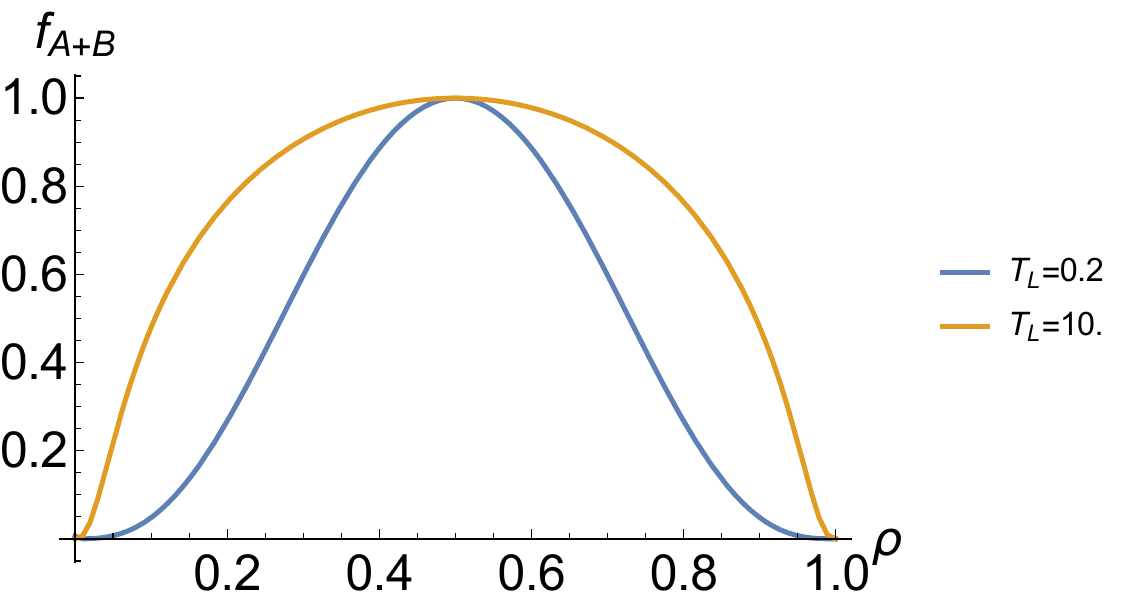} \\
\includegraphics[width=.49\textwidth]{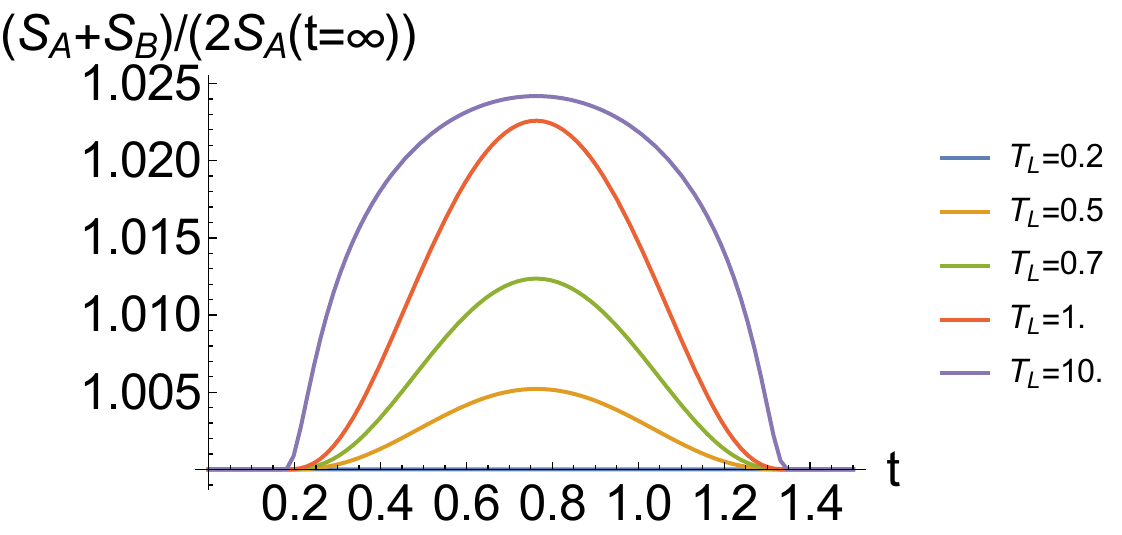} &\includegraphics[width=.49\textwidth]{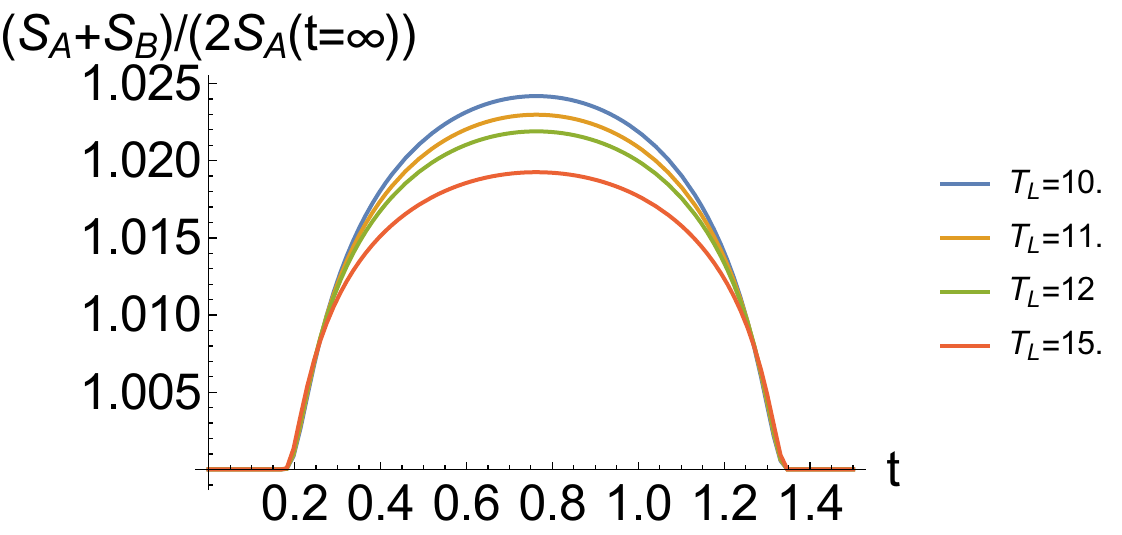}
\end{tabular}
\caption{{\bf Top left:} Sum of $S_A$ and $S_B$ for two different, yet close, values of $T_L$, shifted by the asymptotic ($t=\infty$) value of that sum. The blue plot is the case studied in section \ref{sec:numerics}, as previously the deviation from constancy is of order $10^{-7}$ while the sum is of order $10^{-1}$. The yellow plot shows a similar quantity for temperature $T_L=0.21$, just slightly higher. The deviation is now order of magnitude bigger. {\bf Top right:} Normalised deviation $f_{A+B}$ for different temperatures $T_L$. The blue curve has been already shown in figure \ref{fig:SASBpeak} to follow the universal behaviour $[4\rho(1-\rho)]^3$. For much bigger temperature $T_L=10$, the shape of the curve changes considerably. {\bf Bottom:} Ratio of sum $S_A+S_B$ and asymptotic value of that sum as a function of boundary time. The deviation of the sum from its asymptotic value reaches around $2.5\%$, but seems to be bounded even when one increases the temperature -- compare left and right figures which are in the same scale. The bigger the temperature difference, the more the curve resembles a semi-circle. All lengths in units of \AdS{}~radius ($L=1$).}\label{fig:NonConservation}
\end{figure}

In this way we obtain entanglement entropy for a wide range of temperatures. To compare with results of the previous section, we consider properties of entanglement entropy of the same boundary regions, namely
\begin{equation}
A=[0.175, 1.35],~B=[-1.35,-0.175].\label{eq:intervals}
\end{equation}
All our data is generated using \AdS{}~radius as unit, $L=1$. We are also going to take various temperature differences. In that subsection by $t$ we denote the \emph{boundary time}, to stick to the conventions of the previous section. The first result, shown in figure \ref{fig:EntanglementEntropies}, indicates that both our methods (of this and the preceding section)  yield the same results for the same initial data. That ensures us about the correctness of our results, as numerical techniques used in both approaches are substantially different. 
Now, let us analyse the universal formula for normalised entanglement entropy \eqref{eq:fA}. Using the joining method we are not only able to prove the universal formula (see section \ref{sec:AnalyticalResults}), but also find in what range it is broken. The results on the universal formula for $f_A$ can be seen on figure \ref{fig:UniversalFormula}.
Finally, we reconsider the question of non-conservation of the sum $S_A+S_B$ with the alternative numerical approach of ths section. The results of figure \ref{fig:SASBpeak} pass convergence tests, however the peak is tiny compared to the value of entropy (difference in 6-th decimal). 
Here we confirm the observed behaviour of $S_A+S_B$ in the alternative numerical approach.
Our findings are presented on figure \ref{fig:NonConservation}. An interesting fact is that we numerically find that the normalised sum
\begin{equation}
\frac{S_A(t)+S_B(t)}{2S_A(\infty)}
\end{equation}
is bounded from above by a value of approximately $1.025$. So, the maximal deviation from a constant appears to be at most 2.5\% of the value of entropy --which is however too much to attribute it to a numerical error.


\section{Analytical results}
\label{sec:AnalyticalResults}

Some of the numerical results presented in the two preceding sections
may actually be obtained analytically, at least in some limits. This applies in particular to
the result \eqref{eq:fAuniversal} for the time evolution of the
entanglement entropy. Moreover, we derive a bound on the increase rate for the entanglement entropy. 
We begin this section in \ref{sec:velocityintro} by a review of velocity bounds previously obtained for global quenches. We then obtain analytical results for the time evolution of the entanglement entropy for the limit of both heat bath temperatures small in  section  \ref{sec:universalformula} , and of one of the two temperatures vanishing in section 
\ref{sec:zerotemperature}. Finally we obtain a velocity bound on entanglement growth in section \ref{sec:Tsunami}.

\subsection{Review of universal velocity bounds}
\label{sec:velocityintro}

In the past, much interest has been directed towards the holographic study of the time dependent behaviour of entanglement entropy after global quenches. For example, in a series of papers \cite{AbajoArrastia:2010yt,Balasubramanian:2010ce,Balasubramanian:2011ur,Balasubramanian:2011at,Hartman:2013qma,Liu:2013iza,Li:2013sia,Liu:2013qca}\footnote{See also \cite{Alishahiha:2014cwa} for the case of a background geometry with a hyperscaling violating factor.}, it was found that after a global quench, the entanglement entropy of a sufficiently large boundary region would exhibit an initial quadratic growth of entanglement with time,
\begin{align}
\Delta S(t)\propto t^2+...,
\label{quadratic}
\end{align}
followed by a universal linear growth regime where
\begin{align}
\Delta S(t) = v_E s_{eq} A_{\Sigma}t+...\ .
\label{tsunamiformula}
\end{align}
In this formula, $t$ is the time after the quench, $\Delta S$ is the
change in entanglement entropy, $s_{eq}$ is the entropy density of the
(late time) equilibrium thermal state, $A_{\Sigma}$ is the surface
area of the boundary region $\Sigma$ of which the entanglement entropy
is computed\footnote{$\Sigma$ is assumed to be large compared to the
  inverse temperature of the final equilibrium state. In the case
  $d=2$ where $\Sigma$ only consists of two endpoints of an interval
  (for a connected region), one sets $A_{\Sigma}=2$
  \cite{Calabrese:2005in,Liu:2013qca}.} and $v_E$ is a velocity that
depends on the final equilibrium state. In the case of an
AdS-Schwarzschild black hole as final state, it was found that  \cite{AbajoArrastia:2010yt,Liu:2013iza,Li:2013sia,Liu:2013qca}
\begin{align}
v_E=\frac{\sqrt{d}(d-2)^{\frac{1}{2}-\frac{1}{d}}}{\left(2(d-1)\right)^{1-\frac{1}{d}}},
\label{vE}
\end{align} 
which is referred to as the \textit{entanglement velocity} or \textit{tsunami velocity}. The reason for this nomenclature is that the behaviour \eqref{tsunamiformula} can be understood in terms of a heuristic picture in which the entanglement growth is caused by entangled quasi-particles that were created by the global quench and are propagating at the speed \eqref{vE}, forming the \textit{entanglement tsunami}. See also \cite{Bai:2014tla,Leichenauer:2015xra,Ziogas:2015aja,Tanhayi:2015cax,Hartman:2015apr,Casini:2015zua,Mezei:2016zxg,Mezei:2016wfz} for further work on this topic. A related concept is the so-called \textit{butterfly velocity} \cite{Shenker:2013pqa,Roberts:2014isa}
\begin{align}
v_B=\sqrt{\frac{d}{2(d-1)}}
\label{vB}
\end{align}
for the spatial propagation of chaotic behaviour in the boundary theory. This speed is also connected to the growth of operators in a thermal state \cite{Shenker:2013pqa,Roberts:2014isa}. From \eqref{vE} and \eqref{vB}, it is obvious that
\begin{align}
1\geq v_B\geq v_E,
\end{align}
and the case $d=2$ is the special case where $1= v_B= v_E$. Interestingly, the velocities seem to play an important role in the description of entanglement spreading not only for global quenches, but also for local quenches \cite{Rangamani:2015agy,Astaneh:2014fga,Rozali:2017bco}.

\subsection{Time evolution of entanglement entropy: Both temperatures small}
\label{sec:universalformula}

Based on the matching procedure outlined in section
\ref{sec:Matching}, it is easy to prove the universal formula
\eqref{eq:fAuniversal} for the time evolution of entanglement entropy as an approximation for small $T_L$ and $T_R$. In order to do so, we simply replace $T_L\rightarrow \delta\cdot T_L$, $T_R\rightarrow \delta\cdot T_R$ and expand the expressions for $\partial_{z_j}d_R$ and $\partial_x d_R$ in \eqref{eq:extremise} in the small quantity $\delta$.\footnote{This means that in this section, we assume that $T_L$ and $T_R$ are both small (compared to the interval length $\ell$), but of the same order of magnitude.} Similarly, we expand  the analogous function $d_L$ and its derivative when $x_{max}\leq0$. To lowest non-trivial order in $\delta$, we  find
\begin{align}
\partial_x d_R\propto \partial_{z_j}d_L\propto(\ell-t)(\ell+2t-4x)t-(\ell-2t)z_j^2,
\\
\partial_{z_j}d_R\propto \partial_x d_L \propto (\ell-t)(t-2x)(\ell+t-2x)t+z_j^4.
\end{align}
The condition $\partial_{z_j}d_R=\partial_x d_R=0$ (respectively $\partial_{z_j}d_L=\partial_x d_L=0$) has then the simple solution  
\begin{align}
x=t, z_j=\sqrt{(\ell-t)t}.
\end{align}
This may be inserted into $d_R$ in equation \eqref{eq:RegLenRight} and the similar expression for $d_L$, giving the entanglement entropy $S(t)$, and this in turn can be inserted into \eqref{eq:fA}. Expanding again in small $\delta$ as above, we then find for both the left- and right-side the analytic result
\begin{align}
f(\rho)=3\rho^2-2\rho^3
\label{universalderived}
\end{align}
at order $\delta^0$. Here, $\rho$ is again the rescaled time defined in \eqref{eq:fA}. It is interesting to note that the small $T_L,T_R$ expansions leading to \eqref{universalderived} loose their analytic validity at an order of magnitude of $T_L,T_R$ at which our numerics are still well approximated by \eqref{universalderived}. It might hence be interesting to do the $T_L,T_R$ expansions in a more systematic way and to study the higher orders in more detail. This might also help to understand the range of validity of equation \eqref{eq:fA2}.

It is worth stressing that the universal dynamics of entanglement entropy is not only of purely academic interest. It is known that the low-energy spectrum of excitations of some models (i.e systems with ballistic conductance, possessing quasiparticle description, see \cite{Bernard:2016nci}) are governed by effectively conformal theories. The regime in which this approximation is valid for thermal states is indeed when both of the temperatures are low, so the highest lying parts of spectrum are not largely populated in the thermal state -- which is also the range of validity of our universal formula \eqref{eq:fAuniversal}. This means that in the limit of small temperatures our universal evolution of entanglement entropy should be also valid in ballistic regimes of real, i.e electronic systems.
It is therefore an interesting possible direction of investigation to
compute other quantities, as correlation functions, in this
low-temperature limit and compare them against expectations from other
theories (i.e. lattice models).

\subsection{Entanglement Entropy: Limit of zero temperature for one of the heat baths}
\label{sec:zerotemperature}

In addition to the case where \textit{both} temperatures $T_L$ and $T_R$ are small (studied in section \ref{sec:universalformula}), there is another situation where the matching equations derived in section \ref{sec:Matching} can be solved analytically: The case where $T_R=0$ with arbitrary $T_L$ (or the analogous case $T_L=0$ with arbitrary $T_R$, which we will not consider separately).   

First of all, let us reassure ourselves that this case is actually physical. Setting $T_R\rightarrow0$ in equations \eqref{BBB}-\eqref{theta} has the consequences
\begin{align}
T&\rightarrow 0
\\
\chi&\rightarrow+\infty
\\
\beta&\rightarrow+1
\\
\theta&\rightarrow+\infty.
\end{align}
Despite the divergence of the rapidity $\theta$, we see that for $d=2$, the line element \eqref{eq:boosted} of the boosted black hole has a well-defined limit
\begin{align}
ds^2&\rightarrow\frac{L^2}{z^2}\left(dz^2+(-1+\pi^2T_L^2z^2)dt^2+(1+\pi^2T_L^2z^2)dx^2-2\pi^2T_L^2z^2dtdx\right).
\label{BBB0}
\end{align}
Similarly, instead of \eqref{eq:RegLenRight}, we find the expression
\begin{align}
d_R(z_j,x)=&\log \Big{[}4 \pi ^3 \left(\ell^2-2 \ell x-t^2+2 t x+z_j^2\right) 
\nonumber
\\
&\times
\left(\pi  T_L z_j^2 \cosh \left(\pi  t T_L\right)-(t-2 x) \sinh \left(\pi  t T_L\right)\right)\Big{]}
\label{drR0}
\\
&-\frac{1}{2}  \log \left(16 \pi ^8 T_L^2 z_j^4\right)
\nonumber
\end{align}
which can be shown to be extremised for
\begin{align}
x= \frac{\pi  t T_L (\ell-t) \coth \left(\pi  t T_L\right)+\ell+t}{2+2\pi  T_L (\ell-t) \coth \left(\pi  t T_L\right)},\ 
z_j=\sqrt{\frac{\ell (\ell-t)}{1+\pi  T_L (\ell-t) \coth \left(\pi  t T_L\right)}}.
\end{align} 
This yields the analytic solution
\begin{align}
S(\ell)\propto\log \left(\ell \sinh \left(\pi  t T_L\right)\right)+\log\left(1+\pi  T_L (\ell-t) \coth \left(\pi  t T_L\right)\right)-\log (\pi  T_L).
\label{TR0solution}
\end{align}
Expanding around $T_L=0$ yields to lowest order the universal formula \eqref{eq:fAuniversal} again, however \eqref{TR0solution} is an analytical result for the entanglement entropy which is valid for \textit{any} $T_L$. For large $T_L$, we obtain
\begin{align}
S(\ell)\approx \frac{L}{4G} \pi t T_L=s_{eq}t,
\label{TR0TLlarge}
\end{align}
where we have used \eqref{seq} with $T_R\equiv0$. This result shows that for large $T_L$, the entanglement entropy will increase in an approximately linear way, saturating the bound to be discussed in section \ref{sec:Tsunami}. Note the discrepancy of a factor $A_{\Sigma}=2$ between \eqref{TR0TLlarge} and equation \eqref{tsunamiformula} for $d=2$, where $v_E=1$. This may be because in a global quench, the entanglement tsunami enters the interval of interest from both sides, while in our local quench like setup, the shockwave enters the interval only from one side. However, we should stress that in contrast to the entanglement tsunamis studied in global quenches in \cite{AbajoArrastia:2010yt,Hartman:2013qma,Liu:2013iza,Li:2013sia,Liu:2013qca,Bai:2014tla,Leichenauer:2015xra,Ziogas:2015aja,Tanhayi:2015cax,Hartman:2015apr,Casini:2015zua,Mezei:2016zxg,Mezei:2016wfz},
we do not have a heuristic quasi-particle like picture explaining the entropy increase \textit{and decrease} experienced by different intervals in our inhomogeneous setup even qualitatively. We will discuss the possible bounds on the entropy increase or decrease rate of different intervals in our system for general $T_L,T_R$ in section \ref{sec:Tsunami}.  

\subsection{Bounds on entropy increase rate}
\label{sec:Tsunami} 

As shown in section \ref{sec:numerics}, for the full time-dependent background \eqref{eq:FG} we expect that for an interval of length $\ell$, the time dependent entanglement entropy $S(\ell,t)$ will be constant before the shockwave enters the interval, evolve with time $t$ while the shockwave passes through the interval, and be constant again after the shockwave has left the interval. This is precisely the behaviour seen in figure \ref{fig:Suniversal}, for example. Furthermore, normalising the entropy such as to obtain a dimensionless quantity, we have observed in section \ref{sec:numerics} that for small temperatures $T_L,T_R$, the time dependence of the entanglement entropy can be approximated by the formula \eqref{eq:fAuniversal}, which we analytically proved in section \ref{sec:universalformula}. In this section, we will have a closer look at the rates of entropy increase that we observe in the time dependent entanglement entropies $S(\ell,t)$. 
\\

We begin by analytically deriving some useful expressions.  For a sharp shockwave moving at the speed of light, the change in the entanglement entropy of an interval with length\footnote{Again, in this section we assume that the interval is completely to the left or to the right of the origin $x=0$.} $\ell$ will occur over a time period $\Delta t = \ell$. From \eqref{eq:boostedEE} and \eqref{eq:sbtz} (with 
$T\rightarrow T_L$ for example, as appropriate when the interval is entirely on the left) we then easily find the average entropy increase rate
\begin{align}
v_{av}\equiv\frac{\Delta S}{\Delta t}=\frac{L}{4G\ell}\log\left(\frac{T_L \sinh(\pi\ell T_R)}{T_R \sinh(\pi\ell T_L)}\right).
\end{align}
In particular, we find
\begin{align}
0\leq |v_{av}|\leq\frac{L}{4G} \pi |T_R-T_L|
\label{vav}
\end{align}
with $\lim_{\ell\rightarrow0}v_{av}=0$ and $\lim_{\ell\rightarrow\infty}v_{av}=\frac{L}{4G}\pi |T_R-T_L|$. This is interesting, because it implies that for fixed $T_L$ and $T_R$, $v_{av}$ is bounded. By choosing $T_L$ and $T_R$, however, we can  make $v_{av}$ as large as we want. 

In \eqref{tsunamiformula}, the rate of entropy increase was normalised by the entropy density $s_{eq}$ of the final state. Taking the limit $\ell\rightarrow\infty$ in \eqref{eq:boostedEE}, we find that
\begin{align}
s_{eq}=\frac{L}{4G}\pi(T_L+T_R).
\label{seq}
\end{align}
Hence, motivated by the comparison with the literature on entanglement tsunamis \cite{AbajoArrastia:2010yt,Hartman:2013qma,Liu:2013iza,Li:2013sia,Liu:2013qca,Bai:2014tla,Leichenauer:2015xra,Ziogas:2015aja,Tanhayi:2015cax,Hartman:2015apr,Casini:2015zua,Mezei:2016zxg,Mezei:2016wfz}, we can define the normalised average entanglement entropy increase rate 
\begin{align}
\tilde{v}_{av}\equiv\frac{v_{av}}{s_{eq}},
\label{tildevav}
\end{align}
and hence we find the bound
\begin{align}
|\tilde{v}_{av}|\leq\left|\frac{T_R-T_L}{T_R+T_L}\right|\leq1.
\label{averagerate}
\end{align}
We consequently see that, when normalising in a specific physical way, both the average increase and decrease rate of entanglement entropy in the formation of the steady state will be bounded by a similar bound as observed for two dimensional entanglement tsunamis or the local quenches of  \cite{Rangamani:2015agy}.

\begin{figure}[htbp]
	\centering
	\includegraphics[width=0.75\textwidth]{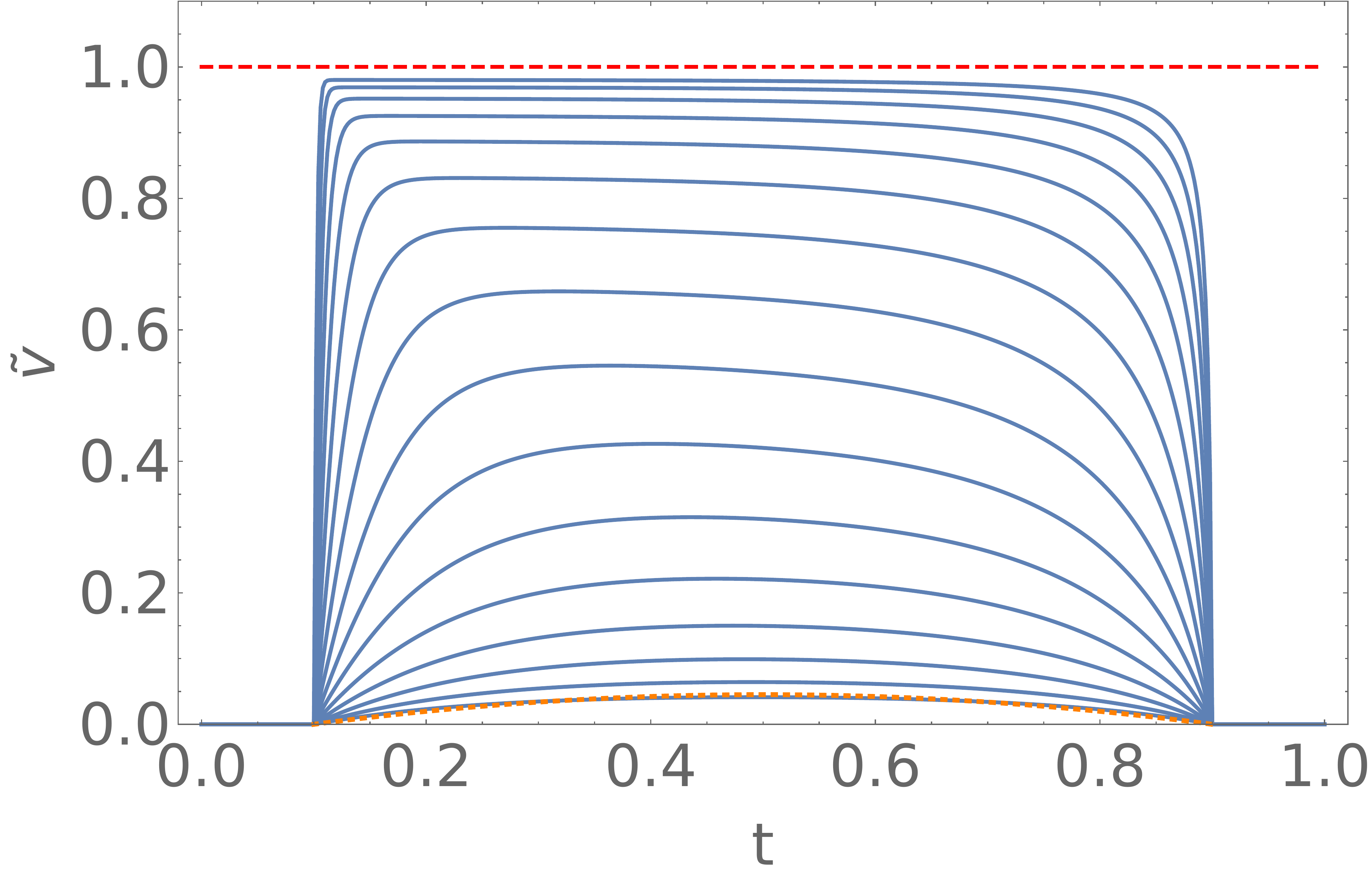}
	\caption{Entropy increase rates \eqref{rate} for an interval from $x_{min}=0.1$ to $x_{max}=0.9$ as a function of time $t$. We have chosen a fixed $T_R=1$ and $T_L=1+10^\alpha$ with $\alpha=-1,-0.8,-0.6,...,2$, where $\alpha$ increases from the lowest solid curve in the figure to the highest one. The dotted (orange) line represents the curve expected from the universal low temperature formula \eqref{eq:fAuniversal} for $T_R=1,\ T_L=1.1$. The dashed (red) line signifies the bound \eqref{bound2}.   }
	\label{fig::ratebound}
\end{figure}

It should be noted however that the bound \eqref{averagerate} is only a bound on the \textit{averaged} increase rate of entanglement entropy. If the universal formula \eqref{eq:fAuniversal} would hold for \textit{any} choice of $T_L,T_R,\ell$, then the momentary entanglement entropy increase rate could violate the bound \eqref{averagerate} by up to $50\%$ at the moment when the shockwave is in the middle of the interval. But as discussed earlier the formula \eqref{eq:fAuniversal} is not valid for any choice of $T_L$,$T_R$ and $\ell$, and numerically we find that the momentary normalised entropy increase (or decrease) rate
\begin{align}
\tilde{v}\equiv\frac{1}{s_{eq}}\frac{dS(\ell,t)}{dt}
\label{rate}
\end{align}
still satisfies the bound 
\begin{align}
|\tilde{v}|\leq1
\label{bound2}
\end{align}
in all examples that we have explicitly checked. See, for example, figure \ref{fig::ratebound}. Furthermore, using the analytical result \eqref{TR0solution} of section \ref{sec:zerotemperature}, we can compute the momentary increase rate
\begin{align}
\tilde{v}(T_R=0)=\frac{1}{\frac{1}{ \pi \ell T_L-\pi t T_L}+\coth(\pi t T_L)}.
\end{align}
For this result, it can be analytically shown that for any parameters $T_L,\ell$ and $t$ the bound \eqref{bound2} is satisfied. This is in contrast to the results of \cite{Kundu:2016cgh}, where it was explicitly found in a different setup that the momentary increase rate for small regions, far away from the tsunami regime, \textit{can} indeed violate the velocity bound \eqref{bound2}. See also \cite{OBannon:2016exv,Lokhande:2017jik} for further discussions of entanglement entropy growth for small subsystems in different setups. 

A bound of the type \eqref{averagerate} is especially interesting when compared to other velocities that are related to the spread of entanglement or other disturbances on the boundary of AdS$_{d+1}$, such as the \textit{entanglement velocity} \eqref{vE} and the \textit{butterfly effect velocity} \eqref{vB}. As said before, the case $d=2$ is the special case where $1= v_B= v_E$, and hence the bound \eqref{averagerate} can be expressed in terms of $v_E$ and/or $v_B$. As we will see in section \ref{sec:Higher-DimensionsAnalytical}, this may however not be the case for higher dimensions any more.

Nevertheless, we can attempt to interpret our findings for $2+1$ bulk dimensions in terms of the intuition provided by the study of the entanglement tsunami phenomenon. As noted in section \ref{sec:zerotemperature} in discussing the result \eqref{TR0TLlarge}, in the limit where the entanglement entropy increases linearly ($\ell\gg T_L^{-1}$) with a rate saturating the bound \eqref{bound2}, the shockwave seems to take the role that the entanglement tsunami had for a global quench. As the shockwave enters the interval only from one side instead of from both sides, the increase rate in \eqref{TR0TLlarge} is only half of the one calculated in a global quench according to \eqref{tsunamiformula}, where $A_{\Sigma}=2$ and $v_E=1$. As pointed out in section \ref{sec:velocityintro}, the linear increase \eqref{tsunamiformula} is only valid when looking at large enough boundary regions (compared to the inverse of the temperature). Our analytical results \eqref{TR0TLlarge} and especially \eqref{eq:fAuniversal} then show how this linear behaviour is modified when moving away from this limit: The linear increase of entropy characteristic of the entanglement tsunami is replaced by a much smoother S-shaped curve. This might, in analogy with the tsunami picture, be called an \textit{entanglement tide}. It should be pointed out that in our matching procedure of section \ref{sec:Matching}, the shockwave is always assumed to be infinitely thin, hence this modification is \textit{not} a result of a finite shockwave size. Also, other works where the evolution of entanglement entropy away from the tsunami regime was studied are \cite{Kundu:2016cgh,OBannon:2016exv,Lokhande:2017jik}, with somewhat contrasting results, as explained above.


\section{Entanglement entropies of systems with many disconnected components}
\label{sec:Inequalities}

\begin{figure}[htbp]
	\centering
	\includegraphics[width=0.3\textwidth]{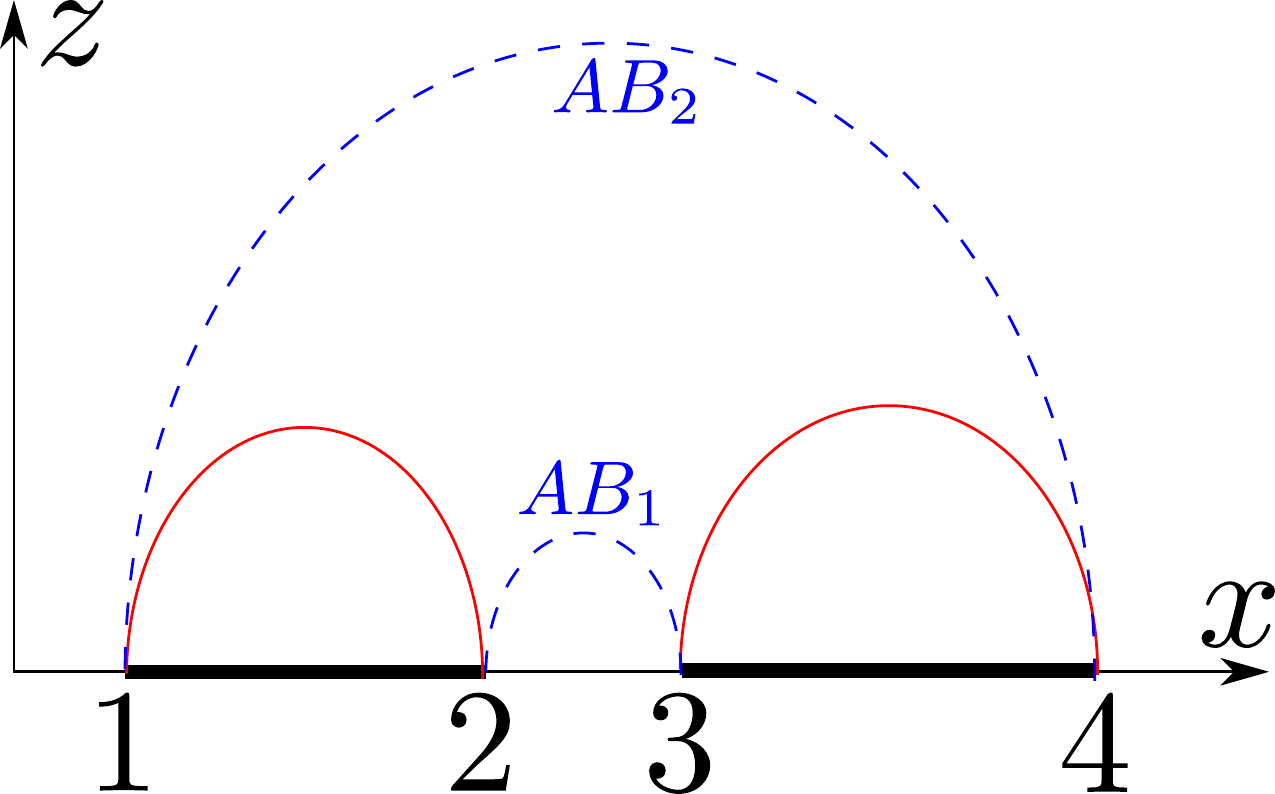}
	\caption{For the union of the two intervals $A$ (from 1 to 2) and $B$ (from 3 to 4), there are two possible ways how 
		to calculate the entanglement entropy $S(A\cup B)$. One is by adding the entanglement entropies of the two intervals 
		$A$ and $B$ (given by the solid red curves), the other is by adding the two curves $AB_1$ and $AB_2$, depicted as 
		dashed blue curves. }
	\label{fig::n2case}
\end{figure}

When working in $1+1$-dimensional CFTs, the subsystems for which
entanglement entropy may be calculated are either isolated intervals
or unions of $n$ intervals. It is known that for example in the case two intervals $A$ and $B$, there are two possible (physical, see section \ref{sec::manyintervals}) configurations for calculating the entanglement entropy for the union of these two intervals, $S(AB)\equiv S(A\cup B)$, as shown in figure \ref{fig::n2case} \cite{Headrick:2007km,Headrick:2010zt,Allais:2011ys}.  By the HRT proposal \cite{Hubeny:2007xt}, the 
entanglement 
entropy is given by the minimal possible configuration of extremal curves,
\begin{align}
S(AB)=\min\left\{S(A)+S(B),S_{AB_1}+S_{AB_2}\right\}.
\label{eq::n2case}
\end{align}
As the parameters defining $A$ and $B$ are varied, there may be phase transitions between these two configurations, and 
we will consequently refer to these configurations as \textit{phases}. Interestingly, the entanglement entropies of the 
subsystem $A\cup B$ and its (sub)subsystems may be required to satisfy certain inequalities, which in the $n=2$ case 
discussed here are only the \textit{subadditivity} (SA) \cite{Araki:1970ba}
\begin{align}
S(AB)\leq S(A)+S(B),
\label{SA}
\end{align}
following immediately from the holographic prescription \eqref{eq::n2case}, and the \textit{triangle} or \textit{Araki-Lieb inequality} \cite{Araki:1970ba}
\begin{align}
S(A B)\geq |S(A)-S(B)|.
\label{ArakiLieb}
\end{align} 

While this is well-known and straightforward for the $n=2$ case just
discussed, some interest has recently emerged \cite{Ben-Ami:2014gsa,	Alishahiha:2014jxa,Bao:2015bfa,Mirabi:2016elb,Bao:2016rbj} for similar concepts for situations involving $n>2$
disconnected intervals. Here we present our study of this case, and
apply it to the steady-state spacetime in subsection
\ref{sec::InequalitiesAnalysis}. The \textit{Wolfram Mathematica} code
that we use for this analysis is uploaded to the arXiv as an ancillary
file together with this paper and with a sample of the numerical results
that is obtained from the matching procedure of section
\ref{sec:Matching}.\footnote{Please note that this file can be opened
  with the free \textit{CDF Player software} \cite{CDFPlayer}. } There
is some overlap between the issues addressed in this section
(especially subsection \ref{sec::manyintervals}) and the ones
investigated in \cite{Bao:2016rbj}, which was published after most of
this section was completed. Although we are working in a covariant
(time-dependent) setting, the findings of \cite{Bao:2016rbj} 
suggest that the code used in our ancillary file may still be optimized.
However, it nevertheless produces the desired results.

\subsection{The phases of the union of $n$ intervals}
\label{sec::manyintervals}

As shown in figure \ref{fig::n2case},  for two intervals ($n=2$) there are two possible phases or 
configurations describing the entanglement entropy of the union of these intervals. Suppose we are given 
$n\in\mathbb{N}$ intervals, how many possible phases are there, and how do they look like? For a simplified situation where the lengths of all intervals are equal, as well as the distances between them, this has already been studied in \cite{Ben-Ami:2014gsa}. We are however interested in the general case here. Of course, for any given $n$, the above question can simply be answered by drawing all possible configurations by hand. However, in 
this subsection we provide an algorithm that for any $n$ enumerates the possible phases in a consistent 
manner, without omitting any solution or counting it twice, and which
can easily be implemented  (see the corresponding ancillary file). We
do not  assume a translation invariant spacetime, 
however we will assume a spacetime with  simple topology, such as \Poincare AdS or a flat black brane, excluding 
possible phenomena such as entanglement plateaux \cite{Hubeny:2013gta}, see also \cite{Bao:2016rbj}. Our task then essentially boils down to finding the \textit{noncrossing partitions} of a set with $n$ elements, a well known combinatorial problem related to the \textit{Catalan numbers} $C_n$ \cite{Catalan}. We will, however, still present our solution to this problem in detail, as this exposition also serves to explain our notation and the inner workings of our ancillary file.

In a $1+1$ dimensional CFT, the $n$ intervals 
under consideration (which we assume to be all part of a specified  equal time slice on the boundary) are all lined up one after the other, and we can enumerate their start- and end-points from $1$ to $2n$, as was already done in figure \ref{fig::n2case}. Note that this is only an enumeration, and not meant to indicate the lengths of the different intervals or the coordinates of the boundary points for example. 

\begin{figure}[htbp]
	\centering
	\includegraphics[width=0.4\textwidth]{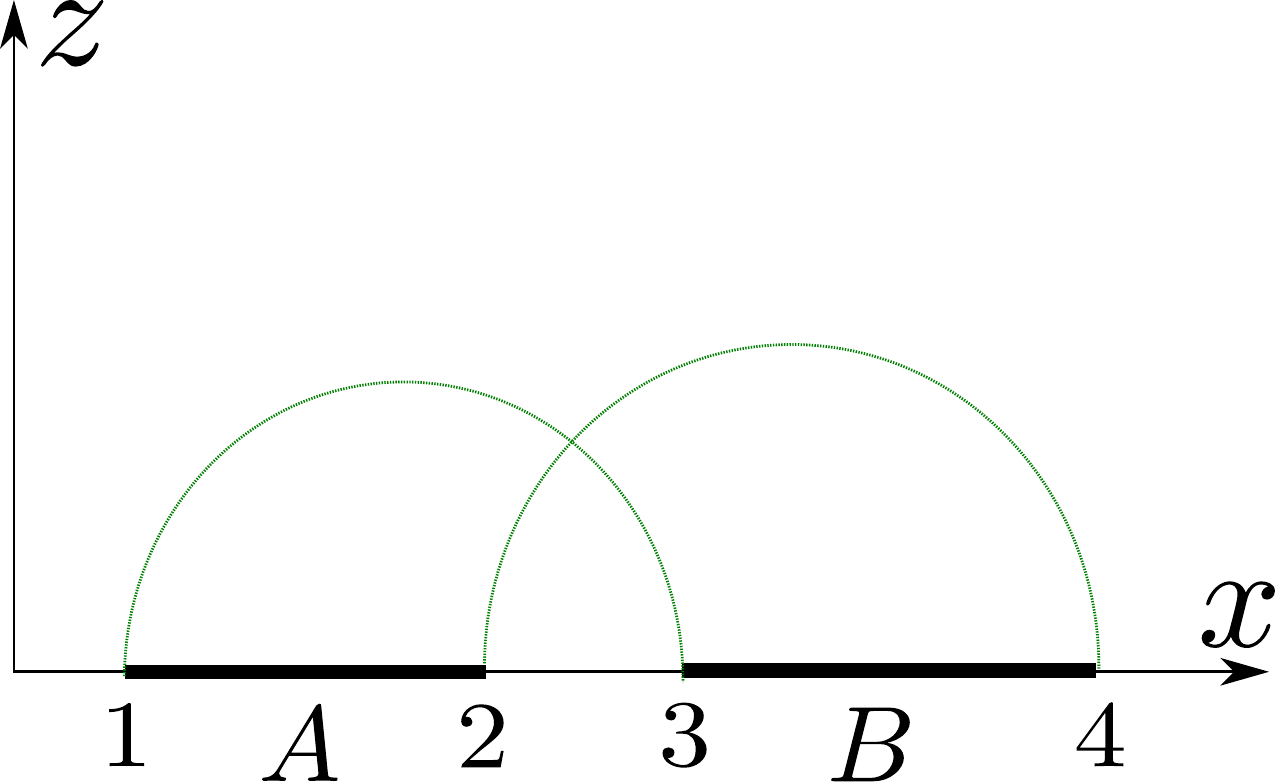}
	\caption{An unphysical configuration for $S(A\cup B)$.}
	\label{fig::n2case2}
\end{figure}

Naively, in the $n=2$ case, we could have also drawn a configuration
as the one depicted in figure \ref{fig::n2case2}, with two curves
crossing each other \cite{Allais:2011ys}. Such a configuration is,
however, considered to be unphysical for various reasons. First, in
the static case where the RT prescription holds, it can easily be
shown that this type of configuration can never yield the lowest
values for the entanglement entropy, hence can be ignored in
\eqref{eq::n2case} \cite{Headrick:2007km,Allais:2011ys}. Second, in
a time dependent (HRT) case the two curves may not actually cross any
more. However,  the co-dimension one surface spanned between them  and
the boundary intervals would then become null or timelike at some
point. As pointed out in \cite{Hubeny:2013gta}, the co-dimension one
surface required by the homology condition has to be restricted to be
spacelike everywhere in the HRT prescription. Hence the configuration
of figure \ref{fig::n2case2} is also  excluded in the dynamic case. Third, it has been discussed in \cite{Allais:2011ys} that configurations of this intersecting type do not play a role when the (regularised) entanglement entropy of an interval is monotonously increasing  with the interval length. 

When enumerating the possible phases of the entanglement entropy of the union of $n$ intervals,
we therefore aim at excluding phases with  curves  intersecting when
projected into the same plane, as  shown in figure \ref{fig::n2case2}.
Due to our labeling of the boundary points, it is clear that each
interval begins at a point labeled by an odd number and ends at an even one. In figure 
\ref{fig::n2case}, for $n=2$ we find one phase where bulk curves connect the points 1 to 2 and 3 to 4, 
and one phase where the bulk curves connect the points 1 to 4 and 2 to
3. In the unphysical case shown in figure 
\ref{fig::n2case2} however, the points 1 to 3 and 2 to 4 are connected
by bulk curves. We hence realize that in order to 
avoid intersections as the one shown in \ref{fig::n2case2}, each odd
label has to be connected to an even label by a  bulk curve. This means that each viable configuration for any $n$ has to be a mapping of the set of odd numbers 
$\{1,3,...,2n-1\}$ to the set of even numbers $\{2,4,...,2n\}$. The number of these possible mappings is given by $n!$, 
the number of possible permutations of the set $\{2,4,...,2n\}$,
\begin{align}
\text{phase 1:\ }\left(
\begin{array}{c}
1 \rightarrow 2 \\
3 \rightarrow 4 \\ 
5 \rightarrow 6 \\ 
...\\
2n-1 \rightarrow 2n
\end{array}
\right),
\ 
...,
\ 
\text{phase i:\ }\left(
\begin{array}{c}
1 \rightarrow \sigma_i(2) \\
3 \rightarrow \sigma_i(4) \\ 
5 \rightarrow \sigma_i(6) \\ 
...\\
2n-1 \rightarrow \sigma_i(2n)
\end{array}
\right),
...
\end{align}
Here, $\sigma_i$ is the $i$-th out of the $n!$ possible permutations of the set $\{2,4,...,2n\}$. Returning to the specific example of $n=2$, we hence obtain the two phases
\begin{align}
S(AB)=S(A)+S(B)\ \Leftrightarrow\  \left(
\begin{array}{c}
1 \rightarrow 2 \\
3 \rightarrow 4
\end{array}
\right) \ \text{``disconnected phase'' 
	,}
\\
S(AB)=S(AB_1)+S(AB_2)\ \Leftrightarrow\  \left(
\begin{array}{c}
1 \rightarrow 4 \\
2 \rightarrow 3
\end{array}
\right)\ \text{``connected phase'' 
	,}
\end{align}
where e.g.~
$1\rightarrow 2$ stands for the curve connecting the points 1 and 2. 

\begin{figure}[htbp]
	\centering
	\includegraphics[width=0.9\textwidth]{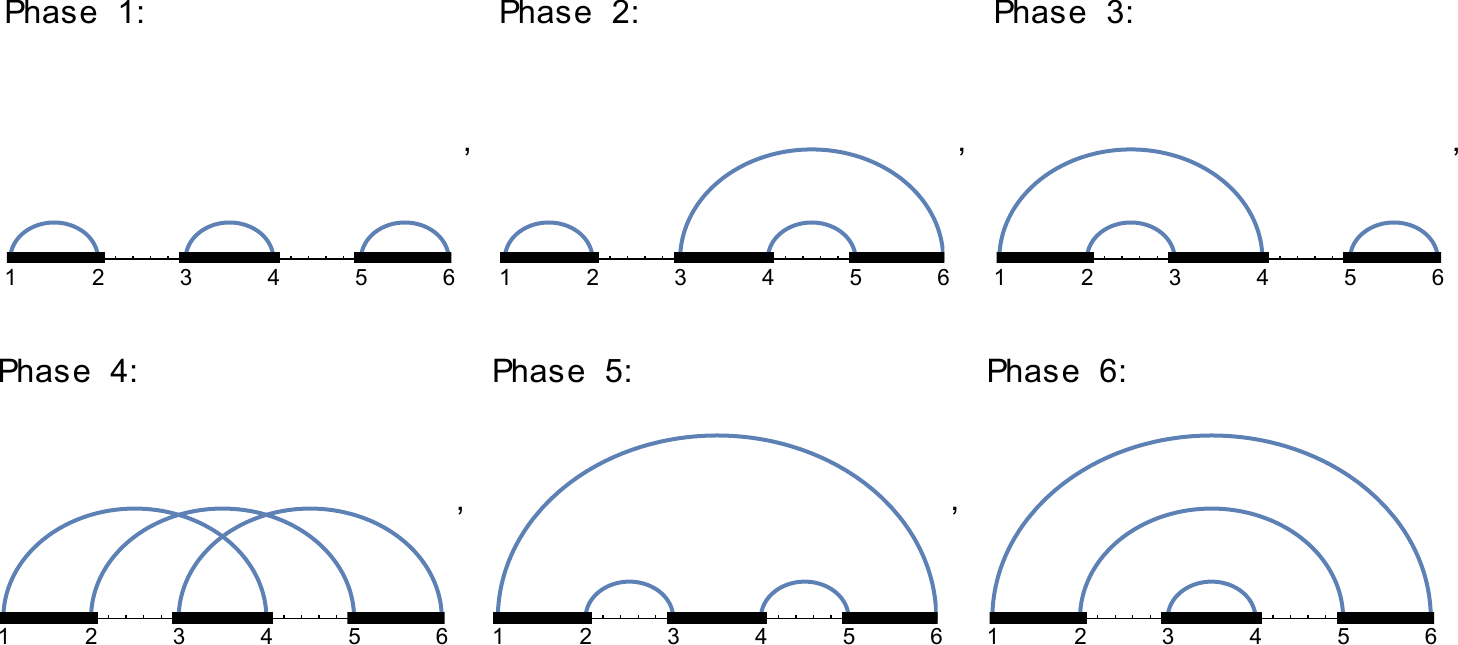}
	\caption{$3!=6$ preliminary configurations for $n=3$. Note that the 4th configuration is likely unphysical. According to the nomenclature of \cite{Ben-Ami:2014gsa}, the phase 6 is referred to as \textit{engulfed} phase.}
	\label{fig::n3case}
\end{figure}

All the possible phases obtained this way for $n=3$ are shown in figure \ref{fig::n3case}. Clearly, there are $3!=6$ of them, 
however we see that there is still one involving intersections. Of course, when sketching these six possible 
configurations by hand, it is easy to identify the one involving
intersections and to discard it. However, from our point 
of view of automatizing this process, we need to formulate and implement the criterion that 
distinguishes the unphysical phase 
\begin{align}
\left(
\begin{array}{c}
1 \rightarrow 4 \\
3 \rightarrow 6 \\ 
5 \rightarrow 2
\end{array}
\right)
\label{unphysicalexample}
\end{align}
from the physical phases such as e.g. 
\begin{align}
\left(
\begin{array}{c}
1 \rightarrow 4 \\
3 \rightarrow 2 \\ 
5 \rightarrow 6
\end{array}
\right).
\label{physicalexample}
\end{align}
To do so, we exclude all configurations in which there are two intervals spanned by the endpoints of bulk curves that intersect in such a way that the intersection is an interval that is not either empty or one of the intervals spanned by the bulk curves. For example, in the unphysical example \eqref{unphysicalexample} the curves span the intervals $[1,4]$, $[3,6]$ and $[2,5]$\footnote{Remember that the numbers 1 to 6 serve here as labels of (ordered) boundary points, and not necessarily as coordinates on the $x$-axis.}. The first and the last of these intersect in $[2,4]$ which is not one of the spanned intervals, hence this configuration is excluded as unphysical. In contrast, in the example \eqref{physicalexample} the curves span the set of intervals $\{[1,4], [2,3], [5,6]\}$, and apart from the empty interval the only intersection between these intervals is $[2,3]$, which is an element of the above set. Hence this phase is considered physical. See the ancillary file for a concrete implementation.

\begin{figure}[htbp]
	\centering
	\includegraphics[width=0.9\textwidth]{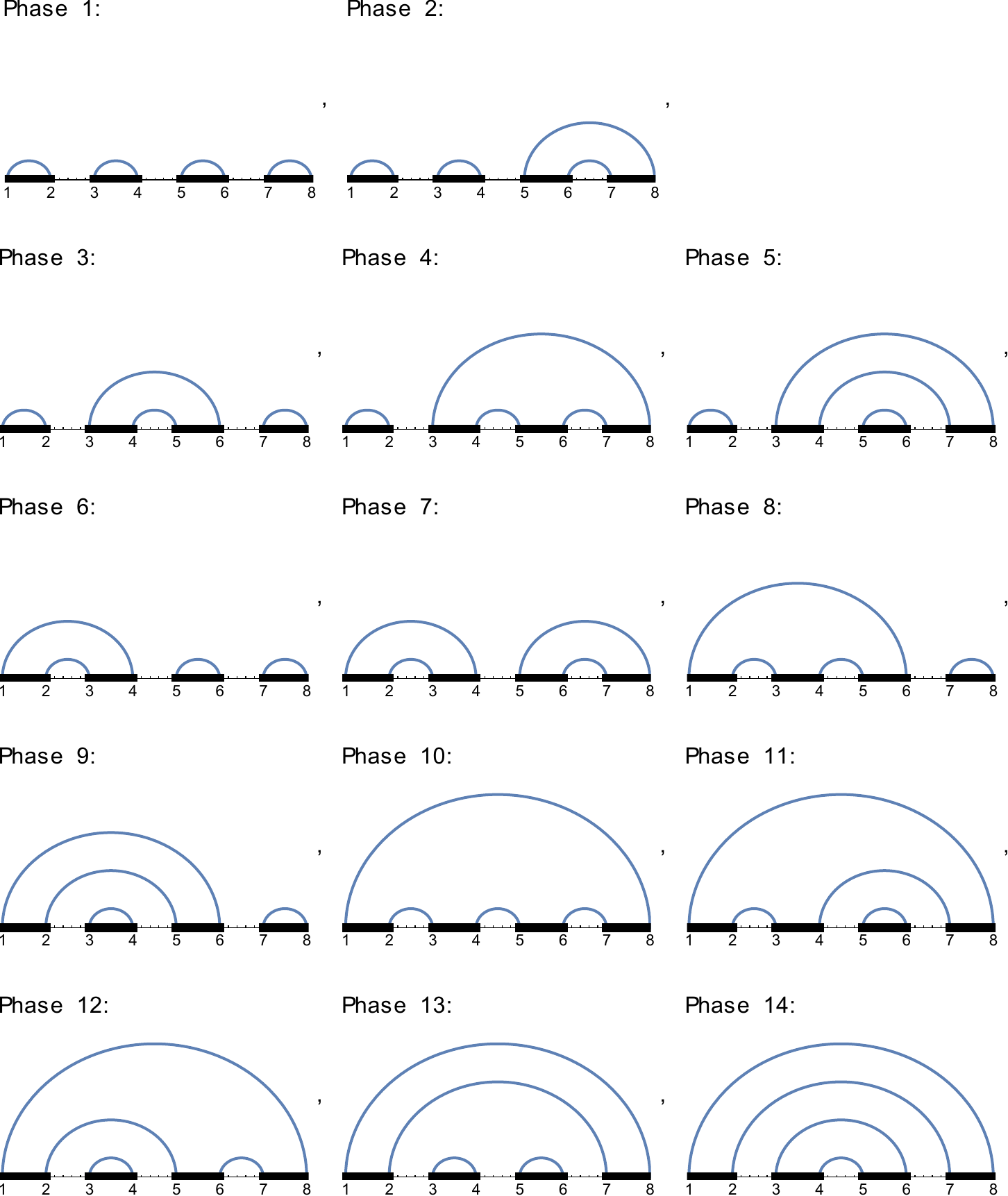}
	\caption{ Physical configurations for $n=4$ intervals. Note that out of the $4!=24$ configurations that our counting would naively have suggested, there are only 14 physical ones, i.e.~14 ones where the curves do not intersect when projected to the same plane. }
	\label{fig::n4case}
\end{figure}

This approach allows us to implement a general algorithm that gives us all possible phases for the entanglement entropy of a set of $n$ disconnected intervals, with any $n$. For the case $n=3$, we then have to exclude phase 4 in figure \ref{fig::n3case}, and are left with the five physical phases already identified in \cite{Allais:2011ys}. For $n=4$ for example, the 14 physical phases are shown in figure \ref{fig::n4case}. For general $n$, the number of these physical phases is given by the $n$-th Catalan number \cite{Catalan}
\begin{align}C_n=\frac{1}{n+1}\left(
\begin{array}{c}
2n \\
n
\end{array}
\right),
\end{align}
which grows as $C_n\sim \frac{4^n}{n^{3/2}\sqrt{\pi}}$ for large $n$. However, with a more optimized code, it may not be necessary to compute all the values of this number of phases \cite{Bao:2016rbj}.

\subsection{Inequalities for the union of $n$ intervals}

Our interest in this section is to study the entanglement inequalities that can be formulated when working with $n>2$ intervals. 
\\

At the level $n=3$, the most well-known inequality that entanglement entropies are expected to satisfy is the \textit{strong subadditivity} (SSA) \cite{doi:10.1063/1.1666274}, commonly stated as
\begin{align}
&S(AB) + S(BC) - S(ABC) - S(B) \geq 0.
\label{SSA}
\end{align}
A different inequality often associated with SSA is \cite{doi:10.1063/1.1666274}
\begin{align}
&S(AB) + S(BC) - S(A) - S(C) \geq 0,
\label{SSAalt}
\end{align}
see \cite{Allais:2011ys,Hayden:2011ag} for a discussion of the relation between \eqref{SSA} and \eqref{SSAalt} in the holographic case. For the static holographic cases in which the Ryu-Takayanagi prescription \cite{Ryu:2006bv,Ryu:2006ef} applies, these two inequalities were proven in \cite{Headrick:2007km}. For the case of time-dependent bulk spacetimes where the HRT prescription \cite{Hubeny:2007xt} applies, a proof of \eqref{SSA} was given in \cite{Wall:2012uf} using the null curvature condition, see also the review \cite{Prudenziati:2015cva}.
\\

Similarly at $n=3$, we encounter what is known as the \textit{monogamy of mutual information}\footnote{See \cite{Lancien:2016nvr} for an illuminating discussion of the concept of \textit{monogamy} for entanglement measures. } or alternatively \textit{negativity of tripartite information}
\begin{align}
&I_3(A:B:C)\equiv S(A) + S(B) + S(C)- S(AB) - S(BC) - S(AC) + S(ABC) \leq 0.
\label{I3}
\end{align}
This was proven for the static RT case in \cite{Hayden:2011ag}, and for the time dependent HRT case in \cite{Wall:2012uf}. There are many more possible inequalities that entanglement entropies for $n\geq3$ intervals have to satisfy \cite{Linden2005,6134666}, which can be seen to follow from \eqref{I3} and the other inequalities discussed so far for $n\leq3$ \cite{Hayden:2011ag}. Hence these inequalities are also proven to hold in holography, assuming appropriate energy conditions. 
\\

The study of entanglement entropy inequalities in AdS/CFT is hence of very high importance for the understanding of holography. On the one hand, if it can be shown that holographic prescriptions satisfy certain entanglement inequalities that do not hold in general quantum theories, this would help distinguish quantum theories that can in principle have a simple holographic dual from those that cannot. On the other hand, if energy conditions in the bulk can be used to prove certain entanglement inequalities that have to hold in the dual, then conversely, it may be possible to derive novel bulk energy conditions from boundary entanglement entropies \cite{Lashkari:2014kda}.

In the following, we will hence use the entanglement entropies that we have calculated in our time dependent background metric \eqref{eq:FG} using the matching procedure of section \ref{sec:Matching} to test, for the manifestly time dependent HRT case, the validity of some of the entanglement inequalities derived in \cite{Bao:2015bfa} for the static RT case. It should however be noted that as the metric \eqref{eq:FG} is a vacuum solution to Einstein's equations everywhere, it trivially satisfies all common energy conditions, and is hence considered to be a physical spacetime. We hence do not expect any of the inequalities of \cite{Bao:2015bfa} to be violated, however as their proof is only valid in the static case, it is interesting to test this expectation thoroughly. At the level of $n=5$ boundary intervals, these inequalities read
\begin{align}
&S(ABC) + S(BCD) + S(CDE) + S(DEA) + S(EAB) - S(ABCDE)
\nonumber
\\
&-S(BC) - S(CD) - S(DE) - S(EA) - S(AB)\geq 0
\label{Oo1}
\end{align}
\begin{align}
&2S(ABC) + S(ABD) + S(ABE) + S(ACD) + S(ADE) + S(BCE) + S(BDE)
\nonumber
\\
&-S(AB) - S(ABCD) - S(ABCE) - S(ABDE) - S(AC) - S(AD) - S(BC)
\nonumber
\\
&- S(BE) - S(DE) \geq 0
\label{Oo2}
\end{align}
\begin{align}
&S(ABE) + S(ABC) + S(ABD) + S(ACD) + S(ACE) + S(ADE)+S(BCE)
\nonumber
\\
&+S(BDE)+S(CDE)-S(AB) - S(ABCE) - S(ABDE) - S(AC)
\nonumber
\\
& - S(ACDE) - S(AD) - S(BCD)-S(BE)-S(CE)-S(DE)
\geq 0
\label{Oo3}
\end{align}
\begin{align}
&S(ABC) + S(ABD) + S(ABE) + S(ACD) + S(ACE) + S(BC) + S(DE)
\nonumber
\\
&-S(AB)-S(ABCD)-S(ABCE)-S(AC) - S(ADE) - S(B) - S(C)
\nonumber
\\
&-S(D)-S(E)
\geq 0
\label{Oo4}
\end{align}
\begin{align}
&3S(ABC) + 3S(ABD) + 3S(ACE) + S(ABE) + S(ACD) + S(ADE)+S(BCD)
\nonumber
\\
&+S(BCE)+S(BDE)+S(CDE)-2S(AB) - 2S(ABCD) - 2S(ABCE) 
\nonumber
\\
&- 2S(AC) - 2S(BD) - 2S(CE) - S(ABDE)-S(ACDE)-S(AD)-S(AE)
\nonumber
\\
&-S(BC)-S(DE)
\geq 0
\label{Oo5}
\end{align}

A further quantity of interest is the $n$-partite information,
\begin{align}
&I_n(A_1:A_2:A_3:...:A_n)\equiv \sum_{i=1}^{n}S(A_i) - \sum_{i<j}^{n}S(A_i\cup A_j) + \sum_{i<j<k}^{n}S(A_i\cup A_j\cup A_k)
\nonumber
\\
& \mp... - (-1)^n S(A_1\cup A_2\cup ... \cup A_n),
\label{In}
\end{align}
generalising the concept of three-partite information introduced in \eqref{I3}. In a holographic context, quantities such as four- and five-partite information where studied for example in \cite{Alishahiha:2014jxa,Mirabi:2016elb}. In fact, based on the examples studied in those papers, the authors proposed the entanglement inequalities 
\begin{align}
&I_4(A:B:C:D)\geq0
\label{I4}
\end{align}
and 
\begin{align}
&I_5(A:B:C:D:E)\leq0.
\label{I5}
\end{align}
While the inequalities \eqref{I4} and \eqref{I5} may be true for the special cases studied in \cite{Alishahiha:2014jxa,Mirabi:2016elb}, where all intervals have the same length and distance from their neighboring intervals, it was already stated in \cite{Hayden:2011ag} that \eqref{I4} and \eqref{I5} do not hold in general holographic setups. In fact, using the numerical data for the time dependent backgrounds studied in this paper or simply data valid for static backgrounds such as the BTZ metric \eqref{eq:btz} and feeding this data into our ancillary file, it is possible to find explicit examples for sets of four or five intervals that will lead to violations of the proposed inequalities \eqref{I4} and \eqref{I5}.

\subsection{Symmetries of entanglement inequalities}
\label{sec::Permutations}

In section \ref{sec::manyintervals} we have described how we can
systematically enumerate all the possible phases that the entanglement
entropy of the union of $n$ intervals can have. In order to check the
validity of inequalities for entanglement entropy using this counting
procedure, it is also important to consider the symmetries of the inequalities under consideration. Take as the simplest example the strong subadditivity inequality \eqref{SSA}, 
valid for the combination of the $n=3$ intervals $A,B,C$. Comparing to
our enumeration introduced in section \ref{sec::manyintervals}, see
also figure \ref{fig::n3case}, it is clear that a priori there may be
five different physical configurations determining the quantity
$S(ABC)$. However, there are different ways to assign the labels
$A,B,C$ to the intervals $[1,2],[3,4],[5,6]$ in figure
\ref{fig::n3case}. If we impose a strict alphabetical ordering $A=[1,2], B=[3,4], C=[5,6]$, the inequality \eqref{SSA} will not be equivalent for example to the inequality 
\begin{align}
S(AB)+S(AC)-S(ABC)-S(A)\geq 0,
\end{align}  
which we obtained by relabeling $A$ and $B$. On the other hand, \eqref{SSA} is invariant under interchanging $A$ and $C$. This means that entanglement inequalities involving many intervals may have a non-trivial amount of symmetry (or asymmetry) under permutation (or renaming) of intervals $A,B,C,...$ which we will have to take into account when employing a strict enumeration of intervals from left to right as we do in section \ref{sec::manyintervals} and in our numerical code for technical reasons. 

In our ancillary file, we solve this problem as follows: Take an inequality, e.g.~\eqref{SSA}, in a form where the intervals are denoted $A,B,C,...$ without assuming a specific ordering of them on the boundary. We then write the inequality with all elements to the left of an $\geq$ sign, and represent it as a set of sets of elements, e.g.
\begin{align}
\{\{A,B\},\{B,C\},\{-B\},\{-A,-B,-C\}\}.
\label{SSAset}
\end{align}  
It is then easy to apply all possible permutations to this set, and filter out the ones that act non-trivially, i.e.~that do not leave it invariant. In the end, we are left with a list of sets of the form \eqref{SSAset}, which correspond to inequalities which are \textit{inequivalent} when using a strict alphabetical ordering $A=[1,2], B=[3,4],...$ of the intervals along the boundary. In this context, it is interesting to note that the degree of symmetries uncovered this way varies from one inequality to the other. For example, while by permuting the intervals $A,B,C,D,E$ (which in this subsection are now assumed to be ordered alphabetically on the boundary) gives us 10 inequivalent inequalities following from \eqref{Oo3}, this number is 60 for \eqref{Oo5}.

\subsection{Analysis and Results}
\label{sec::InequalitiesAnalysis}

We now have almost all prerequisites that are needed in order to check the validity of entanglement inequalities such as \eqref{Oo1} - \eqref{Oo5} in the time dependent system holographically described by the bulk metric \eqref{eq:FG}. As the matching-procedure outlined in section \ref{sec:Matching} can only be applied when the geodesics cross the shockwave once, we have to restrict ourselves to the study of intervals for which all boundary points have $x$-coordinates either larger than zero or smaller than zero. We  generally assume all boundary points of intervals under investigation to be located at equal boundary time. 

We have carried out this analysis for various choices of temperatures
$T_L$ and $T_R$, for various values of the boundary time $t$, and for
intervals to the left and to the right of $x=0$. As the results were
qualitatively similar in all these cases, we will in the following
only discuss the example where we chose $T_L=9,T_R=1$ (hence
$\beta=\frac{4}{5}$) and the boundary time slice to be at $t=1$, with
intervals in the range $x>0$. According to our discussion in section
\ref{sec::manyintervals} there will be $42$ possible phases that the
entanglement entropy of $n=5$ intervals can take. Furthermore, in a
non-homogeneous system, $n$ intervals are defined by $2n$ boundary
points. As our calculations will be numerical, we are faced with a
severe problem: Even if we are able to check the validity of
inequalities such as \eqref{Oo1} - \eqref{Oo5} for any given set of
five intervals, how can we make sure that we find all potentially
interesting cases? After all, we have to cover a $2n$ dimensional
parameter space, on which phase transitions between $42$ different
phases can occur. Numerically, we will of course only be able to check
a finite number of examples. Naively, the best idea 
appears to be  to take a finite number of evenly spaced points on the boundary,
\begin{align}
x=0.1,0.2,0.3,...,2.0
\label{Equalspacing}
\end{align}
and to calculate the entanglement entropy for any interval formed by
any two of these points. From this data, we may then calculate the entanglement entropy of the union of any possible set of $n$ intervals that can be formed from the given boundary points, and subsequently check the validity of all entanglement inequalities of interest. 

However, we can do better than this. In the study of mutual
information for \Poincare backgrounds, where there are only two phases
as shown in figure \ref{fig::n2case}, it is known that the phase will
depend on the relation between the sizes of the two intervals and the
distances between them. In our attempt to cover the relevant phase
space for $n\leq5$ intervals, it will hence be advantageous to allow
for the distances between boundary points to vary between as many
orders of magnitude as possible. Instead of using equally spaced boundary
points such as in \eqref{Equalspacing}, the idea is thus to use points
which are positioned in a fractal-like way\footnote{Indeed, the
  inspiration for this comes from the concept of \textit{fractal
    antennas}, which in antenna technology can be used when attempting
  to transmit in a broadband characteristic, compared to standard
  dipole antennas. We thus aim at choosing the boundary points in such a way that they form a metaphorical `fractal antenna' for the structure of entanglement entropy and $n$-partite information over many length scales in the quantum system that we are studying holographically.}, e.g.
\begin{align}
x=
0,1-\frac{2}{\alpha },1-\frac{4}{\alpha ^2},1-\frac{8}{\alpha ^3},...,1+\frac{8}{\alpha ^3},1+\frac{4}{\alpha ^2},1+\frac{2}{\alpha },2,
\label{fractal}
\end{align}    
where we have found that the choice $\alpha=9/2$ gives a good trade-off between the orders of magnitude of length scales covered and the overall number $N$ of points, which for a reasonable runtime of our numerics we would like to keep at $N=20$.  

\begin{figure}[htb]
	\centering
	\includegraphics[width=0.9\textwidth]{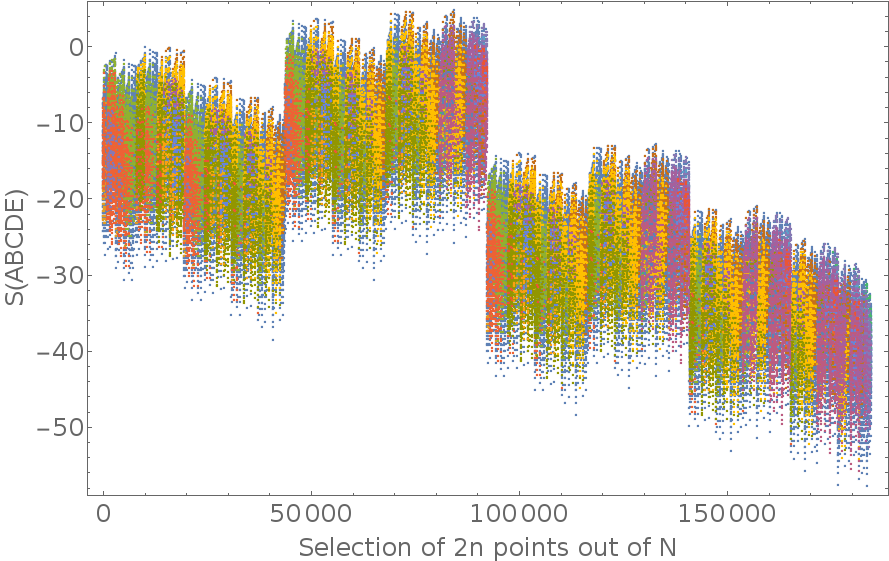}
	\caption{Entanglement entropies for all $184756$ possible unions of $n=5$ non-adjacent intervals formed out of the $N=20$ boundary points \eqref{fractal} at $t=1$, for $T_L=9$ and $T_R=1$. The value of the entanglement entropy is dependent on the explicit cut-off or renormalisation scheme used, so the overall shift of the vertical axis is of no relevance, only linear combinations of entanglement entropies in which the cut-off dependence cancels are physical. The colour represents the different phases that the entanglement entropy of the union of five intervals can be in. The intricate structure of the data over the horizontal axis is due to the lexicographic order in which the $184756$ possible unions of $n=5$ intervals are enumerated and the placement of the boundary points according to \eqref{fractal}.   
	}
	\label{fig::n5EE}
\end{figure}

Using the matching prescription explained in section \ref{sec:Matching}, we have calculated the renormalised lengths of the geodesics connecting any two of the $N=20$ boundary points under consideration. In our case at hand, this requires $\frac{1}{2}N(N-1)=190$ calculations. As a next step, for some $n\leq5$ we want to form $n$ boundary intervals by selecting $2n$ boundary points out of the $N$ available points\footnote{As we are selecting $2n$ \textit{distinct} boundary points, the intervals under investigation will never be adjacent, i.e.~they will never share a boundary point.}. Obviously, there are 
\begin{align}
\left(
\begin{array}{c}
N \\
2n \\
\end{array}
\right)=\frac{N!}{(2n)!(N-2n)!}
\end{align}
ways to do so. Given our $N=20$ points positioned on the boundary time slice $t=1$ according to the sequence \eqref{fractal}, we are hence for example able to study $184756$ distinct unions of $n=5$ non-adjacent boundary intervals. For all these $184756$ different cases, it is then possible to calculate the entanglement entropy, see figure \ref{fig::n5EE}. 
Interestingly, in the case at hand we find that out of the $184756$
available unions of intervals, $100177$  are in the \textit{totally
  disconnected phase} in which $S(A_1\cup
A_2\cup...)=S(A_1)+S(A_2)+...$, and in which hence the entanglement
inequalities \eqref{I3}-\eqref{I5} are trivially
saturated. Consequently, only the remaining $84579$ cases will be of
further interest. It should also be noted that the overall value of
the entanglement entropy for a given union of intervals is dependent
on the explicit cutoff used in our numerics. However, the linear
combinations of entanglement entropies appearing in the inequalities
\eqref{SSA}, \eqref{I3} but also \eqref{Oo1}-\eqref{Oo5}, are always
balanced in such a way that the cut-off dependence of the individual
terms cancels, such that the result is physical.\footnote{For \eqref{SSAalt} this will not be the case unless the intervals $A$, $B$ and $C$ share some of their endpoints. We will  not study this inequality in this paper.}

\begin{figure}[htb]
	\centering
	\includegraphics[width=0.9\textwidth]{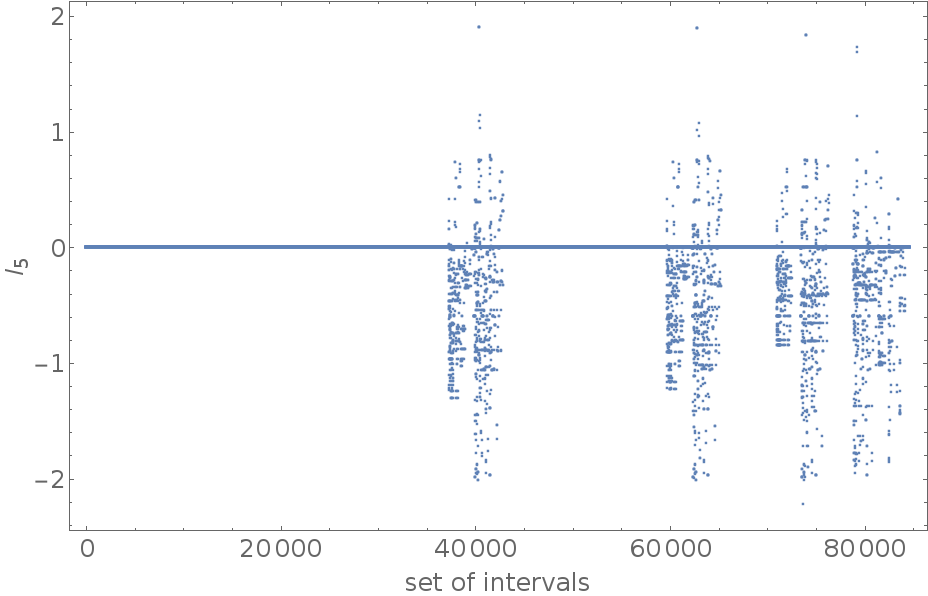}
	\caption{Five-partite information for the $84579$ sets of five intervals (out of initially $184756$) for which the total entanglement entropy is not in the totally disconnected phase. Clearly there are multiple violations of the proposed inequality \eqref{I5}. As in figure \ref{fig::n5EE}, we are here displaying results for the example where $T_L=9$, $T_R=1$ and the boundary time $t=1$.       
	}
	\label{fig::I5}
\end{figure}

\begin{figure}[htb]
	\centering
	\includegraphics[width=0.9\textwidth]{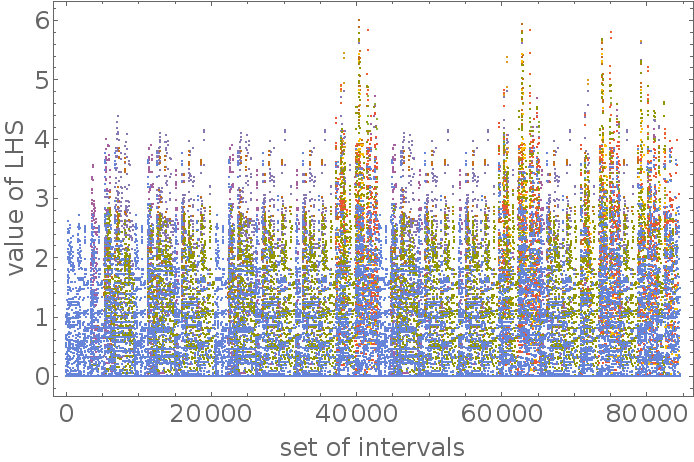}
	\caption{The left-hand sides of the inequality \eqref{Oo1} for the $84579$ sets of five intervals (out of initially $184756$) for which the total entanglement entropy is not in the totally disconnected phase. Clearly there are no violations of the inequality \eqref{Oo1}. The different colors in the figure stand for different permutations, as explained in section \ref{sec::Permutations}. As in figure \ref{fig::n5EE}, we are here displaying results for the example where $T_L=9$, $T_R=1$ and the boundary time $t=1$.       
	}
	\label{fig::Oo1}
\end{figure}

Now, we have all the necessary ingredients together to check the validity of various entanglement inequalities, as well as of their permutated versions as discussed in section \ref{sec::Permutations}. The results are as follows:
\begin{itemize}
	\item At $n=3$, both strong subadditivity \eqref{SSA} and monogamy of mutual information \eqref{I3} are satisfied, as expected based on \cite{Wall:2012uf,Prudenziati:2015cva}. In contrast to \cite{Ben-Ami:2014gsa}, we find that even the \textit{engulfed phase} (see figure \ref{fig::n3case}) can be the minimal one in specific examples. Generically, this seems to happen when the middle interval is very small compared to the gap between the other two intervals. 
	
	\item At $n=4$, the only inequality that we are checking is the positivity of four-partite information \eqref{I4}. In fact, in contrast to \cite{Alishahiha:2014jxa,Mirabi:2016elb}, we find a number of examples for sets of four intervals where this inequality is violated. However, as it was pointed out in \cite{Hayden:2011ag} and was explicitly checked by us, this is not a particular feature of the time dependent case, and happens already in holographic systems with static bulk-spacetime duals.  
	
	\item At $n=5$, we find numerous violations of the negativity of five-partite information \eqref{I5}, see the similar discussion for $n=4$. In fact, out of the $184756$ total and $84579$ nontrivial sets of five intervals under investigation, we find a violation of \eqref{I5} in $417$ cases. It is also noteworthy that even for the $84579$ cases where five-partite information does no \textit{have to} vanish a priori, the result that we obtain vanishes within numerical accuracy for $81183$ cases, see figure \ref{fig::I5}.	
	
	Furthermore, we check the inequalities \eqref{Oo1}-\eqref{Oo5}
        as well as all their relevant permutations, see section
        \ref{sec::Permutations}. The result is that we find
        \textit{not a single case} in which any of these inequalities
        is violated, neither for the specific example currently at
        hand ($T_L=9,T_R=1,t=1$) nor for any other example that we
        studied. See for example figure \ref{fig::Oo1}. We view this as a clear indication that the inequalities \eqref{Oo1}-\eqref{Oo5}, although so far only proven in the static case, will generally also hold in physical time-dependent cases.
\end{itemize}

\section{Comments on higher dimensions}
\label{sec:Higher-Dimensions}

\subsection{State of the art in $d>2$}
\label{sec:Higher-DimensionsDisclaimer}

After investigating the one-dimensional case in  detail, it is natural to ask about a generalisation to higher dimensions. That case is, however, much subtler. It has already been addressed in various works. Let us briefly summarize the current state of discussion about the higher-dimensional case. In \cite{Bhaseen:nature}, a straightforward generalization of the 1+1 dimensional model was suggested, namely a solution consisting of two shockwaves, not necessarily travelling with identical velocities, and a non-equilibrium steady state between the shocks. Such a solution was numerically found in the hydrodynamic regime \cite{Bhaseen:nature}.  Later, a similar solution beyond the hydrodynamic approximation  was found in \cite{Amado:2015uza}, in the framework of gauge/gravity duality. However, at the hydrodynamical level an inconsistency between the non-equilibrium steady-state (NESS) conjecture of \cite{Bhaseen:nature} and thermodynamics was pointed out in \cite{Spillane:2015daa}. The issue is as follows:  The setup of two heat baths put in contact at an initial time is essentially a classical Riemann problem of solving partial differential equations.\footnote{In full generality,  the Riemann problem is a initial value problem for a non-linear PDE with non-continuous, piecewise-constant initial data.} For this type of problem, there is a so-called entropy condition. This  is a requirement that  characteristic lines of the differential operator involved, i.e.~curves along which the initial condition is transported, always end rather than begin on the shock wave.\footnote{In the characteristic formulation of a PDE, the presence of a shockwave is manifested by the intersection of characteristics. On a characteristic line, one direction is distinguished by the fact that the initial condition is evolved forward in time. So, when there is an intersection  of characteristics, it is possible to distinguish whether the line 'begins' or 'ends' in that point.} The name `entropic condition' comes from the fact that if characteristics end at some point, the information about the initial state is lost and hence the entropy is produced. If the characteristics started at the discontinuity, the system would require fixing an initial condition on the shockwave, such that  information would be  produced and entropy would decrease. A detailed analysis of this condition for higher dimensions leads to the conclusion that while a shockwave moving from the region of higher temperature to the colder one is a entropically valid solution, a shock moving in the opposite direction is not (see section 3 of \cite{Spillane:2015daa} for details). The results of \cite{Spillane:2015daa} imply that the solution involving two shockwaves is valid in $d=2$ only, when the velocities of the shocks are identical and equal to one. In higher dimensions, to stay in agreement with entropic considerations, we have to replace the unphysical shockwave by a new solution -- the \emph{rarefaction wave} -- which is continuous but not smooth and much wider than the shockwave. Let us stress   that the double-shock solution is not \emph{mathematically} incorrect since for complicated non-linear PDEs, uniqueness of solutions is not always guaranteed for arbitrary types of boundary or initial conditions. The double shockwave is however \emph{non-physical} due to the  entropic reasons mentioned.  The physical solution is unique in the sense that the shock-rarefaction solution is realised in nature. Let us however emphasize that as shown in \cite{Spillane:2015daa},  the double-shock solution is a valid, physically correct and unique solution to the initial value problem of our non-linear equation in $d=2$.

An important question about the shock-rarefaction solution in higher dimensions is whether it does support the existence of a NESS, defined as a region with constant energy current that can be obtained by boosting a static thermal state with some effective temperature. There are two possibilities: Either the rarefaction solution extends over a large enough region and reaches the existing shock, excluding the formation of NESS, or the rarefaction wave is relatively compact and a NESS is formed between the rarefaction and the shock wave on the other side. In \cite{Lucas:2015hnv}, the authors argue in favour of the latter, based on numerical studies for hydrodynamical setups. Moreover they discover that for most conditions, the  quantitative difference of observables obtained in a non-physical dual-shock solution and those obtained in the thermodynamically favoured rarefaction-shock solution is of order of a  few percent.  The specific properties of the steady state remain similar to the universal behaviour of the NESS in \cite{Bhaseen:nature}.\\
A further question is whether the dual gravity description allows for a physical rarefaction-shock solution. An example of such a solution was found in \cite{Herzog:2016hob} in the limit of large dimensions $d\rightarrow \infty$. However, obtaining a clear, numerical shock-rarefaction solution in the gauge/gravity framework in $d=3$ or $4$ dimensions and testing its properties such as stability is still an open problem. It is worth noting that the authors of \cite{Amado:2015uza} found a full numerical solution of an `almost' Riemann problem where the initial condition is a smooth approximation of a step function, as our \eqref{eq:fLfR} in the framework of AdS/CFT. Since, as discussed previously, the values of the observables obtained from the shock-rarefaction solution and the double shock solution differ only by a few percent, it is not clear which of them is the holographic dual of the hydrodynamic solution. The authors of the previously mentioned paper themselves state that their solution seems to agree with the double-shock conjecture, the work however was published before the entropic issues of the double shock solutions were pointed out in \cite{Spillane:2015daa}.

The arguments mentioned here ensure that a qualitative analysis  of the entanglement entropy in higher dimensions can be carried out,  based on the simple NESS model of \cite{Bhaseen:nature}. We devote this section to this analysis.



\subsection{Analytical considerations}
\label{sec:Higher-DimensionsAnalytical}

Here we present analytical results for the higher-dimensional cases.
These are obtained by assuming that the dual-shock solution is valid
at least approximately. While the higher dimensional analogue of the
time-dependent metric \eqref{eq:FG} is not known analytically any more, we still know the boosted black-brane line element generalising \eqref{eq:boosted} to higher dimensions \cite{Bhaseen:nature}\footnote{Note that in contrast to \cite{Bhaseen:nature}, we are here using a notation in which the dimensionality of the bulk AdS space is $d+1$. Hence the case investigated so far was the one for $d=2$.}
\begin{align}
ds^2&=\frac{L^2}{z^2}\left(\frac{dz^2}{f(z)}-f(z)\left(\cosh\theta dt-\sinh\theta dx\right)^2+\left(\cosh\theta dx-\sinh\theta dt\right)^2+dy_{\perp}^2\right),
\nonumber
\\
f(z)&=1-\left(\frac{z}{z_H}\right)^{d},\ z_H=\frac{d}{4\pi T}.
\label{BBB}
\end{align}
Setting $\theta=0$ and $T=T_{L}$ or $T=T_{R}$, we recover the metrics of the initial static black branes. For the late time steady state, the boost parameter (or \textit{rapidity}) $\theta$ is given by \cite{Bhaseen:nature}
\begin{align}
T&=\sqrt{T_L T_R},
\\
\chi&= \left(\frac{T_L}{T_R}\right)^{\frac{d}{2}},
\\
\beta&=\frac{\chi-1}{\sqrt{\left(\frac{1}{d-1}+\chi\right) (d-1+\chi)}},
\\
\theta&=\arctanh\beta,
\label{theta}
\end{align}
where $\beta$ is the \textit{boost velocity}. It is also important to
note that  in higher dimensions, the two shockwaves move with
different velocities, \cite{Bhaseen:nature}
\begin{align}
u_L=\frac{1}{d-1}\sqrt{\frac{\chi+d-1}{\chi+\frac{1}{d-1}}},\ \ u_R=\sqrt{\frac{\chi+\frac{1}{d-1}}{\chi+d-1}}.
\label{uLuR}
\end{align}
Interchanging $T_L$ and $T_R$ means $\chi\rightarrow1/\chi$ and under this transformation $u_L\leftrightarrow u_R$.

\begin{figure}[htbp]
	\centering
	\includegraphics[width=0.8\textwidth]{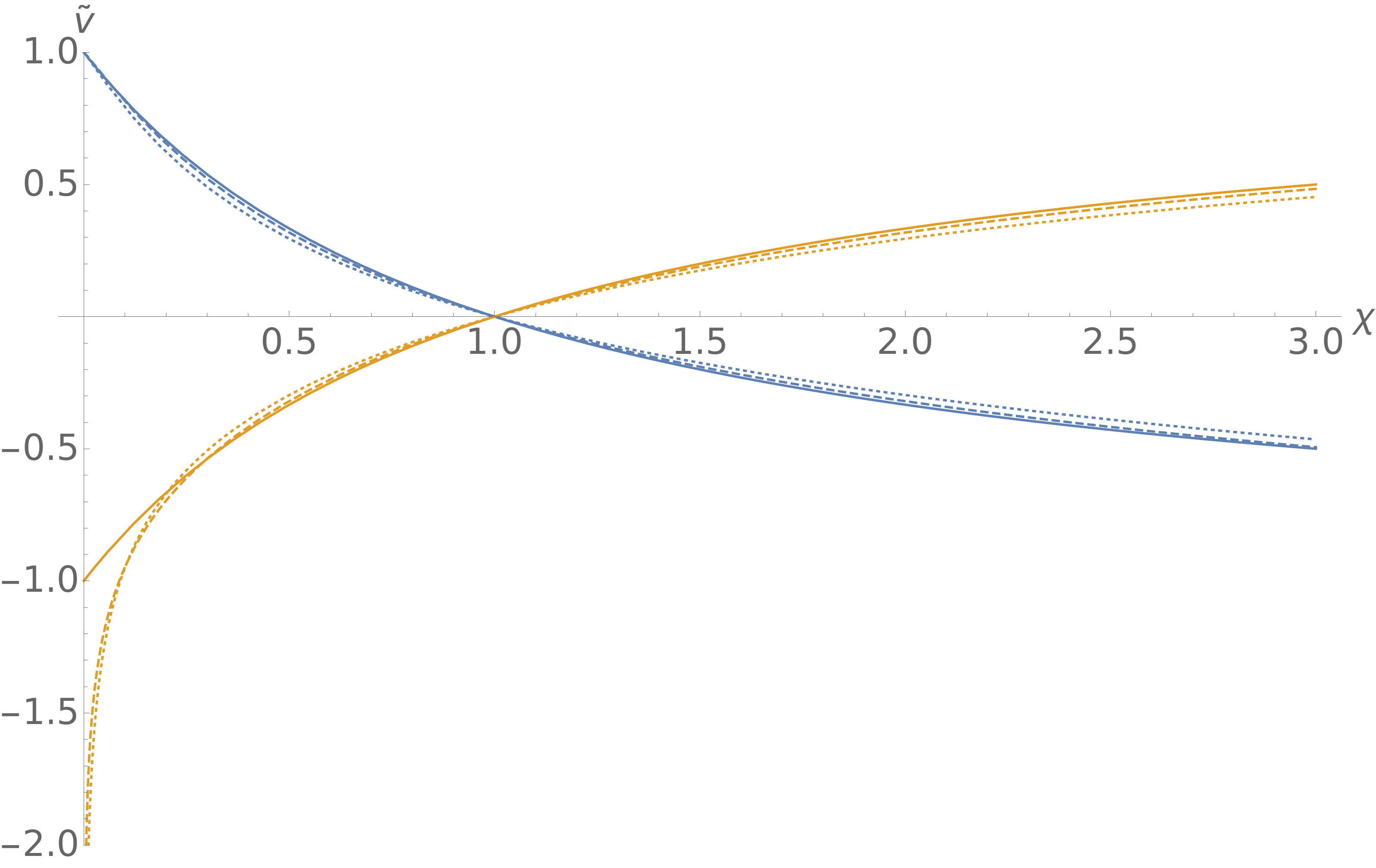}
	\caption{Normalised entropy increase and decrease rates \eqref{vL},\eqref{vR} for $d=2$ (solid), $d=3$ (dashed) and $d=4$ (dotted) as a function of $\chi$. Note that in the formulas \eqref{vL},\eqref{vR}, $\tilde{v}_L>0$, $\tilde{v}_R<0$ for $\chi>1$ and $\tilde{v}_L<0$, $\tilde{v}_R>0$ for $\chi<1$. While $1\geq\tilde{v}_{L,R}\geq-1$ for $d=2$, we see that $1\geq\tilde{v}_{L,R}$ with no lower bound for $d>2$. }
	\label{fig::velocity}
\end{figure}

Although the entanglement entropy $S(\ell)$ for a strip of width $\ell$ and infinite extent in the $dy_{\perp}$ directions cannot be calculated analytically in the background \eqref{BBB}, we find that in the limit $\ell\rightarrow \infty$ where the entanglement entropy becomes extensive, the entropy density analytically reads
\begin{align}
s_{eq}&=\frac{1}{4G}\left(\frac{4\pi TL}{d}\right)^{d-1}\cosh \theta.
\\
&=\frac{(4 \pi L)^{d-1}}{4Gd^{d}} \sqrt{\left(\left(\frac{T_L}{T_R}\right)^{\frac{d}{2}}+d-1\right) \left((d-1) \left(\frac{T_L}{T_R}\right)^{\frac{d}{2}}+1\right)} \left(\frac{T_R}{T_L}\right)^{\frac{d}{4}}(T_L T_R)^{\frac{d-1}{2}}
\end{align}
Let us consider the question whether meaningful statements, similar to
\eqref{vav}, about the (average) entropy increase rates of a strip in
this setup may also be found for higher dimensions. Due to the form of
the velocities \eqref{uLuR}, we see that the time the shockwave takes to pass through a strip\footnote{We here assume a strip with finite extent in the $x$ direction and infinite extent in the $y_{\perp}$ directions.} of width $\ell$ is
\begin{align}
\Delta t_{L/R}=\frac{\ell}{u_{L/R}}.
\end{align}
Just as in section \ref{sec:Tsunami}, we may calculate the average
increase/decrease rate of 
entanglement entropy. We assume for now that the entanglement is only
influenced by the shockwaves, 
and not the light cones. We find
\begin{align}
v_{av,\,L/R}=\frac{\Delta S_{L/R}}{\Delta t_{L/R}}\rightarrow \frac{1}{4G}\left(\frac{4\pi L}{d}\right)^{d-1}u_{L/R}\left(T^{d-1}\cosh\theta-T_{L/R}^{d-1}\right)
\end{align}
for $\ell\rightarrow\infty$. Consequently, in analogy to \eqref{tildevav}, 
\begin{align}
\tilde{v}_{av,\,L}&=\frac{v_{av,\,L}}{s_{eq}}=u_L\left(1-\frac{\chi^{\frac{d-1}{d}}}{\cosh\theta}\right)=\frac{\sqrt{(d-1+\chi ) ((d-1) \chi +1)}-d\chi ^{\frac{d-1}{d}+\frac{1}{2}}}{\sqrt{d-1} ((d-1) \chi +1)},
\label{vL}
\\
\tilde{v}_{av,\,R}&=\frac{v_{av,\,R}}{s_{eq}}=u_R\left(1-\frac{\chi^{\frac{1-d}{d}}}{\cosh\theta}\right)=\sqrt{\frac{\frac{1}{d-1}+\chi }{d-1+\chi }} \left(1-\frac{d\chi ^{\frac{1}{d}}}{\sqrt{\chi  (d-1+\chi ) ((d-1) \chi +1)}}\right).
\label{vR}
\end{align}
As a consistency check, we see that under $T_L\leftrightarrow T_R$, $\tilde{v}_{av,\,L}\leftrightarrow \tilde{v}_{av,\,R}$. Also, for 
$d=2$ this exactly reproduces our findings from section \ref{sec:Tsunami}. Interestingly, in contrast to \eqref{averagerate}, we find that while these formulas imply an \textit{upper bound} 
\begin{align}
\tilde{v}_{av,\,L,R}(\chi)\leq 1
\label{rateboundhigherd}
\end{align} 
on the normalised average entropy \textit{increase}, we do not find a \textit{lower bound} on $\tilde{v}_{av,\,L,R}$ limiting the entropy \textit{decrease} for $d>2$. In figure \ref{fig::velocity}, the two functions $\tilde{v}_{L/R}(\chi)$ are plotted for $d=2,3,4$. We see that in higher dimensions, due to \eqref{rateboundhigherd}, $\tilde{v}_{av}$ may exceed both $v_E$ and $v_B$ defined in section \ref{sec:Tsunami}. However, only a full numerical solution of the higher dimensional case will produce a clearer picture of the relation between the entropy increase rate for a given intervals and other bounds or quantities such as \eqref{vE} or \eqref{vB}. Such a numerical solution will also allow to address the impact of considering a shock or a rarefaction wave in relation to the absence of a lower bound on $\tilde v_{av}$  in $d>2$. This may be relevant for a general
discussion of whether choosing a gravity solution that decreases the thermodynamic
entropy has unphysical consequences for the entanglement entropy.

\subsection{Numerical considerations}
\label{sec:Higher-DimensionsNumerical}

\begin{figure*}[h]
\begin{tabular}{cc}
\includegraphics[width=7.0cm]{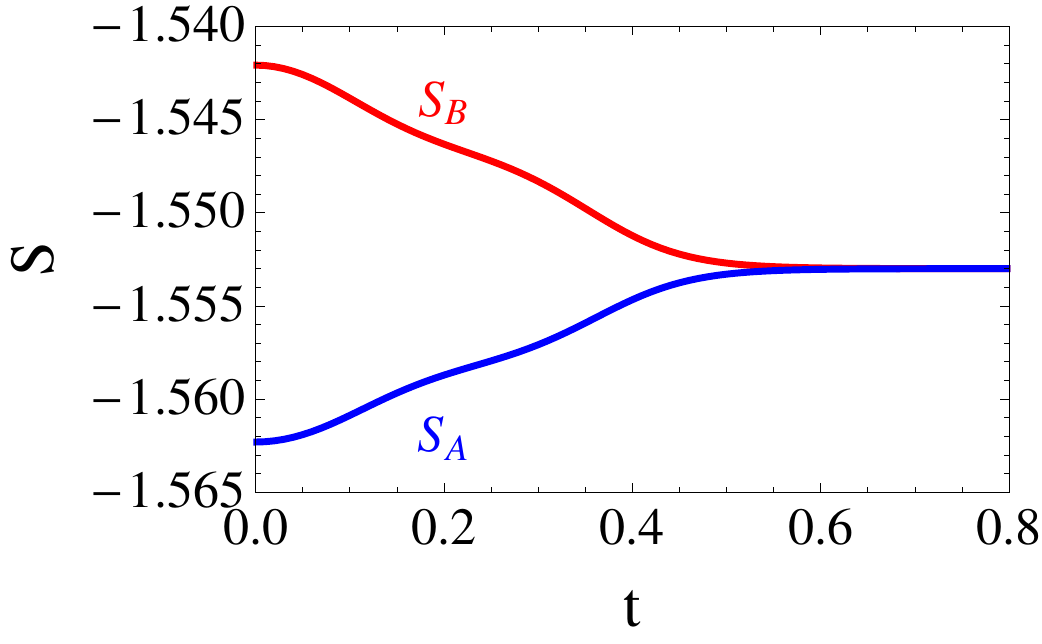} & 
\includegraphics[width=7.0cm]{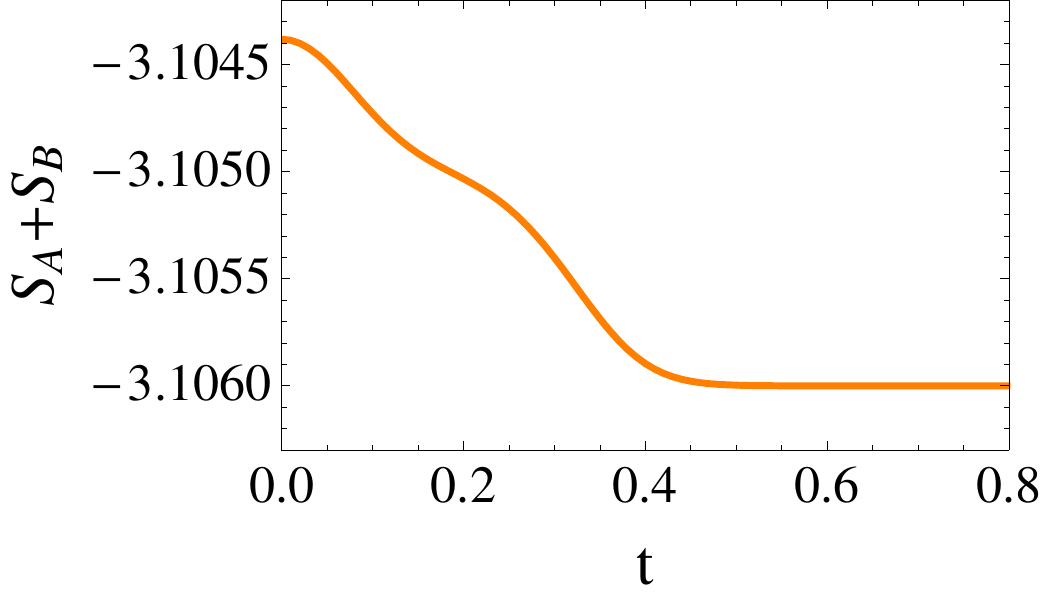} \\
\end{tabular}
\caption{\it (Left) Renormalized entanglement entropies $S_A$ and $S_B$ as a function of time, in a system with $d=3$. (Right) Renormalized entanglement entropy $S_A + S_B$ as a function of time. In both figures we have considered  the linearized background computed in Refs.~\cite{Megias:2015tva,Megias:2016vae}, the intervals $x^A \in [0.05,0.275]$ and $x^B \in [-0.275,-0.05]$, and temperatures $T_L = 0.6$ and $T_R=0.5$. We have set $G=1$ and $L=1$.}
\label{fig1:d2}
\end{figure*}

\begin{figure*}[h]
\begin{tabular}{cc}
\includegraphics[width=7.0cm]{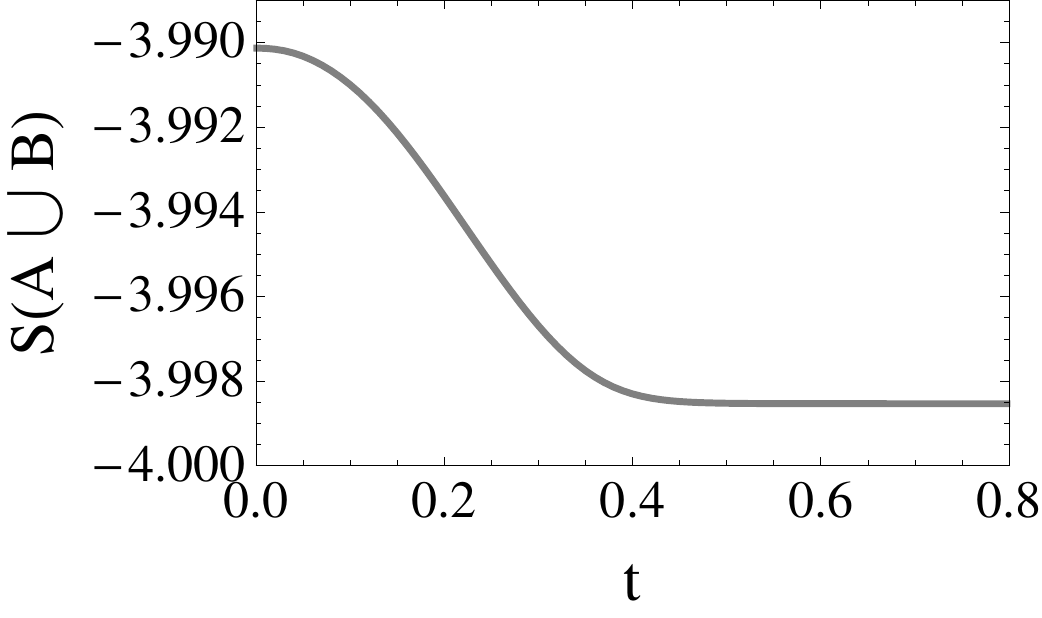} & 
\includegraphics[width=7.0cm]{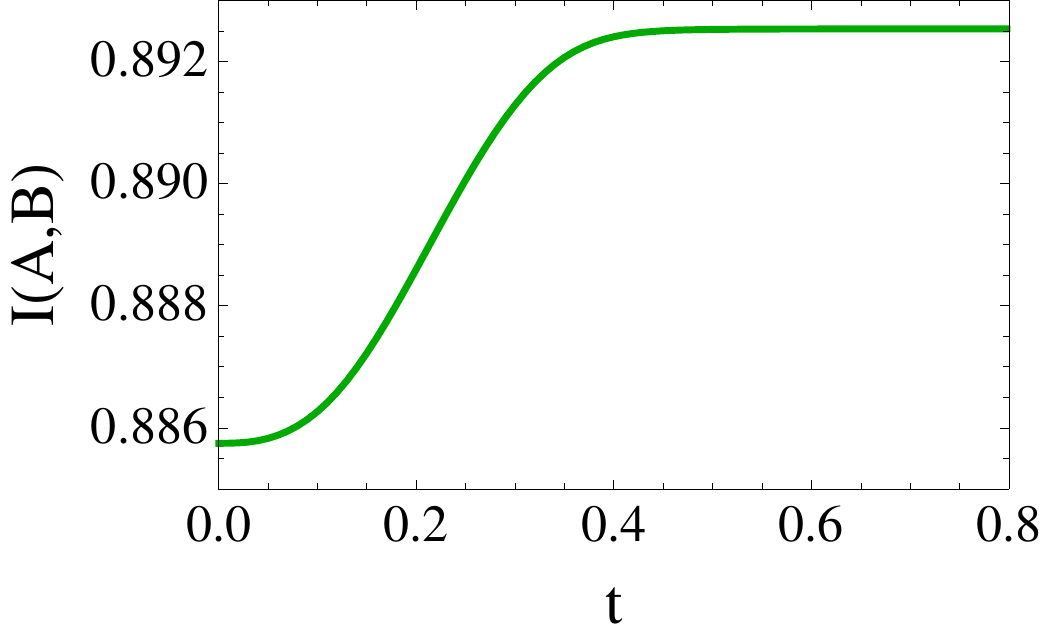} \\
\end{tabular}
\caption{\it (Left) Renormalized entanglement entropy of $A\, \cup B$ as a function of time. (Right) Mutual information of $A$ and $B$ as a function of time. 
We consider a system with $d=3$ within the linearized background of Refs.~\cite{Megias:2015tva,Megias:2016vae}, with the same configuration as in Fig.~\ref{fig1:d2}.}
\label{fig2:d2}
\end{figure*}

Refs.~\cite{Megias:2015tva,Megias:2016vae} give a solution of the background equations of motion on the gravity side in the case $d\ge 2$ by considering a linearization of the system. This approach turns out to be equivalent to linearized hydrodynamics, as it is valid as long as $|T_L - T_R| < |T_L + T_R|$. Using this background, we may compute numerically the entanglement entropy for any number of dimensions by following the procedure of Sec.~\ref{sec:numerics}. The result for $d=3$ and moderate values of $\ell$ is shown in Figs.~\ref{fig1:d2} and \ref{fig2:d2}.~\footnote{We have computed the renormalised entanglement entropy for $d=3$ with the subtraction of the divergent term $S^{div} = \frac{1}{2G\epsilon}$.} These figures display that in  contrast to the case $d=2$ studied in Sec.~\ref{sec:numerics}, the {\it `conservation'} of entanglement entropies between $t=0$ and $t=\infty$ given by (\ref{eq:SASBconservation}) turns out to be not valid for $d=3$, at least within the linearization procedure chosen.~\footnote{Note, however, that the deviations from conservation displayed in Fig.~\ref{fig1:d2} (right) may potentially be explained as a linearization artefact.} As a consequence, the possible existence of a universal behavior for the time evolution of entanglement entropies analogous to (\ref{eq:fAuniversal}) is not obvious in this case. However, the increase of the mutual information with time $\partial_t I(A,B) \ge 0$, cf. Eq.~(\ref{eq:dtI}), appears to be a robust property valid also for $d=3$, and the same can be said for the decrease of $S(A\, \cup B)$ with time. A more detailed study of these  issues will be addressed elsewhere.

\section{Conclusion and outlook}
\label{sec:Conclusion} 

In this work we have studied a holographic model for far-from-equilibrium dynamics that describes the time-dependent properties of energy flow and information flow of two thermal reservoirs initially isolated. In this system, a universal steady state develops, described by a boosted black brane.  A relevant observable that provides physical insight into the evolution of the system is the entanglement entropy, which measures the information flow between two subsystems. By using the exact solution for $d=2$ provided in~\cite{Bhaseen:nature}, we have studied the time evolution of the entanglement entropy, and characterized some universal properties of the quenching process. We also studied the time evolution of mutual information and found it to monotonically grow in time. 

In section \ref{sec:AnalyticalResults}, after a brief overview of velocity bounds for entropy spread and increase, we have investigated the matching procedure outlined in section \ref{sec:Matching} in more detail, showing that in certain circumstances an analytical solution is possible. This allowed us to prove the validity of the universal formula \eqref{eq:fAuniversal} in the appropriate low temperature limits. In subsection \ref{sec:Tsunami}, we then investigated the increase rates of entanglement entropy obtained using the numerical and analytical results of the previous sections. We find that both averaged and momentary entanglement entropy increase and decrease rates are bounded by the speed of light \eqref{bound2}. While this bound is close to being saturated for intervals that are large compared to the scale set by the temperature, this is not the case for smaller intervals, where the universal formula \eqref{eq:fAuniversal} becomes a good approximation, see again figure \ref{fig::ratebound}. This indicates that the shockwave in our setup, which in many ways is similar to a local quench, mimics an entanglement tsunami for large interval sizes $\ell$, leading to a linear entropy increase with the appropriate rate. For small $\ell$ however, the universal behaviour \eqref{eq:fAuniversal} with its characteristic S-shape takes over. We refer to this as an `entanglement tide'. As discussed in section \ref{sec:Higher-Dimensions}, it will be very interesting to study these questions for analogous systems in higher dimensions, where the speed of light, the entanglement velocity $v_E$ and the butterfly velocity $v_B$ are not equivalent any more. This may help to get a better understanding of the mechanisms related to entanglement tsunamis.

Apart from the monogamy of mutual information and strong subadditivity, other inequalities involving a large number of subsystems have been proven in the static case, see \cite{Bao:2015bfa}. In section \ref{sec:Inequalities}, we have studied various  entanglement entropy inequalities, which were proposed for up to $n=5$ intervals, in the present 
time-dependent system. What we found was that the inequalities proven in \cite{Bao:2015bfa} also hold in the time-dependent system under consideration in this paper, at least in all cases that we numerically checked. However, we found that the signs of four- and five-partite information are not definite in this holographic system, in contrast to the results of  \cite{Alishahiha:2014jxa,Mirabi:2016elb}. As the bulk metric investigated in this paper is a vacuum solution everywhere, and hence trivially satisfies the most common energy conditions, we did not have any a priori reason to expect encountering a violation of the entanglement entropy inequalities of \cite{Bao:2015bfa}. It 
may hence be an interesting possibility for future research to check the validity of these inequalities for time-dependent bulk spacetimes that violate, for example, the null energy condition (NEC), similarly to what was done for strong subadditivity in \cite{Prudenziati:2015cva,Allais:2011ys,Callan:2012ip,Caceres:2013dma}. With this paper, we also upload the numerical code used to obtain the results of section \ref{sec:Inequalities} to the arXiv. We hope that this will facilitate future research in this direction.

One of the possible further directions of investigation is suggested by the elegant analytical behaviour of the entanglement entropy in the small temperature limit. It is known that low-energy behaviour of ballistic, quantum-mechanical models is well described by conformal field theories. For a thermal state this means that in the low-temperature regime of lattice model may be approximated by a thermal state of a CFT since that is a situation in which lower part of energy spectrum determines properties of the theory, as more excited states are not occupied. Therefore, we presume that the simple universal evolution of the entanglement entropy we observe should be as well visible in lattice (i.e.~tensor network or exact diagonalisation) calculations. It will be interesting to compare to that kind of models, as local quenches in such systems have recently drawn some attention, see for example \cite{Bohrdt:2016vhv}. Moreover, it is conceivable that further physically observables can be computed in that low-temperature limit.

Finally, comparisons to non-equilibrium hydrodynamics may provide further useful information. Recent work on this includes \cite{Romatschke:2017vte}.  Universal structures in a holographic model of non-equilibrium steady states, which are spatial analogues of quasinormal modes, have recently been considered in \cite{Sonner:2017jcf}.

\section*{Acknowledgements}
We would like to thank Irene Amado, Martin Ammon, Christian Ecker, Daniel Grumiller, Veronika Hubeny, Razieh Pourhasan, Koenraad Schalm, San-Jin Sin, Henk Bart, \'Alvaro V\'eliz-Osorio and Amos Yarom for enlightening discussions. The work of EM is supported by Spanish MINECO under Grant FPA2015-64041-C2-1-P, by the Basque Government under Grant IT979-16, and by the Spanish Consolider Ingenio 2010 Programme CPAN (CSD2007-00042). The research of EM is also supported by the European Union under a Marie Curie Intra-European Fellowship (FP7-PEOPLE-2013-IEF) with project number PIEF-GA-2013-623006, and by the Universidad del Pa\'{\i}s Vasco UPV/EHU, Bilbao, Spain, as a Visiting Professor. MF was supported by NCN grant 2012/06/A/ST2/00396. DF was supported by an Alexander von Humboldt Foundation fellowship. PW would like to thank the Faculty of Physics of Jagiellonian University in Cracow, where large parts of this work were done, for its hospitality.


\bibliographystyle{JHEP}
\bibliography{library}

\end{document}